\documentclass[11pt]{article}
\usepackage{comment}
\usepackage[hidelinks]{hyperref}
\usepackage{enumitem}
\usepackage{amsmath,amssymb,epsf,cite,graphicx,subfigure}
\usepackage{amsmath,amsthm, tikzsymbols, dsfont, amsfonts}
\usepackage{algorithm}
\usepackage{algorithmic}
\usepackage{forloop}
\usepackage{comment}
\usepackage{xcolor}
\usepackage{graphicx}
\usepackage{dsfont}
\usepackage{soul}
\newcommand{\R}{\mathbb{R}}

\allowdisplaybreaks

\newcommand{\Grad}{\nabla}

\newcommand{\EE}{\mathbb{E}}
\usepackage{comment}
\usepackage{xcolor}

\DeclareMathOperator*{\argmin}{arg\,min}

\numberwithin{equation}{section}

\definecolor{airforceblue}{rgb}{0.36, 0.54, 0.66}
\newcommand{\beq}{\begin{equation}}
\newcommand{\eeq}{\end{equation}}

  \theoremstyle{definition}

\usepackage{scalerel}
\usepackage{stackengine,wasysym}

\newcommand\reallywidetilde[1]{\ThisStyle{%
  \setbox0=\hbox{$\SavedStyle#1$}%
  \stackengine{-.1\LMpt}{$\SavedStyle#1$}{%
    \stretchto{\scaleto{\SavedStyle\mkern.2mu\AC}{.5150\wd0}}{.6\ht0}%
  }{O}{c}{F}{T}{S}%
}}

\begin{document}
\baselineskip=15.5pt
\pagestyle{plain}
\setcounter{page}{1}
\begin{center}
{\LARGE \bf Renormalizing Diffusion Models}
\vskip 1cm

\textbf{Jordan Cotler$^{1,a}$ and Semon Rezchikov$^{2,3,b}$}

\vspace{0.5cm}

{\it ${}^1$ Society of Fellows, Harvard University, Cambridge, MA 02138, USA \\}
{\it ${}^2$ Department of Mathematics, Princeton University, Princeton, NJ 08544, USA\\}
{\it ${}^3$ Institute for Advanced Study, Princeton, NJ 08540, USA\\}

\vspace{0.3cm}

{\tt  ${}^a$jcotler@fas.harvard.edu, ${}^b$semonr@princeton.edu\\}

\medskip

\end{center}

\vskip1cm

\begin{center}
{\bf Abstract}
\end{center}
\hspace{.3cm} 
We explain how to use diffusion models to learn inverse renormalization group flows of statistical and quantum field theories. Diffusion models are a class of machine learning models which have been used to generate samples from complex distributions, such as the distribution of natural images. These models achieve sample generation by learning the inverse process to a diffusion process which adds noise to the data until the distribution of the data is pure noise. Nonperturbative renormalization group schemes in physics can naturally be written as diffusion processes in the space of fields. We combine these observations in a concrete framework for building ML-based models for studying field theories, in which the models learn the inverse process to an explicitly-specified renormalization group scheme. We detail how these models define a class of adaptive bridge (or parallel tempering) samplers for lattice field theory. Because renormalization group schemes have a physical meaning, we provide explicit prescriptions for how to compare results derived from models associated to several different renormalization group schemes of interest. We also explain how to use diffusion models in a variational method to find ground states of quantum systems. We apply some of our methods to numerically find RG flows of interacting statistical field theories.  From the perspective of machine learning, our work provides an interpretation of multiscale diffusion models, and gives physically-inspired suggestions for diffusion models which should have novel properties.

\newpage

\tableofcontents

\section{Introduction}
\label{sec:intro}

\subsection{Preliminaries}

Generative machine learning algorithms try to learn a complex, high-dimensional distribution 
\begin{equation}
\mu = p(x) \,dx\,,\quad x \in \R^D\,,
\end{equation}
by finding optimal parameters $\theta^*$ for a family of distributions 
\begin{equation}
p_\theta(x) \,dx\,,
\end{equation}
which is defined either explicitly in terms of a density $p_\theta(x)$ or implicitly in terms of a sampling scheme.  Typically the distribution $p(x)$ to be learned is in a sense `naturally occurring', for instance a distribution over natural images. For many such distributions, each sample $x$ organizes into a \emph{field}; for example, an image is a function $\vec{\phi}: \{1,...,N\}\times\{1,...,N\} \to \R^3$, and $x = \{\vec{\phi}_{i,j}\}_{i,j=1}^N$ are the values of $\phi$ on a lattice embedded in $\R^2$.  Here each lattice site can be viewed as a pixel.  Thus, in the case of images, a distribution $\mu_N$ over such $\vec{\phi}$\,'s may be viewed as a finite-dimensional approximation to a hypothetical infinite-dimensional distribution 
\begin{align}
\label{E:R2toR31}
\mu = \frac{1}{Z} \,P[\,\vec{\phi}\,]\,\mathcal{D}\phi = \frac{1}{Z}\,e^{-S[\,\vec{\phi}\,]} \,\mathcal{D}\vec{\phi}\,,\quad \vec{\phi}: \R^2 \longrightarrow \R^3\,,
\end{align}
where a fixed $\vec{\phi}(y) = \big(\phi_1(y), \phi_2(y), \phi_3(y)\big)$ for $y \in \mathbb{R}^2$ provides an image at infinite resolution.  For simplicity, it is easier to think about black and white images which are naturally described by a single field $\phi(y)$ and would have an associated hypothetical infinite-dimensional distribution
\begin{align}
\label{E:R2toR1}
\mu = \frac{1}{Z} \,P[\phi]\,\mathcal{D}\phi = \frac{1}{Z}\,e^{-S[\phi]} \,\mathcal{D}\phi\,,\quad \phi : \R^2 \longrightarrow \R\,.
\end{align}

In statistical field theory, one studies infinite-dimensional distributions akin to the one above, where the field $\phi(y)$ for $y \in \mathbb{R}^2$ corresponds to e.g.~the magnetization of a magnet at position $y$, the temperature of a metal at position $y$, the density of a plasma at position $y$, etc.  Then the `action' $S[\phi]$ would be derived from a physical model.  For example, a model called ``scalar $\phi^4$ theory'' which has action
\begin{align}
\label{E:O1}
S[\phi] = \int_{\mathbb{R}^2} dy \, \left(\frac{1}{2}\,\nabla \phi \cdot \nabla \phi + \frac{1}{2}\,m^2 \phi^2 + \frac{\lambda}{4!} \,\phi^4 \right)
\end{align}
is central to the study of statistical properties of magnets.

In lattice field theory, one constructs computational methods for sampling from finite-dimensional analogs of distributions like $P[\phi] = \frac{1}{Z}\, e^{-S[\phi]}$ for $S[\phi]$ in~\eqref{E:O1} in order to compute expectation values of observables of interest, such as the average magnetization of a magnet over certain lattice sites. Sampling from such distributions is challenging due to their multi-modal and high-dimensional nature, and there have been attempts to apply machine-learning techniques, e.g.~flow-based generative modeling, to improve sampling \cite{boltzmann_generators_science, flows_for_field_theory, equivariant_gauge_sampler, miranda_sampling}. In the related setting of quantum field theory, one wishes to gain access to expectation values in the quantum \emph{ground state}, which in certain physically relevant cases turns out to be representable by a non-negative distribution (see Section~\ref{Subsec:realvalued}).  In this quantum ground state setting an explicit form of the density is generally unavailable, but one can search for the quantum ground state via a minimization problem directly analogous to the problem of variational inference~\cite{ferminet_1}.

Due to the high-dimensional nature of all of the above problems, one is required make approximations, and to incorporate prior knowledge about the distribution $p(x)$ or $P[\phi]$ into the structure of the model. In generative machine learning, it has been found that one can successfully learn high-dimensional distributions by leveraging a stochastic differential equation (SDE)
\begin{equation}
dx = f(x)\,dt + \sigma(x)\,dB_t
\end{equation}
which induces an on evolution $p_t(x)$ from the desired complex distribution $p_0 = p$ to a simpler distribution $p_\infty$, which is typically Gaussian or uniform.  There is an associated \emph{inverse} SDE 
\begin{equation}
dx = \left(-f(x) - \sigma^T \sigma(x) \,s(x)\right) dt + \sigma(x)\, dB_t
\end{equation}
where $s(x)$ is the \emph{score function}
\begin{equation}
s(x) = \Grad p_t(x)\,,
\end{equation}
which induces a flow from $p_\infty$ to $p_0$. In \emph{diffusion modeling}, one variationally parameterizes the \emph{score function} $s(x)$ via a neural network with parameters $\theta$, so that $s_\theta(x)$ is used as a proxy.  For a fixed $\theta$, one uses the approximate score function and the inverse SDE to generate approximate samples from $p_0$ given easy samples from $p_\infty$. By minimizing a functional computed in terms of these samples, one can improve upon the initial guess of $\theta$ so as to make $s_\theta(x)$ closer to the true $s(x)$, and thus improve the sampler. Here, implicit priors about the distribution $p(x)$ are encoded into the diffusion SDE, the construction of the neural network $s_\theta$, and the particular details of the training scheme, which involves e.g.~a choice of numerical SDE solver.

In the statistical physics of fields, it is well-established that the \emph{renormalization group} provides important analytical and numerical insights into the study of the pertinent distributions~\cite{Zinn-Justin2013-az}.  The basic conceptual insight is that our description of a physical system is contingent on the precision of our measurement apparatus.  In particular, if we can only probe a system in a coarse-grained manner which is insensitive to system properties that are sufficiently small, then our \textit{effective} description of the system can neglect those short-distance properties.  A useful conceptual example is a fluid; if your measurement apparatus cannot resolve individual atoms, then the equations you use to describe what you can measure about the fluid need not model the individual atoms but rather may treat the system as a continuum.  At a more practical level, the question the renormalization group seeks to answer is: given a probability distribution $P[\phi] = \frac{1}{Z}\,e^{-S[\phi]}$ over fields $\phi(y)$, how do we find a \textit{coarse-grained} description of the distribution adequate to accurately reproduce expectation values of observables smoothed over large distances?  We emphasize that for various physical models of interest, fundamental notions such as phase transitions and critical exponents are studied or even defined in terms of the renormalization group, and a wealth of physical intuition is available about the behavior of the renormalization group in the context of specific models (see~\cite{Cardy2015-qt} for a modern treatment of the subject).

Since the renormalization group instantiates a type of coarse-graining process, it may come as no surprise that it can be described in terms of a heat-flow-like equation for probability densities on the space of fields, and also via a stochastic-differential equation for the field variables $\phi$. There are different choices of \emph{renormalization group scheme}, corresponding to different ways of implementing the coarse-graining process; essentially all reasonable countinuous-space renormalization group schemes are characterized by their associated SDE, as we show in Section~\ref{sec:lattice-rg}. To illustrate, consider the \emph{Carosso scheme}~\cite{Carosso2020}
\begin{equation}
\label{E:C1}
d\phi_t(y) = \Delta \phi_t(y)\,dt + dB_t(y)\,,
\end{equation}
where $t$ is a `time' that parameterizes how much we have coarse-grained our system, $\Delta$ is the (spatial) Laplacian, 
and $dB_t(y)$ describes a $t$-dependent Gaussian random field.  In fact~\eqref{E:C1} induces a deterministic flow $P_t[\phi]$ where $P_0[\phi] = P[\phi]$.  We will elaborate on the details and meaning of these equations later on, but for now it suffices to say that $P_t[\phi]$ for $t > 0$ provides our desired coarse-grained description of the physical system described by $P[\phi]$.

How should we conceive of the coarse-grained probability functional $P_t[\phi] = \frac{1}{Z_t}\,e^{-S_t[\phi]}$\,?  Going back to the action in~\eqref{E:O1}, a $t$-dependent version would look like
\begin{align}
\label{E:O1flowed1}
S_t[\phi] = \int_{\mathbb{R}^2} dy \, \left(\frac{a(t)^2}{2}\,\nabla \phi \cdot \nabla \phi + \frac{1}{2}\,m(t)^2 \phi^2 + \frac{\lambda(t)}{4!} \,\phi^4 + \cdots \right)
\end{align}
where the mass $m$ and $\lambda$ are now functions of $t$, along with a new multiplicative factor $a$ which sets the scale of the kinetic term, and the $\cdots$ terms represent new kinds of interaction terms that we pick up in the process of the flow.  The functions $a(t)$, $m(t)$, and $\lambda(t)$ (as well as those suppressed in the $\cdots$ terms) are quantities of physical interest, since they capture information about what we can actually measure with a particular precision of our measurement apparatus.  They comprise the $t$-dependence of the Taylor coefficients of the log-density of $\mu_t = P_t[\phi]\,\mathcal{D}\phi$. Terminologically, the distribution $\mu_0 = P_0[\phi]\,\mathcal{D}\phi$ is called the \emph{UV distribution}, while the distribution $\mu_\infty = P_\infty[\phi]\,\mathcal{D}\phi$ is called the \emph{IR distribution}.  These names derive from the fact that UV light is short-wavelength, and IR light is long-wavelength; accordingly the UV distribution captures short-distance physics and the IR distribution captures long-distance physics.

Putting some of the above ideas together, if we wish to sample from the distribution associated to a statistical field theory, we can use a finite-dimensional discretization of the SDE associated to a renormalization group scheme to define a diffusion model which \emph{learns the inverse SDE to the renormalization group flow}. In this paper we will describe in detail how to design such models. Crucially, many of the parameters of these models have a direct physical interpretation, a property that is difficult to achieve in typical black-box applications of machine learning methodologies to physics. In particular, we will show how one can \emph{compare} results from different choices of models in this class which are based on different RG schemes (i.e.~different SDEs) by deriving appropriate ways of renormalizing the fields (i.e.~rescaling model variables) such that comparisons can be made. We believe that the physical interpretability of these machine learning models will make debugging and modifying such models more straightforward in physical applications, and possibly inspire new techniques in more general machine learning domains. We present empirical results for this class of models, and more broadly aim to make clear the precise connection between diffusion models and the renormalization group for the benefit of physicists and machine learning practitioners, in particular so that ideas and techniques can be communicated between these different communities of researchers.

\subsection{A sketch of bridging the physics and ML perspectives}

\begin{figure}[t!]
    \centering
\includegraphics[width = \textwidth]{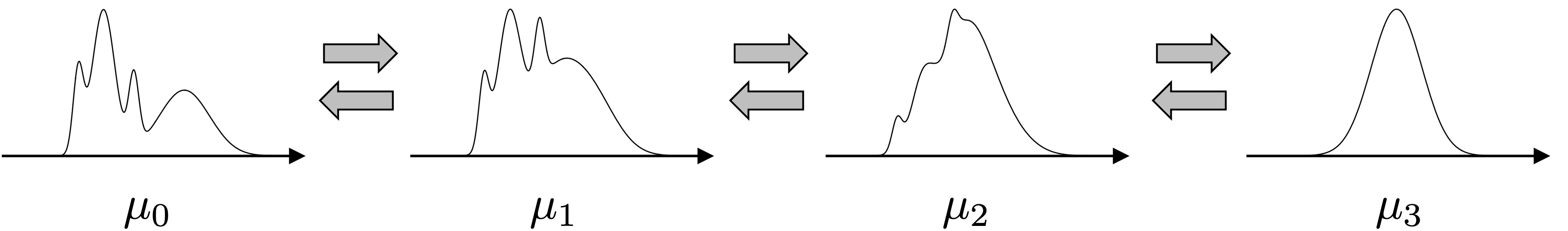}
    \caption{\textit{Parallel tempering}. It is difficult to sample from multimodal distributions like $\mu_0$ , due to slow mixing times of Markov chains. A common strategy is to connect the distribution of interest to a simple unimodal distribution from which sampling is straightforward, and then use coupled MCMC chains (or diffusion models, or normalizing flows) to speed up sampling from $\mu_0$.  The process connecting $\mu_0$ to a simple distribution $\mu_3$ is often a noising or averaging process, and can sometimes be given a physical meaning in the setting of the renormalization group. Multimodality of $\mu_0$ is in physical settings often associated with phase transitions or approximately conserved quantities like the topological charge.}
    \label{fig:paralleltempering}
\end{figure}

From the perspective of lattice field theory, a physicist might have invented diffusion modeling as follows. The distributions $\mu_0$ pertinent for, say, studying quantum chromodynamics, are complicated and often forced to be multimodal due to the existence of physical quantities such as the \emph{topological charge} of a gauge field, which contribute to \emph{critical slowing down} of direct MCMC sampling of $\mu_0$ \cite{WOLFF199093}. A well-known MCMC technique is \emph{bridge sampling} or \emph{parallel tempering} \cite{Swendsen1986, Gelman1998, Neal1996}: one runs parallel MCMC samplers for a family of distributions $\mu_t$\,, $t=0,..., T$ where the \emph{bridge} $\mu_T$ is a simpler (``high-temperature'') distribution which may be unimodal or approximately Gaussian.  See Figure~\ref{fig:paralleltempering} for an illustration.  The method works by using samples from each $\mu_t$ as proposals for each $\mu_{t-1}$, with acceptance ratios chosen such that the joint Markov chain stills samples from $\mu_0 \times \cdots \times \mu_T$. Because $\mu_T$ is simple, e.g.~unimodal, the sampler for $\mu_T$ mixes quickly, which in some cases improves the overall mixing of the sampler from $\mu_0$, despite the additional overhead cost of running several parallel samplers. 

A key problem is engineering a natural sequence of $\mu_t$'s which form a good bridge.  Fortunately, physics gives us a \emph{preferred bridge}: the flow of $\mu_0$ given by the renormalization group (see e.g.~\cite{fisher1998renormalization, rosten2012fundamentals}). Unfortunately, the probability densities of the $\mu_t$ (the images of $\mu_0$ under RG flow) are not known explicitly; this presents a problem since to construct an MCMC sampler for a $\mu_t$, one needs to know the density of that $\mu_t$ somewhat explicitly.  One can make approximate guesses for the densities of $\mu_t$ using analytical techniques, and indeed such methods can be used to improve sampling (see e.g.~\cite{PhysRevD.94.114502}). Alternatively, one can simply \emph{parameterize} the densities of the $\mu_t$ (or at least their score functions, which are sufficient to construct MCMC samplers) and instead perform a \emph{variational, adaptive variant of bridge sampling} where one \emph{learns} the samplers for the distributions in the bridge. Implementing this physically natural idea would lead directly to the models described in this paper.

However, the diffusion SDEs associated to renormalization group schemes have correlations between model variables, in contrast to the original SDEs used in the machine-learning community for image modeling. Indeed, discretizing the Carosso scheme~\eqref{E:C1} on a 2D lattice, we get the SDE
\begin{align}
    \label{eq:discretized-Caroso-SDE}
    d\phi_{i,j} &= (\phi_{i-1,j} + 2 \phi_{i,j} - \phi_{i+1,j})\,dt + (\phi_{i,j-1} + 2 \phi_{i,j} - \phi_{i,j+1})\,dt + (dB_{t})_{i,j} \\ &= (\Delta \phi)_{i,j}\,dt + (dB_t)_{i,j} \nonumber
\end{align}
where here $\Delta$ is the discrete Laplacian.  The above can be written more schematically as
\begin{equation}
\label{E:SDEschematic1}
d\phi_{i,j} = f(\phi_{i,j}, \phi_{i-1,j}, \phi_{i+1,j}, \phi_{i,j-1}, \phi_{i,j+1},\,t) + (dB_t)_{i,j}
\end{equation}
for the variables $\{\phi_{i,j}\}_{i,j}$.  In contrast, diffusion models commonly use SDEs where all variables diffuse independently \cite{ddpg, song2020score}, namely
\begin{align}
\label{E:SDEschematic2}
d\phi_{i,j} = f(\phi_{i,j}, t)\,dt + (dB_t)_{i,j}\,
\end{align}
for some variables $\{\phi_{i,j}\}_{i,j}$.  The key difference between~\eqref{E:SDEschematic1} and~\eqref{E:SDEschematic2} is that in the former equation the dynamics of the variables are coupled via the SDE, whereas in the latter equation each variable evolves according to its own decoupled SDE.

\begin{figure}[t!]
    \centering
    \includegraphics[width = .8\textwidth]{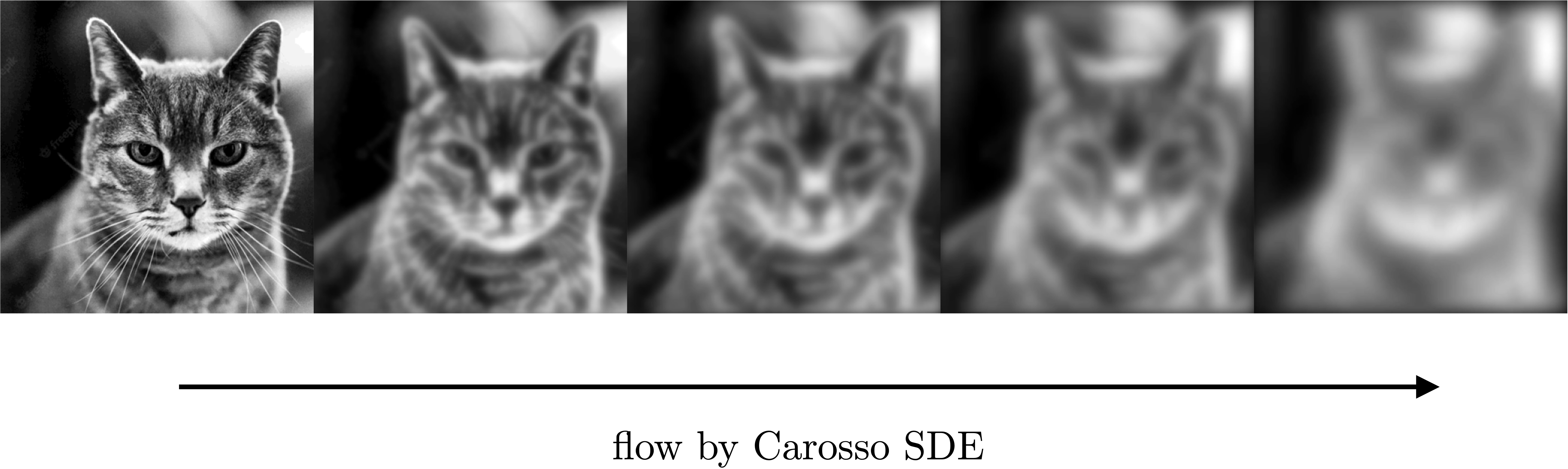}
    \caption{Evolving an image of a cat using the Carosso SDE.}
    \label{fig:catflow1}
\end{figure}

In the discretized Carosso SDE~\eqref{eq:discretized-Caroso-SDE}, the variables diffuse such that high-frequency fluctuations of the discretized field $\phi_{i,j}$ (e.g.~pixels of an image) will be turned to data-independent noise.  As the evolution continues, progressively lower frequencies will in turn be transformed into data-independent noise; see Figure~\ref{fig:catflow1}.  Thus, in the inverse process, low-frequency modes of the field are generated before the high-frequency modes, which may seem intuitively desirable. Remarkably, the image-modeling community has already studied precisely the SDE~\eqref{eq:discretized-Caroso-SDE} for this reason~\cite{blurring_diffusion_models}, and in general, many \emph{multiscale} diffusion models have been developed \cite{subspace_diffusion, mallat_diffusion, cascased_diffusion, iterative_refinrement} and found to improve computational efficiency and state-of-the-art performance. Even in commercial image-generation systems such as Midjourney, one generates high-resolution image models by training a hierarchy of image-conditioned diffusion models which first generate low-resolution images and then repeatedly upsample the images to higher resolution.

We suspect that various architectures of multiscale diffusion models correspond rigorously to implementations of various renormalization group schemes, and in turn the physical understanding of renormalization group schemes may give rise to new designs and diagnostic techniques in machine learning.
From the perspective espoused here, the active use of multiscale diffusion modeling in machine learning suggests that the generative modeling community has independently understood various computational advantages of the renormalization group, and that there are more insights to be mined in this direction. 

\begin{figure}[t!]
    \centering
    \includegraphics[width = \textwidth]{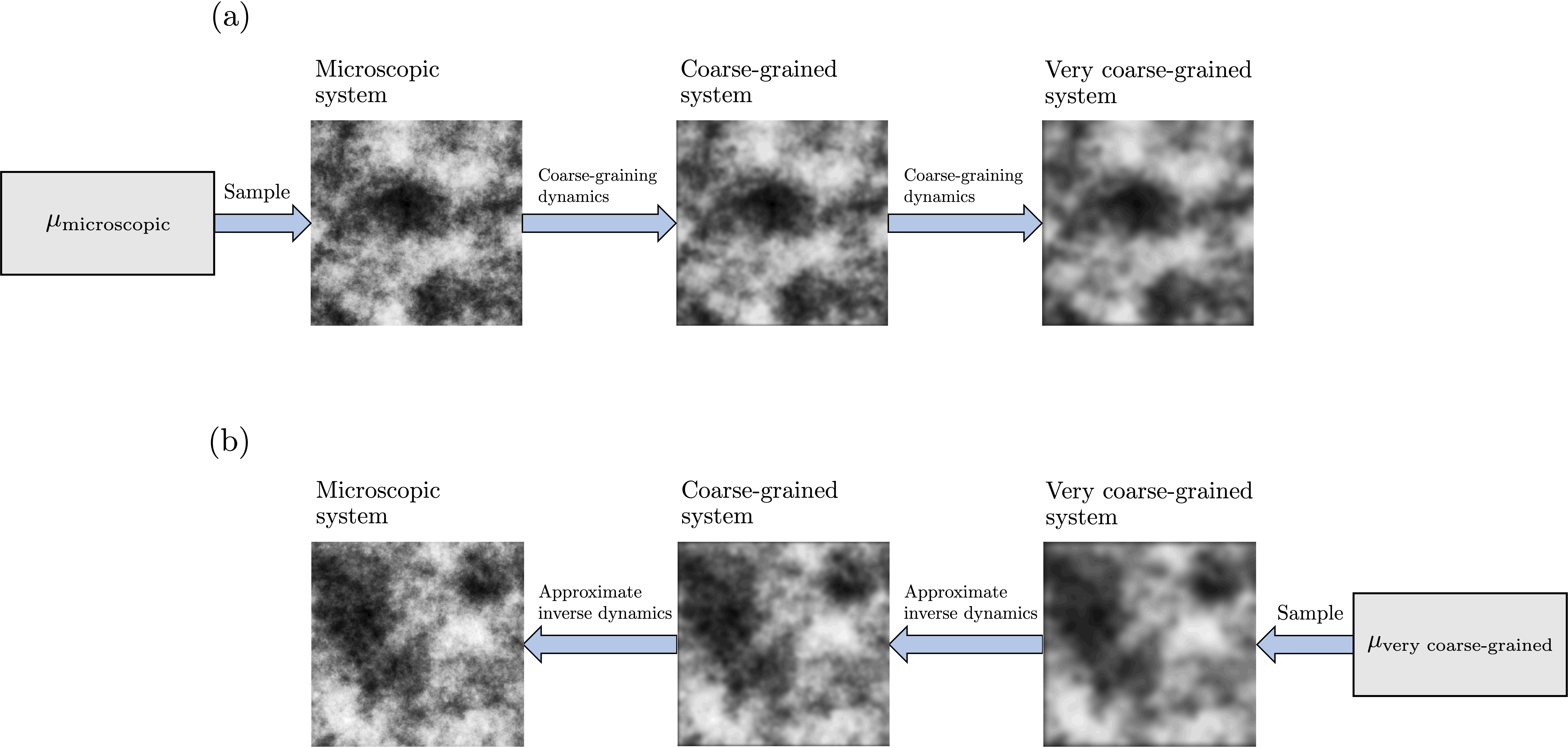}
    \caption{Physical systems tend to be describable by models of varying levels of accuracy (often associated with a hierarchy of length scales). Fine-grained models are accurate but rarely tractable, while coarse-grained models are efficient but have unknown error terms.  (a) In the renormalization group framework, it is possible to instantiate coarse-graining dynamics on a microscopic model to obtain quantifiable coarse-grained systems.  (b) It is also possible to use coarse-grained systems in conjunction with an approximate inverse to coarse-graining dynamics to improve the efficiency of simulations of a microscopic system. Developing a machine learning toolkit to improve coarse-grained physics models in a way that is physically interpretable and justified is a significant problem at the intersection of physics and computer science. This paper studies the case of simple statistical field theories, using the connection between the renormalization group and diffusion models, but the problem is not limited to this setting.}
    \label{fig:simulate_fig}
\end{figure}

We also argue that ideas from diffusion modeling in machine learning can be imported into numerical methods for physics problems. For a typical physical system, one can write down an ansatz for a \emph{microscopic action} or \emph{Hamiltonian}, as well as natural \emph{coarse-graining} procedures (also called \textit{renormalization group flows}). These procedures define \emph{implicit} intermediate-scale models, about which we may know certain pertinent parameters but for which we typically do not know an explicit action or Hamiltonian. In a number of cases, coarse-graining procedures can be chosen to make the simulation of an estimated coarse-grained system \emph{simpler}. Thus, at a very high level, one has two processes that one can run in parallel:
\begin{enumerate}
    \item Simulate the microscopic system, and use the generated dynamics to improve estimates for the dynamics of coarse-grained versions of the system (see Figure~\ref{fig:simulate_fig}(a)); and 
    \item Simulate coarse-grained versions of the microscopic system (more efficiently than the corresponding microscopic system), and use the resulting samples to improve the determination or simulation of the microscopic system. This is done by learning an \emph{approximate inverse} to the coarse-graining procedure (see Figure~\ref{fig:simulate_fig}(b)).
\end{enumerate}
This schematic methodology could theoretically be applied to many modeling problems, such as those in condensed matter, high-energy physics, quantum chemistry, and plasma physics. We hope that the theoretically clean connection between diffusion modeling and the renormalization group for scalar field theories presented in this paper can be expanded (by taking non-relativistic limits, incorporating fermions and gauge symmetries, and elaborating on the connections to Effective Field Theory) to create a robust, physically-interpretable machine learning toolkit for numerical physics problems.  We will begin to develop such a toolkit in this paper.

\subsection{History}

The present paper posits a systematic identification between multiscale diffusion models and the renormalization group. The ideas behind diffusion modeling in machine learning arise from several different, interrelated perspectives \cite{sohl-dickstein_deep_2015, vincent_connection_2011, song2019generative}, in part motivated by the idea of the renormalization group \cite{sohl-dickstein_deep_2015}. These approaches were later unified \cite{ddpg, song2020score} and organized according to a class of SDEs where the model variables diffused in an \emph{uncorrelated} fashion, as in~\eqref{E:SDEschematic2}, and thus do not implement renormalization group transformations in the sense explained in this paper. Later, variants of multiscale diffusion modeling were proposed by many authors~\cite{blurring_diffusion_models, mallat_diffusion, cascased_diffusion, iterative_refinrement, spectral_diffusion_processes} and the topic remains an active area of research.  

On the other hand, in lattice field theory, several lines of work were inspired by L\"uscher's pioneering paper~\cite{luscher_trivializing_maps}, which pointed out that when trying to sample from lattice $SU(N)$ gauge theory, which is a distribution $\mu$ over a compact manifold, there is an implicit characterization of a flow $f_t$ called the \emph{Wilson Flow} such that $(f_t)_*\mu$ limits to a uniform distribution (i.e.~a Haar measure). In fact, this flow ends up being a lattice analog of the gradient flow of the Yang-Mills functional~\cite{atiyah_bott}. As such, it is a PDE for the fields with a gradient flow interpretation which has as its highest-order term the Laplacian operator acting on the fields. It was noticed that this flow smooths out the fields and gives rise to an operation akin to the renormalization group. This led to a significant amount of work applying the `gradient flow' to lattice field theory methods which has had many applications, e.g.~to set the scale of lattice QCD simulations~\cite{gradient_flow_scale_determination}.  The precise meaning of the `gradient flow' was eventually clarified by Carosso~\cite{carosso_thesis}, who explained that there is a valid renormalization group scheme which is a stochastic analog of the gradient flow. In this paper we will take the Carosso scheme~\cite{Carosso2020} as a starting point and use it to illustrate the connection between diffusion modeling and the renormalization group since the scheme has a simple lattice implementation (which indeed turns out to have been studied in the generative modeling community~\cite{blurring_diffusion_models}).  We will then turn to more general schemes.

In a different line of work following L\"uscher's idea, there are a series of papers which apply machine learning methods to attempt to \emph{learn} a trivializing flow in order to sample more efficiently from UV distributions~\cite{flows_for_field_theory, flows_for_gauge_theory, 2211.07541}. These works can be viewed as part of a broader trend to apply black-box machine learning methods to problems in field theory, molecular dynamics, and condensed matter~\cite{carleo2019machine}. Flow-based generative modeling methods have demonstrated improved integrated autocorrelation statistics for samplers~\cite{flows_for_field_theory} and improved exploration of topological charge sectors~\cite{flows_for_gauge_theory}, but evaluation of their effectiveness at larger lattice sizes is still ongoing~\cite{2211.07541}. Crucially, trivializing flows are \emph{not unique}, and in these ML-based methods, the flow learned \emph{has no physical interpretation}, making it more challenging to interpret, debug, and reason about the models.

\subsection{Summary of this paper}

In this paper we propose a framework for multiscale diffusion and flow-based generative modeling techniques, and then specialize to problems in statistical field theory in which the models \emph{learn the inverse renormalization group flow} of a field theory. We begin in Section~\ref{sec:inference-background} with a basic review of Bayesian inference and Monte Carlo sampling. In Section~\ref{sec:diffusion-models}, we review the mathematical ideas and basic implementation details of diffusion models. In Section~\ref{sec:lattice-rg}, we review the basic ideas of lattice field theory, and outline how the renormalization group is required to perform calculations of physical quantities of interest. In Section~\ref{sec:renormalizing-diffusion-models}, we explain how to design a diffusion model based on the Carosso scheme and how to use it to sample from the UV distribution of a lattice scalar field theory. Importantly, we will explain how the renormalization group requires that physical quantities are extracted from \emph{rescalings} of the variables in this model. This idea may be of particular interest to researchers in machine learning.  In \emph{multiscale} diffusion models, the corruption process affects different frequencies at different rates, and the ideology of the renormalization group suggests to \emph{appropriately} rescale all variables to \emph{focus in} on the ``interesting'' part of the distribution. In particular, there is an associated diffusion process for the \emph{renormalized} fields which has qualitatively different behavior from typical diffusion processes, because it can have fixed points which are highly non-Gaussian (although a generic distribution flows to a Gaussian fixed point). For multiscale diffusion models associated to renormalization group schemes, one can derive an approximation to the SDE for the renormalized fields, which can be used in physical modeling applications and may be helpful in applications to generative modeling.

It is a fundamental fact of statistical field theory that certain quantities of physical interest, such as critical exponents, are independent of the choice of renormalization group scheme. Later on in Section~\ref{sec:renormalizing-diffusion-models}, we write down the SDEs for general Wegner-Morris exact renormalization group schemes.  As a special case, we describe a diffusion model based on the more traditional Polchinski exact renormalization group scheme~\cite{polchinski1984renormalization}.  Next we explain how renormalization group ideas indicate how to rescale variables to meaningfully compare results between the schemes. 
 We provide formulae and intuitions for how different multiscale diffusion SDEs should be compared on an equal footing, which are results of potential interest to the ML community.

In Section~\ref{sec:groundstates} we explain how to use diffusion modeling to provide an explicit class of variational methods for learning the ground states of quantum field theories. The basic challenge for variational methods for computing ground states is that, as noted by Feynman~\cite{feynman1987proceedings}, without an inspired parameterization, the gradients of the variational problem blow up because they are primarily sensitive to unimportant high-frequency fluctuations. In low-dimensional quantum field theories, variational methods based on the renormalization group (e.g.~DMRG \cite{white1992density}, MPS~\cite{perez2006matrix}, and MERA~\cite{vidal2008class}) have been helpful for computing quantities of physical interest like correlation functions and critical exponents (see e.g.~\cite{pfeifer2009entanglement}).  However, these methods have not yet been successfully adapted to general higher-dimensional field theories, in part due to computational limitations. The multiscale-diffusion-based ansatz presented in this paper is a novel, physically grounded variational method for finding the renormalization group flow of the ground state of a quantum field theory, and we provide some intuition about why the exploding gradients problem observed by Feynman may be less significant for these methods.

In Section~\ref{sec:numerics}, we show tests of a numerical implementation of one of our renormalization group-based algorithms for sampling from statistical field theories.  Our results show that flow-based methods optimized with objectives derived via the renormalization group can learn the RG flows of basic scalar field theories such as $\phi^4$ theory. 

We end the paper in Section~\ref{sec:discussion} with a discussion of directions for further research, including subjects that might benefit from being studied using multiscale diffusion models. 

In Appendix~\ref{App:diffusion_lit_review}, we provide a brief review of literature on diffusion models.  In Appendix~\ref{App:functionalderivs}, we describe some technical points about the lattice discretization of functional derivatives that we utilize in the body of the text.  Finally, in Appendix~\ref{App:ERGreview}, we give a short review of the Exact Renormalization Group.

\section{Variational inference and sampling}
\label{sec:inference-background}

\subsection{Basic objects}

As we recall from the Section~\ref{sec:intro}, in typical modeling problems we are confronted with a distribution of interest
\begin{equation}
p(x)\,dx = \frac{1}{Z}\,e^{f(x)}\, dx\,,\quad x \in \R^D,
\end{equation}
where $Z$ is a normalization constant such that $\int dx\,p(x) = 1$. We may have access to this distribution for a finite number of empirical samples $\{x_i\}_i$, or through an expression for the log-density $f(x)$; typically, the normalization constant $Z$ is unknown and challenging to compute. 

In variational inference, we try to approximate the distribution of interest $p$ by a variational family of distributions
\begin{equation}
p_\theta(x)\,dx = \frac{1}{Z_\theta}\,e^{f_\theta(x)}\,dx\,,\quad x \in \R^D,
\end{equation}
which we may represent explicitly via $f_\theta(x)$ or implicitly via a method to sample from $p_\theta(x)$. A convenient quantity one can use to represent $p_\theta(x)$ is the 
\emph{score function}
\begin{equation}
    s_\theta(x) = \Grad_x \log  p_\theta(x) = \Grad_x f_\theta(x)\,.
\end{equation}
The score function is a vector-valued function that points in the direction of increasing probability mass. The score function has many convenient properties, including that it is independent of the normalization constant $Z_\theta$, and that many sampling algorithms only require knowledge of the score function to draw samples from $p_\theta(x)$. Models that represent $p_\theta(x)$ via its score function are called \emph{score-based} models.

To find the optimal parameter $\theta$ in the variational family, we must optimize an objective function depending on $\theta$. Many such functions with convenient properties can be built out of the \emph{KL divergence}
\begin{equation}
\label{eq:kl-divergence}
    \text{KL}(p | q) = \int dx\, p(x) \log \!\left(\frac{p(x)}{q(x)}\right) = - H(p) - \mathbb{E}_{x \sim p(x)}\left[\log q(x)\right].
\end{equation}
Here $H(p) := - \int dx \, p \log p$ is the \emph{entropy} of $p$.
The KL divergence is non-negative and only zero when $p=q$. A related quantity is the so-called Fisher divergence\footnote{There are several related but not identical quantities involving expectation values of squares of gradients, including the Fisher information $\int dx \, p(x | \theta) (\partial_\theta p(x | \theta))^2$. The
Fisher divergence of \eqref{eq:fisher-divergence} is connected to the KL divergence by de Bruijn's identity (see \cite[Definition 2.5]{dasgupta2008asymptotic}) and by the Cram\'{e}r-Rao bound in the case of location parameters $\theta$ (see \cite[Section 17.7]{cover1999elements}), and possesses entropy-like monotonicity properties under taking i.i.d.~sums \cite{courtade2016monotonicity}.}
\begin{equation}
\label{eq:fisher-divergence}
D_F(p | q) = \int dx \,p(x) |\Grad \log p(x) - \Grad \log q(x)|^2\,,
\end{equation}
which is the basis of the \emph{score matching} techniques reviewed in the next section. 

Most natural objective functions that can be used for variational inference involve an intractable integral over $x$, and thus require \emph{gradient estimators} for an optimal $\theta$ to be found using (stochastic) gradient descent. For example, if we choose the objective function to be $L_1(\theta) = \text{KL}(p_\theta | p)$, then
\begin{equation}
\label{eq:kl-gradient-identity-1}
    \Grad_\theta \text{KL}(p_\theta | p)  = \mathbb{E}_{x \sim p_\theta}[ \Grad_\theta f(x)(f_\theta(x) - f(x)+1)]
\end{equation}
where we have used the identity $\Grad p = p \,\Grad \log p$.  On the other hand, if we choose the objective function to be $L_2(\theta) = \text{KL}(p | p_\theta)$, then 
\begin{equation}
    \Grad_\theta \text{KL}(p | p_\theta) = - \mathbb{E}_{x \sim p(x)}\left[ \Grad_\theta f_\theta\right]\,.
\end{equation}
Thus to optimize $L_1(\theta)$ using stochastic gradient descent we need access to samples from $p_\theta$, whereas to optimize $L_2(\theta)$ we need access to samples from $p$. Moreover, to optimize $L_1(\theta)$ we need access to $f(x)$ and $f_\theta(x)$, which may or may not be possible depending on the model architecture and the nature of the problem; to optimize $L_2(\theta)$ we need access to $\Grad_\theta f_\theta$ which may also be challenging for certain score-based models. The usage of the backwards or forwards KL divergences $\text{KL}(p_\theta | p)$ and $\text{KL}(p | p_\theta)$ in variational inference tends to have different tradeoffs due to their zero-forcing and zero-avoiding behavior respectively \cite{murphy2012machine}, although each version can be useful in practice. Moreover, there are a wide variety of methods for writing gradient estimators in certain classes of models which may improve the variance of the estimators \cite{mnih2016variational}.

\subsection{Variational lower bounds}
\label{subsec:variationalowerbounds}
\newcommand{\zz}{\mathbf{z}}
 One can use variational inequalities to produce new optimization objectives involving new variables. Introducing a new random variable $z$ with probability density $r(z)$, we have by Bayes' rule
\begin{equation}
p_\theta(x) = \int dz\, p_\theta(x|z) \,r(z)\,.
\end{equation}
In many cases, $z$ can be chosen in a meaningful way, such that the form of $p_\theta(x | z)$ and $r(z)$ can be taken as providing a \emph{definition} of the variational family $p_\theta(x)$.  Then for any auxiliary distribution $q(z|x)$, by multiplying and dividing by $q$ one has 
\begin{equation}
\label{eq:importance-sampling}
p_\theta(x) = \mathbb{E}_{z \sim q(z|x)} \!\left[ \frac{p_\theta(x | z)}{q(z | x)}\right]
\end{equation}
and thus by Jensen's inequality we have the \emph{variational bound}
\begin{equation}
    \label{eq:variational-bound}
    \log p_\theta(x) \geq \mathbb{E}_{z \sim q(z | x)}\!\left[ \log\!\left(\frac{p_\theta(x | z)}{q(z | x)}\right)\right]\,.
\end{equation}
In using Jensen's inequality we have thrown away the term $\text{KL}(q( z | x) | p_\theta( z | x))$, which can thus be thought of as an error term.

One can think of~\eqref{eq:importance-sampling} as the justification for \emph{importance-sampling}: given knowledge of $p_\theta(x | z)$, one can estimate (or \emph{define}) $p_\theta(x)$ by sampling $x$ with some importance weighting distribution $q$. Now, by~\eqref{eq:kl-divergence}, one sees that optimizing $\text{KL}(p | p_\theta)$ is equivalent to optimizing $\mathbb{E}_{x \sim p}[\log p_\theta(x)]$. Unfortunately, Jensen's inequality implies that the importance-sampling  estimator for $\log p_\theta(x)$ is biased downwards, with an error term proportional to how well $q( z | x)$ approximates $p_\theta( z | x)$. In variational inference one typically optimizes the variational lower bound of \eqref{eq:variational-bound} (i.e.~the right-hand side) instead of $\mathbb{E}_{x \sim p}[\log p_\theta(x)]$, using judicious choices of $q( z | x)$ which may themselves be chosen variationally from a family $q_\theta$.

\subsection{Sampling algorithms}
\label{subsec:sampling}

After producing an estimate for the log-probability density $\log p_\theta$ or the score function $s_\theta$ via variational inference, for applications one will need to sample from the corresponding distribution $p_\theta(x)\,dx$. In all methods which facilitate sampling, one defines a stochastic dynamical system such that its states at various time steps are approximate samples from the distribution $p_\theta(x)\,dx$ of interest.  Below we review several prominent samplings methods for sampling from $p_\theta(x)\,dx$\,; in our exposition we omit the $\theta$ subscript since it will not play a role in the sampling methods.

\paragraph{Markov Chain Monte Carlo (MCMC).}
A fundamental method, the \emph{Metropolis-Hastings algorithm}, can be used when one has access to the log probability density $\log p(x)$. The algorithm is parameterized by the choice of \emph{proposal distribution} $Q(x | y)$. The method is slightly simplified when $Q(x | y) = Q(y | x)$, namely the \emph{symmetric} case. In this case, one chooses an initial point $x_0$ from some background distribution, and then runs the following algorithm: \\ [.4cm]
\textbf{initialize} $x_0,\,i=0$ \\
\textbf{if} $i < i_{\max}$ \textbf{do} \\
\indent sample $x' \leftarrow Q(x'|x_i)$ \\
\indent set $x_{i+1} = x'$ \\
\indent with probability $\alpha = \exp\!\big(f(x') - f(x_i)\big) = p(x')/p(x_i)$, replace $i \to i+1$ \\
\textbf{end if} \\
\textbf{return} $x_{i_{\max}}$
\\ \\
When $Q$ is not symmetric, one simply modifies the formula for the \emph{acceptance ratio} $\alpha$ slightly. For large $i_{\max}$\,, under very general conditions \cite{murphy2012machine} the distribution of the variable $x_{i_{\max}}$ will be very close to $p(x)\,dx$. For example, choosing $Q(x | y)$ to be a Gaussian centered at $y$ guarantees convergence independent of $p(x)$; here, one can think of the variance of the Gaussian as a ``temperature'' parameter.

However, while convergence occurs for most choices of a proposal distribution, the \emph{rate of convergence} (how large $i_{\max}$ needs to be for $x_{i_{\max}}$ to be close in distribution to a sample from $p(x)$) and the \emph{decorrelation rate} (a measurement of how how large $k$ needs to be so that $x_i$ and $x_{i+k}$ are independently distributed) are both highly sensitive to the choice of proposal distribution $Q$, and both grow  quickly with the dimension and ``complexity'' of the target distribution $p$. Both of these quantities affect the accuracy and variance of estimates of expectation values $\mathbb{E}_{x \sim p(x)}[g(x)]$ produced via sampling. Thus, the challenge is to adapt the choice of proposal distribution $Q$ to $p$ in order to make sampling fast enough to be practical. 

\paragraph{Score function-based sampling.}
It is possible to sample from a distribution $p(x)\,dx$ without having access to $\log p(x)$, but only relying on the score function $s(x)$.  The basic method in this class is (\emph{unadjusted}) \emph{Langevin sampling}: one initializes $x_0$ from an arbitrary initial distribution, and then forms the stochastic process
\begin{equation}
\label{eq:langevin-sampling}
    x_{i+1} = x_i + \epsilon \,\Grad \log p(x) + \sqrt{2 \epsilon} \,\delta w_i 
\end{equation}
where $\delta w_i \sim \mathcal{N}(0, \mathds{1})$ for each $i$, and uses the fact that under fairly weak regularity conditions $x_i \sim p(x)$ for large $i$.

If there is access to $\log p(x)$, one can perform \emph{adjusted Langevin sampling} which can be viewed as a hybrid between (unadjusted) Langevin sampling and Metropolis-Hastings.  The basic idea is that we can use the stochastic process~\eqref{eq:langevin-sampling} to define a proposal distribution $Q(x|y)$ to be used in the Metropolis-Hastings algorithm.  This proposal distribution is useful because it is tailored to the probability distribution $p(x)\,dx$ of interest.  In particular, we define a $Q(x|y)$ via the following sampling algorithm:
\\ \\
\textbf{initialize} $x_0 = y$ \\
\textbf{for} $i = 0,1,...,N$ \textbf{do} \\
\indent sample $\delta w_i \leftarrow \mathcal{N}(0,\mathds{1})$ \\
\indent set $x_{i+1} = x_i + \epsilon \,\Grad \log p(x) + \sqrt{2 \epsilon}\,\delta w_i$ \\
\textbf{end for} \\
\textbf{set} $x = x_N$ \\
\textbf{return }$x$
\\ \\
Above, $\epsilon$ has to be chosen so that it is sufficiently small, and $N$ has to be chosen so that it is sufficiently large.

\paragraph{Hamiltonian Monte Carlo and related methods.}
The Langevin sampling algorithm is based on the Euler-Maurayama discretization of \emph{overdamped Langevin dynamics}, which is the continuous-time stochastic process defined by the stochastic differential equation (SDE)
\begin{equation}
dx = s(x) \,dt + \sqrt{2} \,dW_t\,.
\end{equation}
Thinking of the log-probability $\log p(x)$ as the negative of a \emph{potential function} $U(x)$, overdamped Langevin dynamics is simply the dynamics of a particle undergoing Brownian diffusion in the potential field $U$. In Brownian diffusion the particle moves diffusively as opposed to inertially, under the potential $U$. One can modify the dynamics to incorporate inertial motion under the potential $U$; the chaotic mixing properties of such dynamics should heuristically improve the convergence rate of the corresponding stochastic process to the equilibrium distribution $p(x)\,dx \propto e^{- U(x)}\,dx$~\cite{neal2011mcmc}, where in this context the score is $s(x) = - \nabla U(x)$. Such a stochastic process is given by the second-order Langevin dynamics: 
\begin{equation}
    \label{eq:second-order-langevin}
    dx = -v \,dt\,,\quad dv = (s(x) - \gamma v) \,dt + \sqrt{2 \gamma} \,dW_t\,.
\end{equation}
Here $\gamma$ is a friction parameter, and one formally recovers overdamped Langevin dynamics by taking $\gamma \to \infty$, where $v$ changes much faster than $x$ and thus equilibrates to a normal distribution centered at $s(x)$ instantaneously. In the above equation, the $-\gamma v \,dt$ term is a \emph{friction} term, the $\sqrt{2 \gamma}\,dW_t$ term is a \emph{noise} term, and the $s(x)\,dt = - \nabla U(x) \,dt$ term corresponds to noiseless \emph{Hamiltonian dynamics} in the potential $U$. The ratio between the noise term and the friction term in~\eqref{eq:second-order-langevin} is precisely chosen such that the stable distribution $\tilde{p}(x, v) \,dx\,dv$ of the dynamics marginalizes to $p(x)\,dx$\,; if the friction term is dropped, this marginalization property no longer holds~\cite{chen2014stochastic}. 

Thus, given only access to the score function $s(x)$, one can sample from $p(x)\,dx$ via \emph{underdamped Langevin sampling} by initializing $x_0$, choosing $v_0$ to be normally distributed around zero, evolving $(x_0, v_0)$ via a discretization of the SDE~\eqref{eq:second-order-langevin} to $(x_N, v_N)$, and then using $x_N$ as an approximate sample. If one has access the log-probability density, one can, as in adjusted Langevin sampling, instantiate a hybrid method with Metropolis-Hastings; this is called the \emph{adjusted underdamped Langevin sampler}. In this hybrid method, the Metropolis acceptance step renders it unnecessary for  
the stationary density of the SDE~\eqref{eq:second-order-langevin} to marginalize to $p(x)\,dx$. Indeed, in this setting one can simply \emph{drop} the friction term (or even both the friction term and the noise term) from the adjusted underdamped Langevin sampler while keeping the Metropolis acceptance step; this gives rise to the \emph{Hamiltonian Monte-Carlo} method which is a mainstay of physical and statistical simulations~\cite{Neal1996, neal2011mcmc}. One can incorporate Riemannian metrics and variable covariance matrices for the noise in attempts to improve convergence of the resulting samplers~\cite{girolami2011riemann}, and there is an evolving and complex theoretical literature comparing variations of these methods (see e.g.~\cite{raginsky2017non, cheng2018underdamped, diffusion-sampling-theory-1}).

\paragraph{Bridge sampling.}
There is a class of methods which, instead of using more complex dynamical systems to define improved proposal distributions $Q(x | y)$, run a series of parallel sampling chains for a collection of different distributions, coupling the chains in some way such that the overall mixing times may be improved. This class of methods includes parallel tempering or replica exchange~\cite{Swendsen1986, marinari1992simulated}, and is connected to bridge sampling~\cite{gelman1998simulating}.  The philosophy behind these methods is as follows.  Consider a sequence or \textit{bridge} of distributions $\mu_0, \mu_1,...,\mu_T$ where $\mu = \mu_0 = p(x)\,dx$ and $\mu_t = p_t(x)\,dx$.  Here $\mu$ the is distribution that we want to sample from, and $\mu_T$ is a distribution that is easy to sample from (e.g.~a unimodal Gaussian), say using an MCMC sampler which converges rapidly and can quickly produce uncorrelated samples.  The distributions $\mu_i$ should in some sense interpolate between $\mu = \mu_0$ and $\mu_T$.  The idea, then, is to use samples from $\mu_T$ to seed a proposal distribution to sample from $\mu_{T-1}$\,, and then to use ensuing samples from $\mu_{T-1}$ to seed a proposal distribution to sample from $\mu_{T-2}$\,, and so on down to $\mu = \mu_0$.

Slightly more explicitly, if $x^{(t)}$ is approximately a sample from $\mu_t$ and if $Q_{t-1}(x|y)$ is a proposal distribution for $\mu_{i-1}$, then we can use $Q_{t-1}(x | x^{(t)})$ as a proposal distribution in MCMC to sample from $\mu_{t-1}$.  This kind of procedure will tend to produce rapidly mixing proposal distributions, insofar as $\mu_{t-1} \approx \mu_t$ for all $t$.  More generally, in a bridge sampling scheme, one runs samplers for all the $\mu_t$ in parallel, transferring samples between the $\mu_t$ according to some appropriate rules such that the stationary distributions of the joint chain marginalizes to $\mu = \mu_0$ upon forgetting samples from the auxiliary distributions in the bridge. The method is improved by choosing an appropriate bridge (i.e.~judicious choices of the $\mu_t$'s) and appropriate proposal distributions $Q_t(x|y)$ (which in this context are called \textit{exchange} distributions) which are adapted to the features of the $\mu_t$'s. It can be helpful to choose $\mu_t$ to have physical meaning: if $\mu_0$ is the distribution of states of a protein at a low temperature, then one can choose the $\mu_t$ to be distributions over states of the protein at progressively higher temperatures as $t$ is increased, so that the whole Markov process simulates denaturation and folding of the protein upon temperature cycling~\cite{sugita1999replica}.

\section{Review of mathematical aspects of Diffusion Models}
\label{sec:diffusion-models}

In this section, we review the fundamental ideas underlying a class of score-function-based hierarchical generative model  called \emph{diffusion models}, which have recently become the state-of-the-art models for image generation, powering commercially available tools like Midjourney and Stable Diffusion~\cite{stablediffusion}.  Some useful reviews include~\cite{song2021maximum, yang_song_blog_1, song2019generative}.

\subsection{Score-based generative modeling}

\subsubsection{General setup}

As mentioned in the previous section, in score-based generative modeling, we attempt to approximate the score $s(x)$ of the true distribution $p(x)\,dx$ via an estimated score $s_\theta(x)$, which suffices to approximately sample from $p(x)$ via unadjusted Langevin sampling. The natural objective function to use in this case is the $\ell^2$ error of the score function under the data, namely
\begin{equation}
\label{eq:fischer-score-divergence}
\frac{1}{2} \,\EE_{x \sim p(x)}\!\left[|s_\theta(x) - \Grad \log p(x) |^2 \right]
\end{equation}
which is just (one half of) the Fisher divergence $D_F(p | p_\theta)$ from~\eqref{eq:fisher-divergence}. However, one does not know the true value of $\log p(x)$ in many practical settings.  Following Hyv\"{a}rinen~\cite{hyvarinen_estimation_2005}, it is prudent to integrate by parts and subtract a $\theta$-independent constant to arrive at the equivalent objective function
\begin{equation}
\label{eq:hyarvinen-loss}
\EE_{x \sim p(x)}\!\left[ \Grad \cdot s_\theta(x) + \frac{1}{2}|s_\theta(x)|^2\right] \approx \frac{1}{N} \sum_{i=1}^N \!\left(\Grad \cdot s_\theta(x_i) + \frac{1}{2}|s_\theta(x_i)|^2\right)\,,
\end{equation}
where $x_i$\,, $i=1,..., N$ are samples from $p(x)$ such as e.g.~natural images or microscopic configurations of the spins of a magnet.  Importantly,~\eqref{eq:hyarvinen-loss} manifestly does not require explicit knowledge of $\log p(x)$.

\subsubsection{Introducing a noise scale}

We would like to ameliorate the common issue of $p(x)$ being concentrated along a submanifold of the data space $\R^D$.  A useful approach is to smooth out the probability density $p(x)$ by noising it, forming the new probability density 
\begin{equation}
\label{eq:adding-noise}
p_{\sigma}(\tilde{x}) = \int dx\,p(x)\,\frac{e^{-\frac{1}{2 \sigma^2}|x - \tilde{x}|^2}}{(2\pi \sigma^2)^{D/2}}\,.
\end{equation}
A convenient way to sample from $p_{\sigma}(\tilde{x})$ is to take empirical samples $x_i \sim p(x)$ and produce synthetic samples $\tilde{x}_i \sim x_i + \sigma z_i$, where $z_i \sim \mathcal{N}(0,\mathds{1})$. One can now try to estimate the score of $p_\sigma$ for some small value of $\sigma$, and use samples from $p_\sigma$ as an approximation to samples from $p$. Score matching for $p_\sigma$ is easier \cite[Figure~1]{song2019generative}, and the corresponding smoothing $p_\sigma^{\text{data}}$ of the empirical data distribution $p^{\text{data}}(x) = \frac{1}{N} \sum_{i=1}^N \delta(x-x_i)$ has a natural gradient estimator. In particular, suppose we parameterize the score function via
\begin{equation}
s_\theta(x) = \frac{1}{\sigma^2} \left( W^T\cdot\text{sigmoid}(W\cdot x+ b) + c - x\right), \,\,\text{ where }\theta = (W, b, c)\,
\end{equation}
and where the dimensions of $W$ are $D' \times D$, the dimensions of $b$ are $D' \times 1$, and the dimensions of $c$ are $D \times 1$.  We suppose that $D' < D$.  Then minimizing~\eqref{eq:fischer-score-divergence} using $\tilde{x} \sim p^{\text{data}}_\sigma(\tilde{x})$ in place of $x \sim p(x)$, and also replacing $p$ with $p_\sigma$ inside the expectation,  i.e.~minimizing
\begin{align}
\label{E:somereplacements1}
\frac{1}{2} \,\EE_{\tilde{x} \sim p_\sigma^{\text{data}}(\tilde{x})}\!\left[|s_\theta(\tilde{x}) - \Grad \log p_{\sigma}(\tilde{x}) |^2 \right],
\end{align}
turns out to be equivalent (in the sense that the gradients agree up to a global rescaling) to minimizing the quantity~\cite{vincent_connection_2011}
\begin{equation}
\label{E:newtominimize1}
\mathbb{E}_{x \sim p(x)}\, \mathbb{E}_{\substack{\!\!\!\!\tilde{x} \,=\, x \,+\, \sigma z \\ z \,\sim\, \mathcal{N}(0, \sigma I)}}\!\left[|W^T\cdot\text{sigmoid}(W\cdot\tilde{x}+b) + c - x|^2\right].
\end{equation}
It is useful to slightly reorganize the above quantity in order to better interpret it.  Let us define the nonlinear map
\begin{equation}
\mathcal{E}(\tilde{x}) := \text{sigmoid}(W\cdot\tilde{x}+b)
\end{equation}
which implements an \emph{encoding} of the noised sample into a lower-dimensional (i.e.~$D'$-dimensional) representation, and further define the nonlinear map
\begin{equation}
\mathcal{D}(y) = W^T \cdot y + c
\end{equation}
which will serve as a \textit{decoding} from the lower-dimenension representation back to the higher-dimensional (i.e.~$D$-dimensional) one.  Then we can rewrite~\eqref{E:newtominimize1} as
\begin{equation}
\mathbb{E}_{x \sim p(x)}\, \mathbb{E}_{\substack{\!\!\!\!\tilde{x} \,=\, x \,+\, \sigma z \\ z \,\sim\, \mathcal{N}(0, \sigma I)}}\!\left[|\mathcal{D} \circ \mathcal{E}(\tilde{x}) - x|^2\right].
\end{equation}
The above is readily interpreted as the objective function for a \emph{denoising autoencoder}: the composition $\mathcal{D} \circ \mathcal{E}$ is a $1$-layer neural network that attempts to \emph{reconstruct} the original sample $x$ from the noised sample $\tilde{x}$ by encoding and subsequently decoding it through a lower-dimensional space. 

More broadly,~\eqref{E:newtominimize1} for general $s_\theta(x)$ is equivalent to the \emph{denoising score function objective}~\cite{vincent_connection_2011} 
\begin{align}
\label{eq:denoising-score-function-objective}
\mathbb{E}_{x \sim p(x)}\, \mathbb{E}_{\substack{\!\!\!\!\tilde{x} \,=\, x \,+\, \sigma z \\ z \,\sim\, \mathcal{N}(0, \sigma I)}}\!\left[ \left|s_\theta(\tilde{x}) - \Grad_{\tilde{x}} \log\!\left(\frac{e^{-\frac{1}{2\sigma^2}|x - \tilde{x}|^2}}{(2\pi\sigma^2)^{D/2}}\right)\right|^2\right] = \mathbb{E}_{x \sim p(x)}\, \mathbb{E}_{\substack{\!\!\!\!\tilde{x} \,=\, x \,+\, \sigma z \\ z \,\sim\, \mathcal{N}(0, \sigma I)}}\!\left[ \left|s_\theta(\tilde{x}) - \frac{1}{\sigma^2}(x - \tilde{x})\right|^2\right]\,.
\end{align}
We see that optimizing over $\theta$ pressures $s_\theta(\tilde{x})$ to \emph{point in the direction of the denoising of }$\tilde{x}$, namely the vector $x - \tilde{x}$. 

\subsubsection{Introducing multiple noise scales}

So far we have considered smoothing out $p(x)$ by introducing a noise scale $\sigma$.  Here we generalize that approach by taking inspiration from bridge sampling.  Specifically we introduce a series of noise scales
\begin{equation}
0 <\sigma_1 < \cdots < \sigma_T
\end{equation}
and corresponding noised distributions 
\begin{align}
p_{\sigma_i}(\tilde{x}) := \int dx\,p(x)\,\frac{e^{- \frac{1}{2 \sigma_i^2}|x - \tilde{x}|^2}}{(2\pi \sigma_i^2)^{D/2}}\,.
\end{align}
If $s_\theta^{(i)}(x)$ is a score function associated with each $p_{\sigma_i}$, we can optimize a weighted sum of score matching losses
\begin{equation}
\label{eq:averaged-denoising-score-function-objective}
\sum_{i=1}^L \lambda_i\, \mathbb{E}_{\tilde{x} \sim p_{\sigma_i}^{\text{data}}(\tilde{x})}\!\left[\left|\Grad_x \log p_{\sigma_i}(\tilde{x}) - s_\theta^{(i)}(\tilde{x})\right|^2\right]
\end{equation}
where the $\lambda_i > 0$ are weighting parameters (usually taken to be $\lambda_i = \sigma^2_i$).  The above equation is a multiscale generalization of~\eqref{E:somereplacements1}.  
Crucially, taking inspiration from~\eqref{eq:denoising-score-function-objective}, one should parameterize $s_\theta^{(i)}(x)$ via a \emph{deep convolutional neural network}; the score functions should be a composition of functions which attempt to find denoisings of noised samples. In practice, one parameterizes the $s_\theta^{(i)}(x)$ via a U-Net with skip connections~\cite{song2019generative, ronneberger2015u}, a neural network architecture which has proven successful in image modeling. Conveniently, one can sample from $p_{\sigma_1}(x)$ via annealed Langevin sampling: one samples from a uniform distribution, which is a good approximation to $p_{\sigma_T}$ for large $\sigma_T$, then takes several updates as in \eqref{eq:langevin-sampling} with the score function $s_\theta^{(T)}(x)$, then with $s_\theta^{(T-1)}(x)$, and so on until $s_\theta^{(1)}(x)$. This adapted bridge sampling method then naturally resolves many problems associated with the slow Langevin sampling of the (typically multimodal and highly oscillatory) distribution $p(x)$ with score function $\Grad \log p(x) \approx s_\theta^{(1)}(x)$.

There is a related class of models, referred to as \emph{denoising diffusion models}, which come from a physics-inspired framework \cite{sohl-dickstein_deep_2015} based on minimizing a KL divergence \eqref{eq:kl-divergence} instead of a score matching objective. In this setting, given an initial sample $x_0 \sim p(x_0)$, one defines noised variables 
\begin{equation}
\label{eq:ou-noise}
x_{i+1} = \sqrt{1 - \beta_i} \,x_i + \sqrt{\beta}_i \,z_i\,, \quad z_i \sim \mathcal{N}(0,\mathds{1})\,.
\end{equation}
The above is a discrete-time Ornstein-Uhlenbeck process such that $x_T$ for large $T$ is approximately distributed according to a mean-zero Gaussian with identity covariance. (This is in contrast with the noising process~\eqref{eq:adding-noise}, which converges to a uniform distribution.)  Slightly overloading notation (albeit in a standard manner), we let $p(x_{i+1}|x_i)$ denote the probability density of $x_{i+1}$ given $x_{i}$.  We still reserve $p(x_0)$ for denoting the probability density we wish to model.  In the above setting of~\eqref{eq:ou-noise}, $p(x_{i+1}|x_i)$ is a Gaussian in $x_{i+1}$.  It turns out that if $\beta_i$ is small then $p(x_{i}|x_{i+1})$ is approximately Gaussian in $x_i$; this is the discrete-time analog of the fact that there exists a reverse SDE for every given SDE, which we will recall in more detail shortly. Writing $p(x_0)$ as
\begin{equation}
p(x_0) = \int dx_1 \cdots dx_{T-1} \, dx_T \, p(x_0 | x_1) p(x_1 | x_2)\cdots p(x_{T-1}|x_T)\, p(x_T)
\end{equation}
where $p(x_T)$ denotes the probability associated with the random variable $x_T$, we observe that $p(x_T)$ is approximately Gaussian (in $x_T$) and each of the $p(x_{i}|x_{i+1})$ is approximately Gaussian (in $x_i$).  This suggests that we can approximate $p(x_0)$ by
\begin{equation}
p_\theta(x_0) = \int dx_1 \cdots dx_{T-1} \, dx_T \, p_\theta(x_0 | x_1) p_\theta(x_1 | x_2)\cdots p_\theta(x_{T-1}|x_T)\, p_\theta(x_T)
\end{equation}
where we have chosen the parameterizations in the following manner.  We let $p_\theta(x_T) = \frac{1}{(2\pi)^{D/2}}\,e^{-\frac{1}{2}|x_T|^2}$ so that it is $\theta$-independent.  We further let $p_\theta(x_i | x_{i+1})$ be a Gaussian in $x_i$ whose mean and variance depend, in a prescribed way, on $x_{i+1}$ and the parameters $\theta$.  This set of $p_\theta(x_i | x_{i+1})$'s comprise a useful variational family, as discussed in Section~\ref{subsec:sampling}.  If we have judiciously chosen the form of the $p_\theta(x_i | x_{i+1})$'s and optimized for the parameters $\theta$, then an approximate sampling algorithm for $p_\theta(x_0) \approx p(x_0)$ is as follows:
\\ [.07cm]
\textbf{sample} $x_T \leftarrow \mathcal{N}(0,\mathds{1}) = p_\theta(x_T)$ \\
\textbf{for} $i=T,T-1,...,1$ \textbf{do} \\
\indent sample $x_{i-1} \leftarrow p_\theta(x_{i-1} | x_i)$ \\
\textbf{end for} \\
\textbf{return} $x_0$
\\ \\
The above is nice since it only entails sampling from Gaussian distributions at every step.

Following the ideas of variational inference, one recalls (as in Section~\ref{subsec:variationalowerbounds}) that minimizing $\text{KL}(p(x_0) | p_\theta(x_0))$ agrees with maximizing $\mathbb{E}_{x_0 \sim p(x_0)} [ \log p_\theta(x_0)]$, which by~\eqref{eq:variational-bound} has the lower bound (see~\cite{sohl-dickstein_deep_2015})
\begin{align}
\mathbb{E}_{x_0 \sim p(x_0)} [ \log p_\theta(x_0)] &\geq \mathbb{E}_{x_0,x_1,...,x_T \,\sim\, p(x_0,x_1,...,x_T)} \! \left[ \log \frac{p_\theta(x_0,x_1, ..., x_T)}{p(x_1,..., x_T | x_0)}\right] \nonumber \\ &= \mathbb{E}_{x_{T-1},x_T \,\sim\, p(x_{T-1},x_T)} \!\left[ \log p_\theta(x_T) + \log \frac{p_\theta(x_{T-1} | x_T)}{p(x_{T} | x_{T-1})}\right]
\end{align}
where $p(x_0, x_1, ..., x_T)$ denotes the joint distribution over all the $x_i$, and $p(x_{T-1}, x_T)$ denotes the joint distribution over only $x_{T-1}$ and $x_T$\,.  Via a rearrangement, the quantity in the second line can be rewritten further as  
\begin{equation}
\label{E:rearrange1}
\mathbb{E}_{x_0,x_1,...,x_T \,\sim\, p(x_0,x_1,...,x_T)} \!\left[\text{KL}(p(x_T | x_0)|p_\theta(x_T)) + \sum_{i=2}^T \text{KL}(p(x_{i-1} | x_i, x_0) | p_\theta(x_{i-1} | x_i)) - \log p_\theta(x_0 | x_1) \right]
\end{equation}
where here $p(x_{i-1}|x_i, x_0)$ denotes the probability of $x_{i-1}$ conditioned on fixed $x_i$ and $x_0$.  A useful feature of the above equation is that each of the terms (except the first, which disappears upon taking gradients with respect to $\theta$ since $p_\theta(x_T)$ is $\theta$-independent) can be computed analytically, since all the quantities involved are Gaussians which can be explicitly computed.  For this purpose is it convenient to use the identity $p(x_{i-1}|x_i, x_0) = \frac{p(x_{i-1},\, x_i|x_0)}{\int dx_{i-1}\,p(x_{i-1},\,x_i|x_0)}$, where $p(x_{i-1}, x_i|x_0)$ is an explicitly computable Gaussian using~\eqref{eq:ou-noise}.

As an explicit example of how to parameterize the $p_\theta(x_{i-1}|x_i)$'s, consider
\begin{align}
p_\theta(x_{i-1} | x_i) &= \frac{1}{(2 \pi \beta_i^2)^{D/2}}\,\exp\!\left(-\frac{1}{2 \beta_i^2}  \left|x_{i-1} - \mu_\theta^{(i)}(x_i)\right|^2\right) 
\end{align}
where $\mu_\theta^{(i)}(x_i)$ is given by
\begin{align}
\mu_\theta^{(i)}(x_i) &= \frac{1}{\sqrt{1-\beta_i}}\left(x_i - \epsilon_\theta^{(i)}(x_i)\right) \prod_{j=1}^{i-1}\frac{1}{1-\beta_j}\,.
\end{align}
Here $\epsilon_\theta^{(i)}(x_i)$ is parameterized by a suitable neural network. 
 Then optimizing over~\eqref{E:rearrange1}, one recovers a particular weighted sum of score matching objectives from a KL-minimization framework. This method led to a practically successful implementation (DDPG) of this paradigm in~\cite{ddpg}.

\subsection{Taking the continuum limit} 

From the previous discussion we have seen that both score matching and variational inference involve the choice of a process which adds noise to a sample (see~\eqref{eq:adding-noise} and \eqref{eq:ou-noise}, respectively), as well as the learning of a corresponding denoising process (parameterized by $s_\theta$ and $\epsilon_\theta$, respectively). In each case, the noising process occurs in discrete time steps.  As such, it is natural to take the continuum limit in time and consider forwards SDEs of the form
\begin{equation}
\label{eq:forwards-sde}
    dx = b(x, t)\,dt  + \sigma(x, t)\,dW_t
\end{equation}
where we will use the \emph{It\^{o}} formulation.

For example, rewriting the noise process \eqref{eq:adding-noise} as 
\begin{equation}
    x_i = x_{i-1} + \sqrt{\sigma^2_i - \sigma^2_{i-1}}\,\,z_{i-1}\,, 
\end{equation}
where we have set $\sigma_0 = 0$, we see that the continuum limit of this process (where $\sigma_i = \sigma(i/N))$ as $N \to \infty$ for some function $\sigma: [0,1] \to \R$)
\begin{equation}
\label{eq:exploding-variance-sde}
dx = \sqrt{ \frac{ d \sigma(t)^2}{dt}} \,dW_t\,.
\end{equation}
On the other hand, the noising process \eqref{eq:ou-noise} has continuum limit\footnote{Here, this continuum limit is taken in the sense that $\beta_i = (1/N) \beta(i/N)$ for some function $\beta: [0,1] \to \R$, as $N \to \infty$. One shows that the continuum limit is~\eqref{eq:bounded-variance} using the Taylor expansion of $\sqrt{1-\epsilon}$ in $\epsilon$.}
\begin{equation}
    \label{eq:bounded-variance}
    dx = -\frac{1}{2} \beta(t)\,x\, dt + \sqrt{\beta(t)}\,dW_t\,. 
\end{equation}

If we set $x(0)$ to be distributed as $p(x) =: p_{t=0}(x)$, and the process $x(t)$ satisfies \eqref{eq:forwards-sde}, then $x(T)$ will be distributed according to some distribution $p_{t=T}(x)$, and more generally the marginal distribution of $x(t)$ will be $p_{t}(x)$ for $t \in [0,1]$. One can also write down a natural reverse SDE with time paramater $\tilde{t} = T-t$, which flows $p_{t=T}(x)$ to $p_{t=0}(x)$, namely~\cite{ANDERSON1982}
\begin{equation}
\label{eq:reverse-sde}
dx = \Big(-b(x, \tilde{t}\,) + \Grad \cdot \left( \sigma(x, \tilde{t}\,) \cdot \sigma^T(x, \tilde{t}\,)\right) + \sigma(x, \tilde{t}\,) \cdot \sigma^T(x, \tilde{t}\,) \Grad \log p_{\tilde{t}}(x)\Big) d\tilde{t}+ \sigma(x, \tilde{t}\,) \,dW_t\,.
\end{equation}
What we mean is that if $x(\tilde{t})$ is a process with $x(0)$ distributed as $p_{\tilde{t}=0}(x) = p_{t=T}(x)$, and $x$ satisfies \eqref{eq:reverse-sde}, then $x(T)$ will be distributed according to $p_{\tilde{t}=T}(x) = p_{t=0}(x)$, and more generally $x(\tilde{t})$ will be distributed acording to $p_{t=T-\tilde{t}}(x)$. Notice that the above process involves the score function $\Grad \log p_t(x)$.  If $b(x,t)$ and $\sigma(x, t) = g(t)$ are scalars, then~\eqref{eq:reverse-sde} simplifies to 
\begin{equation}
dx = \big(-b(x,\tilde{t}) + g(\tilde{t})^2\,\Grad \log p_{\tilde{t}}(x)\big)\,d\tilde{t} + g(\tilde{t}) \,dW_t\,.
\end{equation}

With the above notations at hand, we are prepared to take the continuous-time limit of the objective function~\eqref{eq:averaged-denoising-score-function-objective}, giving us 
\begin{equation}
\label{eq:continuum-loss}
\mathbb{E}_{t \sim [0,T]}\,\mathbb{E}_{x \sim p_t^{\text{data}}(x)}\!\left[\lambda(t) \,|\Grad \log p_t(x) - s_\theta(x,t)|^2\right]
\end{equation}
where $\lambda(t) \geq 0$ is a weighting function. One can optimize this objective with respect to $\theta$ by discretizing time, simulating the forwards process via a discretization, and then estimating \eqref{eq:continuum-loss} using the Hy\"{a}rvinen loss \eqref{eq:hyarvinen-loss}. Then, having found appropriate parameters $\theta$, one can sample from $p(x) \approx p_{\theta, t= 0}(x)$ by simulating the flow of the reverse SDE~\eqref{eq:reverse-sde}. The analog of the identity in~\eqref{eq:denoising-score-function-objective} comes from the fact that we can rewrite~\eqref{eq:continuum-loss}, up to a $\theta$-independent constant, as~\cite{song2021maximum} 
\begin{equation}
\label{eq:continuum-score-matching-objective}
    \mathbb{E}_{t \sim [0,T]}\!\left[\lambda(t) \,\mathbb{E}_{x \, \sim \, p(x)}\mathbb{E}_{x' \, \sim \, p_{t,0}(x'|x)}[\Grad_{x'} \log p_{t,0}(x'|x) - s_\theta(x', t)]\right]
\end{equation}
where $p_{t,0}(x'|x)$ denotes the probability density that the stochastic process produces $x'$ at time $t$ given that it started at $x$ at time $0$.  The kernel $p_{t,0}(x'|x)$ is called the time-$t$ \emph{transition kernel} of the noising process~\eqref{eq:forwards-sde}, and in many cases it can be computed analytically. In such cases, using~\eqref{eq:continuum-score-matching-objective} as the objective has computational advantages, for instance one can make use of an explicit formula for the kernel $p_{t,0}(x'|x)$ as opposed to solving the SDE by other means.

We can provide a variational bound for for $\text{KL}(p_{t=0} | p_{\theta, t=0})$ in terms of~\eqref{eq:continuum-loss} (or equivaently~\eqref{eq:continuum-score-matching-objective}) via an application of the Girsanov theorem~\cite{benton2022denoising, song2021maximum}. For example, if $b(x, t) = b(t)$ and $\sigma(x, t) = g(t)$ are scalars, then letting $\lambda(t) = \frac{1}{2}\,g(t)^2$  in~\eqref{eq:continuum-loss} we have
\begin{equation}
    \text{KL}(p_{t=0} \,|\, p_{\theta, t=0}) \leq \frac{1}{2} \,\mathbb{E}_{t \sim [0,T]}\!\left[ g(t)^2 \,\mathbb{E}_{x \sim p_t^{\text{data}}(x)}\!\left[ |\Grad \log p_t(x) - s_\theta(x,t)|^2\right]\right]+ \text{KL}(p_{t=T} | p_{\theta, t = T})\,.
\end{equation}
where we recall again that $p_{\theta, t = T} = p_{t=\infty}$ is $\theta$-independent.

There is also an equivalent reverse \emph{ODE} which flows $p_T$ to $p_0$ just as~\eqref{eq:reverse-sde} does.\footnote{The form of \eqref{eq:reverse-ode} follows from the solution to the continuity equation for $p_{T-t}$}  In particular, the equation is
\begin{equation}
\label{eq:reverse-ode}
\frac{dx}{d\tilde{t}} = -b(x, \tilde{t}\,) + \frac{1}{2}\left(\Grad \cdot \left( \mathbf{\sigma}(x, \tilde{t}\,) \cdot \sigma^T(x, \tilde{t}\,)\right) + \sigma(x, \tilde{t}\,) \cdot \sigma^T(x, \tilde{t}\,) \,\Grad \log p_{\tilde{t}}(x)\right)
\end{equation}
where $\sigma^T$ denotes the transpose of $\sigma$.

One now wishes to optimize an approximation to~\eqref{eq:reverse-sde} over choices of $s_\theta(x, t)$ as a proxy for $\Grad \log p_t(x)$. To do this, one discretizes time and then solves~\eqref{eq:forwards-sde} with an SDE solver, and then backpropagrates gradients of the time-discretized loss~\eqref{eq:continuum-loss} into $\theta$. The   backwards SDE/ODE is then implicitly used during training in backpropagation or the adjoint method \cite{chen2018neural}, when computing the gradients of the loss with respect to $\theta$. One can then draw samples from $p_0$ by drawing samples from $p_T$, and then solve a numerical discretization of the backwards SDE~\eqref{eq:reverse-sde} or ODE~\eqref{eq:reverse-ode} from $\tilde{t}=0$ to $\tilde{t}=T$. Given that $p_T$ (for large $T$) usually approximately has an explicit log-likelihood (e.g.~$p_T$ is approximately Gaussian), one can get tractable estimates for $\log p_0$ (and its $\theta$-gradients) by using the reverse ODE as well as a formula for the instantaneous change in $\log p_t$ along with the Hutchingson trace estimator~\cite{ffjord}. The practicalities of how one designs and trains the neural networks parameterizing the score function, as well as the methods for tuning the SDE, are outside the scope of this paper, but we provide some initial pointers to the literature for interested readers in Appendix \ref{App:diffusion_lit_review}.

\section{Lattice field theory and the renormalization group}
\label{sec:lattice-rg}

\subsection{Lattice discretization}

Recall that in statistical field theory, we study infinite-dimensional probability distributions over spaces of fields. Formally, these distributions have their log-probability specified by the integral of a local quantity over the domain of the field. In scalar $\phi^4$ theory in $d$ dimensions, for which we previously considered the case $d=2$, the fields are functions $\phi: \R^d \to \R$, and the probability measure is  
\begin{equation}
    \label{eq:path-integral-measure}
 \frac{1}{Z} \,e^{- S[\phi]} \mathcal{D}\phi = \frac{1}{Z} \exp\left\{ - \int_{\R^d} dy \left( \frac{1}{2} \Grad \phi \cdot \Grad \phi + \frac{1}{2} m^2 \phi^2 + \frac{\lambda}{4!}\,\phi^4\right)\right\} \mathcal{D}\phi\,,
\end{equation}
where $S[\phi]$ is called the \emph{action} of the field theory, $\mathcal{D}\phi$ is the formal volume measure on the space of fields, and $Z$ is a normalization constant. To make sense of such an infinite-dimensional distribution, which is not obviously well-defined, it is prudent to construct a finite-dimensional analog by fixing a lattice spacing $\epsilon$ and a lattice size $N$, which we take to be even. Then we define lattice points $\epsilon\, \textbf{n}$ for $\textbf{n} \in \{-N/2 + 1,...,N/2\}^d$.  We will denote $\{-N/2 + 1,...,N/2\}$ (modulo $N$) by $\mathbb{Z}_N$, and accordingly we write $\textbf{n} \in \mathbb{Z}_N^d$.  One then defines real variables $\phi(\textbf{n})$ which are meant to be samples of $\phi(y)$ at $y = \epsilon \, \textbf{n}$. The derivative in \eqref{eq:path-integral-measure} is discretized via
\begin{equation}
\partial_i \phi(y) \longrightarrow \frac{1}{\epsilon}\left(\phi(\textbf{n} + \textbf{e}_i) - \phi(\textbf{n})\right) 
\end{equation}
where we define the vector
\begin{equation}
\textbf{e}_i := (e_i^1, e_i^2,...,e_i^d)\,,\quad e_i^j := \delta_i^j\,.
\end{equation}
Here $\delta_i^j$ is the Kronecker delta.  In this paper we will take the lattice to be periodic (i.e.~it becomes a latticization of the $d$-dimensional torus); one could instead can impose other boundary conditions which would modify the derivative at the boundary.  The integral $\int_{\mathbb{R}^d} dy$ in~\eqref{eq:path-integral-measure} is approximated by the finite volume integral $\int_{(-N/2 + 1)\epsilon}^{(N/2) \epsilon} \cdots \int_{(-N/2 + 1)\epsilon}^{(N/2) \epsilon} dy$, and then approximately discretized via
\begin{equation}
\int_{(-N/2 + 1)\epsilon}^{(N/2) \epsilon} \cdots \int_{(-N/2 + 1)\epsilon}^{(N/2) \epsilon} dy\,f(y) \longrightarrow \sum_{\textbf{n} \in \mathbb{Z}_N^d} \epsilon^d\,f(\epsilon\,\textbf{n})\,.
\end{equation}
The finite-dimensional analog of~\eqref{eq:path-integral-measure} is then 
\begin{equation}
    \label{eq:discretized-path-integral-measure}
   \mu_{\epsilon, m, \lambda}[\phi_N]=  \frac{1}{Z_{N, \epsilon,a, m, \lambda}} \left(\prod_{\textbf{n} \in \mathbb{Z}_N^d} d\phi(\textbf{n})\right) e^{- S_{\epsilon, a,m,\lambda}[\phi_N]}
\end{equation}  
where
\begin{equation}
 S_{\epsilon,a,m,\lambda}[\phi_N] = \epsilon^d \sum_{\textbf{n} \in \mathbb{Z}_N^d} \left[\frac{1}{2}\left\{\sum_{i = 1}^d \frac{a^2}{\epsilon^2}\left(\phi(\textbf{n} + \textbf{e}_i) - \phi(\textbf{n})\right)^2 \right\} + \frac{1}{2}\,m^2 \phi(\textbf{n})^2 + \frac{\lambda}{4!}\,\phi(\textbf{n})^4\right]\,.
\end{equation}
We have written out the above discretization in full detail in order to be explicit.  To obtain the appropriate continuum limit of the action, we would like $a = 1$; however, it will be useful for us to consider the case of general $a$ for reasons we explain later.  Notice that when $\lambda > 0$, the $\phi(\textbf{n})^4$ term dominates the action $S_{\epsilon,a,m,\lambda}[\phi_N]$ for large values of $\phi$, and so the probability distribution in \eqref{eq:discretized-path-integral-measure} is normalizable; accordingly, the normalization constant $Z_{N, \epsilon, a,m, \lambda}$ is finite.

Fixing values of $N$, $\epsilon$, $a$, $m$, and $\lambda$, the log-density of~\eqref{eq:discretized-path-integral-measure} is explicit, so now one can draw samples from the corresponding density using e.g.~Hamiltonian Monte Carlo. This is precisely what is done in lattice field theory calculations in order to study physical systems, for instance to estimate the mass of a proton using QCD (although the relevant analog of~\eqref{eq:path-integral-measure} for QCD problems involves many more complications when discretizing the corresponding infinite-dimensional integral, reviewed in \cite{fodor2012light}). Much work in lattice field theory involves using the structure of probability distributions like~\eqref{eq:discretized-path-integral-measure} in order to build more efficient samplers.

\subsection{Renormalization group}

\subsubsection{The lattice setting}

In order to define~\eqref{eq:discretized-path-integral-measure}, we introduced discretization parameters $N$ and $\epsilon$ such that we are formally sampling $\phi(y)$ over a cube of side length $N\epsilon$ on densely-packed lattice points with lattice spacing $\epsilon$.  However, these discretization parameters do not appear in the formal expression~\eqref{eq:path-integral-measure}. Since the sampling scale $\epsilon$ that we introduced was \emph{arbitrary}, we would like for the physical predictions of the model, at least for distances \emph{much larger} than the sampling scale $\epsilon$, to be independent of $\epsilon$. Let us fix $N\epsilon = L$, and write $\mu_{N, m, \lambda}$ for the distribution~\eqref{eq:discretized-path-integral-measure}; here we have dropped the dependence on $\epsilon$ since it is fixed in terms of $N$ by $\epsilon = L/N$.  Since the dimensionality of the distribution $\mu_{N, a, m, \lambda}$ varies with $N$, in order to \emph{make sense} of the independence of the long-range physics on $N$ (and $\epsilon$) we need to be able to \textit{compare}
between between the distributions $\mu_{N, a, m, \lambda}$ for different values of $N$.  Moreover, in order for the long-range physics to be independent of $N$ (and $\epsilon$), it may be necessary to vary the parameters $a$, $m$ and $\lambda$ with $N$; as such, as write $\mu_{N, a_N, m_N, \lambda_N}$ to allow for this possibility.

Since samples $\phi_N$ from  $\mu_{N, a_N, m_N, \lambda_N}[\phi_N]$ are supposed to be lattice approximations to the continuous field $\phi(y)$, it is natural to implement \textit{comparison maps} by defining some  ``resampling'' or ``field interpolation'' map $f_{N N'}$ for $N > N'$, which (deterministically or stochastically) takes samples $\phi_N$ from $\mu_{N, a_N, m_N, \lambda_N}[\phi_N]$ to samples $\phi_{N'}$ from $\mu_{N', a_{N'}, m_{N'}, \lambda_{N'}}[\phi_{N'}]$.  We would like for such maps $f_{NN'}$ to be \textit{local}.  This means that $[f_{N N'}(\phi_{N})](\textbf{n}')$, which is a field configuration $\phi_{N'}(\textbf{n}')$ evaluated at the location $\frac{L}{N'}\,\textbf{n}' \in \R^d$, should be computed from \emph{nearby lattice sites} at lattice spacing $L/N$; in other words, $[f_{N N'}(\phi_N)](\textbf{n}')$ is primarily a function of $\phi(\textbf{n})$ for $|\frac{L}{N}\,\textbf{n} - \frac{L}{N'}\,\textbf{n}'| \leq c\,L/N$ for some $c \sim O(1)$.  For example, on a 2D lattice if we fix $N_t = 2^t$ for $t = 0, 1, 2, \ldots$, such that as $t$ increases the lattice sites get dyadically subdivided, then it is natural to choose $[f_{N_{t+1}, N_t}(\phi(\textbf{n}))](\textbf{n}')$ to be the the average of the values of $\phi(\textbf{n})$ on the $2\times 2$ square subdivided sites.

The locality of the comparison map enforces that certain long-range properties of correlations are preserved upon resampling.  To be precise, it is useful to define \emph{correlation functions} which are expectation values of a product of fields at different lattice sites as a function of the position: 
    \begin{equation}
    c_{\mu_{N, a_N, m_N,\lambda_N}[\phi_{N}]}(\textbf{n}_1, ... , \textbf{n}_r) := \mathbb{E}_{\phi_N \sim \mu_{N, a_N, m_N, \lambda_N}[\phi_{N}]}[\phi(\textbf{n}_1) \phi(\textbf{n}_2) \cdots \phi(\textbf{n}_r)]\,.
    \end{equation}
It is also convenient to define the resampled correlation functions
    \begin{equation}
    c_{\mu_{N, a_N, m_N,\lambda_N}[\phi_{N}]}^{f_{NN'}}(\textbf{n}_1', \ldots, \textbf{n}_r') := \mathbb{E}_{\phi_N \sim \mu_{N, a_N, m_N, \lambda_N}[\phi_{N}]}\!\Big[[f_{NN'}(\phi_N)](\textbf{n}_1') [f_{NN'}(\phi_N)](\textbf{n}_2') \cdots [f_{NN'}(\phi_N)](\textbf{n}_r') \Big]\,,
    \end{equation}    
which we can think of as capturing correlations of $\mu_{N, a_N, m_N,\lambda_N}$ at intermediate to large distance scales; the resampled correlation functions do not capture correlations at short distance scales since the resampled fields $[f_{NN'}(\phi_N)](\textbf{n}')$ are not strictly local on the lattice.  It follows from the usual properties of pushforwards that
    \begin{equation}
    c_{\mu_{N,a_N, m_N,\lambda_N}}^{f_{NN'}}(\textbf{n}_1', \ldots, \textbf{n}_r') = c_{(f_{NN'})_* \mu_{N,a_N, m_N,\lambda_N}}(\textbf{n}_1', \ldots, \textbf{n}_r')\,,
    \end{equation}
which means that the intermediate to long-distance properties of correlation functions of $\mu_{a_N, m_N,\lambda_N}[\phi_{N}]$ are preserved in $(f_{NN'})_*\mu_{N, a_N, m_N,\lambda_N}[\phi_{N}]$.

In many circumstances, it is useful to approximate $(f_{NN'})_* \mu_{N, a_N, m_N,\lambda_N}[\phi_{N}]$ by $\mu_{N', a_{NN'}, m_{NN'},\lambda_{NN'}}[\phi_{N'}]$ with judiciously chosen couplings $a_{NN'}$, $m_{NN'}$, and $\lambda_{NN'}$.  In particular, we seek parameters $a_{NN'}$, $m_{NN'}$, and $\lambda_{NN'}$ such that
\begin{equation}
\label{E:longdistancecondition1}
c_{(f_{NN'})_* \mu_{N, a_N, m_N,\lambda_N}[\phi_{N}]}(\textbf{n}_1, \ldots, \textbf{n}_r) \approx c_{\mu_{N', a_{NN'}, m_{NN'},\lambda_{NN'}}[\phi_{N'}]}(\textbf{n}_1, \ldots, \textbf{n}_r)
\end{equation}
for $|\frac{L}{N'}\,\textbf{n}_i - \frac{L}{N'}\,\textbf{n}_j|_{\mathbb{Z}_N^d} \gg O(1)$ with $i \not = j$.  The notation $|\textbf{n}-\textbf{m}|_{\mathbb{Z}_N^d}$ means the (shortest) distance between $\textbf{n}$ and $\textbf{m}$ on $\mathbb{Z}_N^d$ viewed as a periodic lattice.  This is all to say that we want the intermediate to long-distance correlation functions of $(f_{NN'})_* \mu_{N, a_N, m_N,\lambda_N}[\phi_{N}]$ and $\mu_{N', a_{NN'}, m_{NN'},\lambda_{NN'}}[\phi_{N'}]$ to approximately match, if it is possible to do so.

The above discussion of how the coupling effectively change on account of the flow is so important that it is worth drawing it out more explicitly.  Let us write $\mathcal{D}\phi_N := \prod_{\textbf{n} \in \mathbb{Z}_N^d} d\phi(\textbf{n})$ for the Euclidean measure over fields sampled at lattice spacing $L/N$. 
 Then we can write $(f_{N N'})_*\mu_{N, a_N, m_N, \lambda_N}$ in a similar form as $\mu_{N', a_{NN'}, m_{NN'}, \lambda_{NN'}}$:
\begin{align}
\label{eq:renormalized-density}
(f_{N N'})_*\mu_{N, a_N, m_N, \lambda_N}[\phi_{N}] &= \frac{1}{Z_{N'}} \exp\!\Bigg(\!-\left(\frac{L}{N'}\right)^{\!d} \sum_{\textbf{n} \in \mathbb{Z}_{N'}^d} \Bigg[\frac{1}{2}\left\{\sum_{i=1}^d \frac{a_{NN'}^2}{\left(\frac{L}{N'}\right)^{\! 2}}\left(\phi(\textbf{n} + \textbf{e}_i) - \phi(\textbf{n})\right)^2 \right\} \\
    & \qquad \qquad \qquad \qquad \qquad \qquad \qquad + \frac{1}{2}\,m_{NN'}^2 \phi(\textbf{n})^2 + \lambda_{NN'}\,\phi(\textbf{n})^4\Bigg] + \cdots \Bigg) \,\mathcal{D}\phi_{N'} \nonumber
\end{align}
where $1/Z_{N'}$ is an appropriate normalization.  Moreover the $\cdots$ in the exponential denote terms in the log probability density contain other polynomials in the $\phi(\textbf{n})$'s (and possibly non-polynomials) which correspond to non-local and higher-order interactions.  Insofar as the $\cdots$ terms are small, we can choose not to include them; then we can more cleanly think of $a_{N}$, $m_{N}$, and $\lambda_{N}$ as flowing to $a_{NN'}$, $m_{NN'}$, and $\lambda_{NN'}$, respectively.  In particular, what it \textit{means} for the $\cdots$ terms to be small is that~\eqref{E:longdistancecondition1} holds for intermediate to long-distance correlators.

Thus, we can think of the \emph{averaging} process $f_{N N'}$ as generating a \emph{dependence} of the coefficients $a_N, m_N, \lambda_N$ on $N$. Alternatively, the requirement \eqref{E:longdistancecondition1}, which expresses that the long-range correlation functions are independent of $N$, forces us to \emph{tune} the parameters $a_N$, $m_N$, $\lambda_N$ as a function of $N$ such that this physical requirement holds.  We conclude that \emph{the coefficients of the model are dependent on the discretization scale of the field.} This dependence of the coefficients of the model on the discretization scale belongs to the framework of the \emph{renormalization group} (or \emph{RG}). A choice of \emph{renormalization group scheme} is essentially a choice of averaging process $f_{N N'}$ for all $(N, N')$ such that $f_{N' N''} \circ f_{N N'} = f_{N N''}$ (the \emph{semigroup property}).

In some circumstances, it is convenient to rescale the $\phi$ field so that the \textit{kinetic} term in the action, namely $\frac{1}{2}\sum_{i = 1}^d \frac{1}{\left(\frac{L}{N}\right)^2}\left(\phi(\textbf{n} + \textbf{e}_i) - \phi(\textbf{n})\right)^2$, has a unit coefficient.  In a sense, this term sets the size of fluctuations of the field $\phi$, and so it may be useful to work in units where such fluctuations are canonically normalized.  Concretely, suppose that at scale $\epsilon = L/N$ we have $a_N = 1$.  Then performing one RG step to scale $\epsilon' = L/N'$, we may obtain an $a_{NN'}$ which does not equal one.  However, implementing the field redefinition $\phi(\textbf{n}) \to \phi(\textbf{n})/a_{NN'}$ and defining $\tilde{m}_{NN'} := m_{NN'}/a_{NN'}$ as well as $\tilde{\lambda}_{NN'} := \lambda_{NN'}/a_{NN'}^4$,~\eqref{eq:renormalized-density} becomes
\begin{align}
\label{eq:renormalized-density_2}
(f_{N N'})_*\mu_{m_N, \lambda_N}[\phi_{N}] &= \frac{1}{\widetilde{Z}_{N'}} \exp\!\Bigg(\!-\left(\frac{L}{N'}\right)^{\!d} \sum_{\textbf{n} \in \mathbb{Z}_{N'}^d} \Bigg[\frac{1}{2}\left\{\sum_{i=1}^d \frac{1}{\left(\frac{L}{N'}\right)^{\! 2}}\left(\phi(\textbf{n} + \textbf{e}_i) - \phi(\textbf{n})\right)^2 \right\} \\
    & \qquad \qquad \qquad \qquad \qquad \qquad \qquad + \frac{1}{2}\,\widetilde{m}_{NN'}^2 \phi(\textbf{n})^2 + \widetilde{\lambda}_{NN'}\,\phi(\textbf{n})^4\Bigg] + \cdots \Bigg) \,\mathcal{D} \phi_{N'} \nonumber
\end{align}
where $\widetilde{Z}_{N'}$ is the new normalization appropriate for the distribution.  The equation above now does have the desired kinetic term, at the cost of modifying the definition of $\phi$.  Then the analog of~\eqref{E:longdistancecondition1} is
\begin{equation}
\label{E:longdistancecondition2}
c_{(f_{NN'})_* \mu_{N, 1, m_N,\lambda_N}[\phi_{N}]}(\textbf{n}_1, \ldots, \textbf{n}_r) \approx \frac{1}{a_{NN'}^r}\,c_{\mu_{N', 1, \widetilde{m}_{NN'},\widetilde{\lambda}_{NN'}}[\phi_{N'}]}(\textbf{n}_1, \ldots, \textbf{n}_r)
\end{equation}
for $|\frac{L}{N'}\,\textbf{n}_i - \frac{L}{N'}\,\textbf{n}_j|_{\mathbb{Z}_N^d} \gg O(1)$ with $i \not = j$, where we have accounted for the rescaling of the field $\phi$ with the factors of $a_{NN'}$.  A way to think about~\eqref{E:longdistancecondition2} is that given a $\mu_{N, 1, m_N,\lambda_N}[\phi_{N}]$, we can ask for parameters $a_{NN'}$, $\widetilde{m}_{NN'}$, and $\widetilde{\lambda}_{NN'}$ such that~\eqref{E:longdistancecondition2} holds for intermediate to long-distance correlators.  The formulation preserves the canonical normalization of the kinetic term in the action after each RG step.

To conclude this subsection, we emphasize that the infinite-dimensional distributions~\eqref{eq:path-integral-measure} are ill-defined, and (like all integrals) need to defined via a limiting procedure. The dependence of the coefficients $a_N, m_N, \lambda_N$ on the discretization scale $N$, is not illusory; indeed, it is required for the \emph{definition} of~\eqref{eq:path-integral-measure} as a limit of measures such as~\eqref{eq:discretized-path-integral-measure} or~\eqref{eq:renormalized-density}. This phenomenon repeatedly arises in works in mathematical physics on constructive statistical field theory, which aim to make rigorous mathematical sense of measures like~\eqref{eq:path-integral-measure} \cite{bauerschmidt2021log, frohlich1976infrared, glimm1985collected}. In many cases, some of the coefficients $a_N$, $m_N$, $\lambda_N$ are forced to go to zero or to infinity as $N\to \infty$ in order to get a well-defined limiting distribution.

\subsection{A toy analog of the renormalization group} 

While the procedure described above may seem complicated, a simple of the same phenomenon already arises in the setting of the simplest continuous-time stochastic process, namely Brownian motion. Letting $X(t)$ be a Brownian motion such that $dX(t) = dB_t$\,, we discretize time into steps of size $\delta t = \epsilon$.  This amounts to replacing $X(t)$ with $X^\epsilon(i)$, and setting $X^{\epsilon}(i+1) = X^{\epsilon}(i) + \sqrt{\epsilon}\,Z_i$ where $Z_i \sim \mathcal{N}(0,1)$. Then the joint distribution over $X^{\epsilon}(i)$ and $X^{\epsilon}(i+1)$ becomes
\begin{align}
\frac{1}{Z} \exp\!\Big(- A_\epsilon X^{\epsilon}(i)^2 -B_\epsilon X^\epsilon(i+1)^2 - C_\epsilon(X^{\epsilon}(i) - X^\epsilon(i+1))^2\Big)\,,
\end{align}
which is analogous to the quadratic part of \eqref{eq:renormalized-density}, 
where $A_\epsilon$, $B_\epsilon$, $C_\epsilon$ are chosen such that
\begin{align}
\mathbb{E}[X^{\epsilon}(i)^2] &= i\,\epsilon \\
\mathbb{E}[X^{\epsilon}(i+1)^2] &= (i+1)\epsilon \\
\mathbb{E}[(X^\epsilon(i+1)-X^{\epsilon}(i))^2] &= \epsilon\,.
\end{align}
We are forced to rescale $A_\epsilon, B_\epsilon, C_\epsilon$ this way with $\epsilon$ such that the $\epsilon \to 0$ limit gives a well-defined continuous stochastic process. Donsker's theorem \cite{stroock2010probability} shows that we can define the same continous-time stochastic process $X(t)$ from the $\epsilon \to 0$ limit of a large family of discrete-time stochastic processes $X^\epsilon(i)$: we only need to require that $X^\epsilon(i+1) - X^\epsilon(i)$ is mean-zero and of variance $\epsilon$, while its Gaussianity is unnecessary.  The independence of the \emph{macroscopic} properties of the model from its \emph{microscopic} specification is a simple analog of the \emph{universality} phenomenon connected to RG which is used in statistical field theory to make sense of phase transitions of physical systems, which we briefly overview in Section~\ref{subsec:phasetransitions}.

\subsection{Interfacing the renormalization group with modeling and simulation}

To make sense of drawing samples from the density~\eqref{eq:path-integral-measure}, we have introduced a discretization parameter $\epsilon$ so that we can instead sample from discrete analogs like~\eqref{eq:renormalized-density_2}. In the context of e.g.~\eqref{eq:renormalized-density_2}, we have argued that given an initial $m_N$ and $\lambda_N$ (and $a_N = 1$) at scale $\epsilon = L/N$, we can determine the flowed parameters $m_{N'} := \widetilde{m}_{NN'}$ and $\lambda_{N'} := \widetilde{\lambda}_{NN'}$ (as well as $a_{N'} := a_{NN'}$) at some other fixed value of $\epsilon' = L/N'$ for $N' < N$.  This is to say that if we know $m_N$ and $\lambda_N$ at some initial scale $\epsilon = L/N$, then we can determine a (approximate) flow of the theory to all larger scales; this means that $m_N$ and $\lambda_N$ at that initial scale serve as initial conditions.

When comparing with a physical system, we would like to use some experimental measurements to set the initial conditions $m_N$ and $\lambda_N$ of our model.  Then the model can be leveraged via simulations (e.g.~RG flow and sampling) to make predictions about the experimental system which can be compared with data.  The fixing of the initial conditions is achieved by \emph{scale-setting}: one computes an expectation value of some quantities $\mathcal{F}_i(\phi_N)$ over $\phi_N \sim \mu_{1, m_N, \lambda_N}[\phi_{N}]$, where the $\mathcal{F}_i$ are experimentally accessible functions of the physical field $\phi(x)$ that can be written in terms of some correlation functions at scale $\epsilon = L/N$.  One then sets the values of $m_N$ and $\lambda_N$ such that the experimentally measured values $\mathcal{F}_{\text{expt},\,i}[\phi]$ are approximately the predicted values of $\mathcal{F}_i(\phi_N)$ using samples from $\mu_{N, 1, m_N, \lambda_N}[\phi_{N}]$. Only a finite number of experimental measurements are thus needed to set the parameters, and once the parameters are set, other experimental predictions can be made from the simulated model. In practice, because much is known about the structure of the renormalization group for physically relevant models, other RG-based methods can used to set the scale, such as the method of \emph{gradient flow} \cite{luscher_trivializing_maps}, which is closely connected to the ideas of this paper.

The quantities $m, \lambda$ in~\eqref{eq:path-integral-measure} are, in the context of other models, interpreted as a ``mass'' and ``interaction strength'', respectively.  Thus, the renormalization group suggests that the experimentally-measured mass of a particle should be dependent on the precision of a measurement used to measure its mass, which in fact occurs experimentally; the same is true of the interaction strength (see e.g.~standard texts like \cite{Peskin:1995ev, schwartz2014quantum}).

\subsection{Comments on RG fixed points and phase transitions}
\label{subsec:phasetransitions}

Here we make comments about a more general class of models.  Suppose that our measure has the form
\begin{align}
\label{E:morecomplicatedmu}
\mu_{N, \{\lambda_{i,N}\}}[\phi_{N}] = \frac{1}{Z_{N}} \, \exp\!\left(- \left(\frac{L}{N}\right)^{\! d}\sum_i\,\lambda_{i,N}\,M_{i,N}[\phi_{N}]\right) \mathcal{D} \phi_N
\end{align}
where the $M_{i,N}[\phi_{N}]$'s given by
\begin{align}
\label{E:Mdef1}
M_{i,N}[\phi_N] = \sum_{\textbf{n} \in \mathbb{Z}_N^d}\prod_{k = 0}^{k_{\max, i}} (\Delta^k \phi(\textbf{n}))^{q_{i,k}}
\end{align}
where each $q_{i,k} \in \mathbb{Z}_{\geq 0}$ and $\Delta$ is the discrete Laplacian operator which acts on fields as
\begin{equation}
\label{eq:discrete-laplacian}
\Delta \phi(\textbf{n}) = \sum_{i = 1}^d \!\frac{1}{\left(\frac{L}{N}\right)^{\! 2}}\!\left(\phi(\textbf{n} + \textbf{e}_i) - 2\,\phi(\textbf{n}) + \phi(\textbf{n} - \textbf{e}_i)\right)\,.
\end{equation}
By $\Delta^k$ we simply mean applying $\Delta$ to a function $k$ times. It will be convenient to define
\begin{align}
D_i := \sum_{k = 0}^{k_{\max, i}} q_{i,k}
\end{align}
which counts the total number of multiplicative $\phi$'s appearing in an $M_{i,N}$.  Note that $M_{i,N}[\phi_N]$ depends on $N$ in two ways: through $\phi_N$ and the sum over $\textbf{n} \in \mathbb{Z}_N^d$, and through the powers of $\frac{L}{N}$ coming from the Laplacians.  When we write $M_{i,N'}[\phi_{N'}]$, we mean that we are changing all of those dependencies on $N$ to $N'$.

\begin{figure}[t!]
    \centering
    \includegraphics[scale = .58]{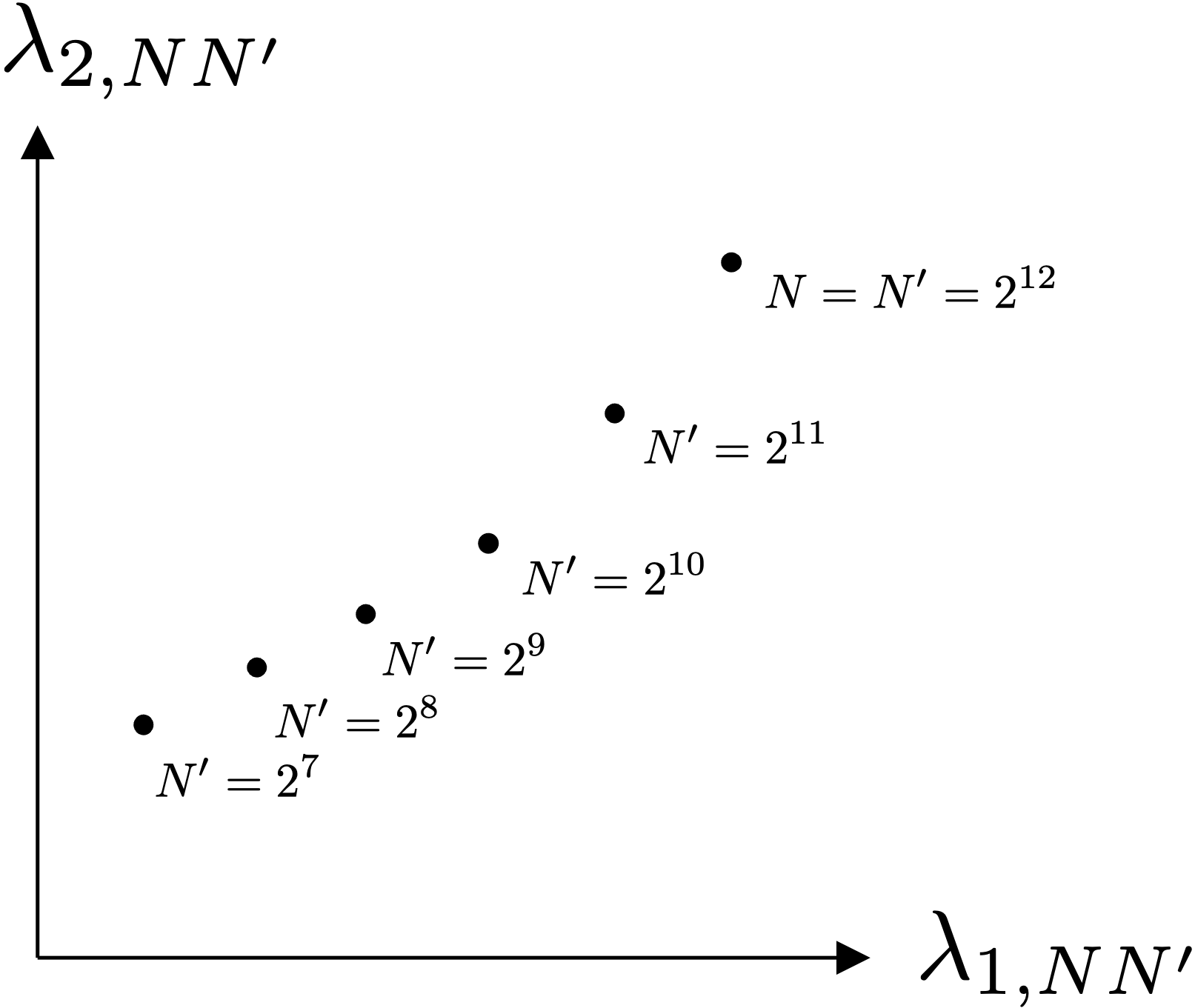}
    \caption{Schematic of the flow of $\lambda_{1,NN'}$ and $\lambda_{2,NN'}$ with varying $N'$, plotted as $\lambda_{2,NN'}$ versus $\lambda_{1,NN'}$.}
    \label{fig:lambda_flow}
\end{figure}

We can (approximately) flow the couplings to smaller values of $N$, namely $N' < N$, using an RG flow map.  In particular, we define the couplings $\lambda_{i,NN'}$ by
\begin{align}
\label{E:morecomplicatedmu_v2}
(f_{NN'})_* \mu_{N, \{\lambda_{i,N}\}}[\phi_{N}] \approx \mu_{N', \{\lambda_{i,NN'}\}}[\phi_{N'}]  = \frac{1}{Z_{N'}} \, \exp\!\left(- \left(\frac{L}{N'}\right)^{\! d}\sum_i\,\lambda_{i,NN'}\,M_{i,N'}[\phi_{N'}]\right) \mathcal{D}\phi_{N'}\,,
\end{align}
where the right-hand side of the $\approx$ should be regarded as the best approximation (in some quantitative way that can be specified) to the left-hand side.
Suppose that for $L$ sufficiently large, we plot the trajectories of the couplings $\{\lambda_{i,NN'}\}_i$ as a function of decreasing $N'$.  As we decrease $N'$ (for $1 \ll N' \ll L$), we obtain a diagram as in Figure~\ref{fig:lambda_flow}, which should be thought of as a plot of the RG flow of the parameters of the model. There can arise \emph{special points} in this plot which appear to be fixed points of the flow, in the following sense: there exists a constant $y$ such that for $1 \ll N'' < N' \leq N$, 
\begin{equation}
\label{E:criticalcondition0}
\lambda_{i,NN'} \approx \left(\frac{N'}{N''}\right)^{(d-y) \cdot D_i }\,\lambda_{i,NN''}\quad\text{for all }i.
\end{equation}
Letting $\hat{\lambda}_i := \lambda_{i,NN'}$, the above equations mean that
\begin{equation}
\label{E:criticalcondition1}
(f_{N'N''})_* \,\mu_{N',\{\hat{\lambda}_{i}\}}[\phi_{N'}] \approx \mu_{N'',\{\hat{\lambda}_{i}\}}[(\frac{N'}{N''})^{d-y}\,\phi_{N''}]\,.
\end{equation}
In words, if we perform RG on $\mu_{N',\{\hat{\lambda}_{i}\}}[\phi_{N'}]$, the distribution has approximately the same couplings, and only the $\phi(\textbf{n})$ fields are rescaled.  Such distributions thus possess a type of scale-invariance, and are called RG fixed points.  In nature, \emph{second-order phase transitions} correspond to RG fixed points. 
This includes critical points of water, at which the distinction between liquid and gaseous water ceases to exist \cite{Sengers2009}.

The scale-invariance of a critical point, as expressed in~\eqref{E:criticalcondition1}, can manifest in correlation functions.  For instance, if the interactions in~\eqref{E:morecomplicatedmu} are approximately isotropic with respect to the lattice, then we expect the two point function of the scalar to approximately satisfy
\begin{equation}
\label{E:2pointpower1}
c_{\mu_{N',\{\hat{\lambda}_i\}}\![\phi_{N'}]}(\textbf{n}_1, \textbf{n}_2) \approx \frac{C}{|\textbf{n}_1 - \textbf{n}_2|_{\mathbb{Z}_{N'}^d}^{2(d-y)}}
\end{equation}
for $1 \ll |\textbf{n}_1 - \textbf{n}_2|_{\mathbb{Z}_{N'}^d} \ll N'$, where $C$ is a constant.  See e.g.~\cite{cardy1996scaling} for an explanation of how this kind of power law scaling arises.  In any case, one often finds that correlation functions at critical points have a power law behavior.

More broadly, in the limit of large $N$ we can think of the class of distributions~\eqref{E:morecomplicatedmu} as comprising the space of \textit{all possible field theories} of a translation-invariant scalar in a fixed number of spatial dimensions.  From this point of view, RG flows can be thought of as implementing dynamics on the space of theories: we start with an initial theory, and move through the space of theories by using the RG flow map.  Although the dynamics through theory space is contingent on the particular choice of RG transformation $f$, the \emph{qualitative} features of the RG flow diagram can be independent of the choice of RG scheme.  It is a remarkable empirical fact, with a theoretical justification that is outside the scope if this paper (see e.g.~\cite{cardy1996scaling, fisher1998renormalization, kardar2007statistical, rosten2012fundamentals}), that for many different models like~\eqref{eq:discretized-path-integral-measure} RG flows possess \textit{quantitative} features that are largely independent of the choice of RG transformation.  This is referred to in the literature as \textit{scheme independence}.  For instance, quantities like $y$ in~\eqref{E:2pointpower1} associated to an RG fixed point are scheme-independent.  Moreover, the general theory of RG flows suggests that we can truncate the sum in the exponential of~\eqref{E:morecomplicatedmu} to a small number of terms, where the choice of those terms is contingent on the initial distribution from which we start the RG flow.  Such methodologies are central to the study of statistical models of extended physical systems.  

\subsection{Continuous-time formalism}
\label{subsec:continuoustime}

So far, our discussion of RG on the lattice has considered discrete RG steps, i.e.~in which we coarse-grain the lattice from $N^d$ sites to $(N')^d < N^d$ sites.  We likened this to a discrete-time dynamical system on the space of lattice theories with a fixed number of sites $N^d$, where time is a proxy for the number of iterations of the RG flow.  However, there is a similar, alternative framework that enables \textit{continuous} RG steps; the induced RG flows can then be viewed as continuous-time dynamical systems on the space of lattice theories.  In this context, time is likewise a proxy for the amount that we have flowed or coarse-grained.

To understand how continuous RG is possible on the lattice, we begin with a high-level discussion before delving into particular continuous RG schemes.  For simplicity, suppose we again have a lattice scalar field on a $d$-dimensional lattice of size $N^d$, which defines lattice points $\epsilon \, \textbf{n}$ for $\textbf{n} \in \mathbb{Z}_N^d$.  We further equip the lattice with periodic boundary conditions.  We can thus think of the lattice as living on a $d$-dimensional torus.  In this setting, we can readily take the Fourier transform of the field $\phi(\textbf{n})$, namely
\begin{equation}
\label{E:latticeFourier1}
\widetilde{\phi}(\textbf{p}) := \frac{1}{N^d} \sum_{\textbf{n} \in \mathbb{Z}_N^d} e^{-i \frac{2\pi}{N}\textbf{p}\cdot \textbf{n}}\,\phi(\textbf{n})
\end{equation}
where $\textbf{p} \in \mathbb{Z}_N^{d}$ (i.e.~$\textbf{p} \in \{-N/2+1,...,0,1,...,N/2\}^d$ (modulo $N$)).  The high-frequency components of $\widetilde{\phi}(\textbf{p})$ correspond to large $|\textbf{p}|$.  For reasons we will discuss later, it will be prudent to consider $\widehat{\textbf{p}}$ which has components
\begin{equation}
\label{E:phat1}
\widehat{p}_i := \frac{2N}{L} \sin\!\left(\frac{2\pi}{N}\,\frac{p_i}{2}\right),\quad i=1,...,d\,,
\end{equation}
so that $|\widehat{\textbf{p}}|^2 = \sum_{i=1}^d \widehat{p}_i^2$.  Using this notation, we note that high-frequency components of $\widetilde{\phi}(\textbf{p})$ also correspond to large $|\widehat{\textbf{p}}|$.

Thus if we want to smooth out $\phi(\textbf{n})$ in real-space, then we can equivalently progressively remove the high-frequency components of $\widetilde{\phi}(\textbf{p})$.

\subsubsection{A first guess}

From the above point of view, an initial guess for a continuous RG map is
\begin{equation}
\label{E:fLambda1}
f_t^{\text{guess}}[\widetilde{\phi}](\textbf{p}) = e^{-|\widehat{\textbf{p}}|^2 t}\,\widetilde{\phi}(\textbf{p})\,,
\end{equation}
for $t \geq 0$.  The map $f_t^{\text{guess}}[\widetilde{\phi}](\textbf{p})$ progressively dampens the higher-frequency modes as we increase $t$.  In particular, modes with $|\widehat{\textbf{p}}| \gg \frac{1}{\sqrt{t}}$ are significantly dampened, and so here $\Lambda_t \approx \frac{1}{\sqrt{t}}$ is our effective \textit{UV cutoff} scale.  Note that the lattice scale itself also provides a maximum cutoff, since $|\widehat{\textbf{p}}|$ can be at most $2 \sqrt{d} N/L$.

However, $f_t^{\text{guess}}[\widetilde{\phi}](\textbf{p})$ is in fact \textit{not} a valid RG scheme.  This fact was emphasized in~\cite{Carosso2020, wetterich1991average}; let us provide some of the essential intuition here.  Suppose we measure a system such that we can only access momentum modes less than a scale $\Lambda$; we call these modes \textit{IR modes}.  Accordingly we have the inability to access momentum modes greater than the scale $\Lambda$; we call these modes \textit{UV modes}.  In the microscopic distribution, there are IR modes which are coupled to the UV modes; as such, any reasonable procedure for marginalizing over the UV modes will affect the residual distribution over IR modes which can measure.  This can be articulated much more concretely: any physical RG map $f_t[\widetilde{\phi}](\textbf{p})$ should be a mixture of momentum modes when $|\widehat{\textbf{p}}|$ is near the cutoff scale.  For instance, maps of the form\footnote{More generally, the mixture of momentum modes could be nonlinear.}
\begin{align}
f_t[\widetilde{\phi}](\textbf{p}) = \sum_{\textbf{q}} Q_t(|\widehat{\textbf{p}-\textbf{q}}|)\,e^{-|\widehat{\textbf{q}}|^2 t}\,\widetilde{\phi}(\textbf{q})
\end{align}
for non-trivial $Q_t$ do mix momentum modes.  This requirement of mixing momentum modes is achieved by block-spin RG, but is \textit{not} achieved by the map $f_t^{\text{guess}}[\widetilde{\phi}](\textbf{p})$ in~\eqref{E:fLambda1} which merely rescales each $\widetilde{\phi}(\textbf{p})$ and hence does not mix momentum modes.

There is a way to augment the map $f_t^{\text{guess}}[\widetilde{\phi}](\textbf{p})$ so that it can become a valid RG scheme.  Suppose that we construct a new map $f_t[\widetilde{\phi}](\textbf{p})$ which is \textit{stochastic}, satisfying
\begin{align}
\label{E:stochanalog1}
\mathbb{E}[f_t[\widetilde{\phi}](\textbf{p})] = e^{-|\widehat{\textbf{p}}|^2 t}\,\widetilde{\phi}(\textbf{p})\,,
\end{align}
where we are averaging over some stochastic noise.  If the stochastic map has non-trivial \textit{variance} in the sense that $\widetilde{\phi}(\textbf{p})$ can be stochastically mapped into other nearby momentum modes, then our stochastic map can comprise a valid RG scheme.  We pursue this approach below.

\subsubsection{A first look at stochastic RG and the Carosso scheme}

Let us formulate a stochastic analog of~\eqref{E:fLambda1} along the lines of~\eqref{E:stochanalog1}, defined implicitly through a systems of stochastic ODEs.  Since the expressions are straightforward in position space (i.e.~using $\textbf{n}$ instead of $\textbf{p}$), we will work in position space here.  In particular, the Carosso scheme~\cite{Carosso2020} considers
\begin{equation}
\label{E:f1eq1}
f_t[\phi](\textbf{n}) =: \phi_t(\textbf{n})
\end{equation}
as the solution to the stochastic differential equation
\begin{equation}
\label{E:Carossointro1}
\partial_t \phi_t(\textbf{n}) = \Delta \phi(\textbf{n}) + \eta_t(\textbf{n})\,,\qquad \phi_0(\textbf{n}) = \phi(\textbf{n})\,,
\end{equation}
for all $\textbf{n} \in \mathbb{Z}_N^d$, where here $\Delta$ denotes the discrete Laplacian defined in~\eqref{eq:discrete-laplacian} and $\eta_t(\textbf{n})$ is a Gaussian random field satisfying
\begin{equation}
\label{E:Carossoexpect1}
\mathbb{E}[\eta_t(\textbf{n})] = 0\,, \qquad \mathbb{E}[\eta_t(\textbf{n}) \eta_s(\textbf{m})] := \left(\frac{N}{L}\right)^{\! d} \delta(t-s)\, \exp\!\left(- \Lambda_0^2\,\epsilon^2|\textbf{n}-\textbf{m}|_{\mathbb{Z}_N^d}^2\right),
\end{equation}
again for all $\textbf{n},\textbf{m} \in \mathbb{Z}_N^d$.  Since $\exp\left(- \Lambda_0^2\,\epsilon^2|\textbf{n}-\textbf{m}|_{\mathbb{Z}_N^d}^2 \right)$ equals $1$ for $\textbf{n} = \textbf{m}$ and is approximately $0$ for $\textbf{n} \not = \textbf{m}$, we can replace the Gaussian by a Kronecker delta $\delta_{\textbf{n},\textbf{m}}$ so that~\eqref{E:Carossoexpect1} becomes
\begin{equation}
\label{E:Carossoexpect2}
\mathbb{E}[\eta_t(\textbf{n})] = 0\,, \qquad \mathbb{E}[\eta_t(\textbf{n}) \eta_s(\textbf{m})] := \left(\frac{N}{L}\right)^{\! d} \delta(t-s)\, \delta_{\textbf{n},\textbf{m}}\,.
\end{equation}
We can think of the Carosso RG flow as smoothing out the scalar field $\phi$ at progressively larger distance scales as $t$ becomes larger.  In momentum space, the Carosso scheme satisfies
\begin{align}
\label{E:stochanalog2}
\mathbb{E}[\widetilde{\phi}_t(\textbf{p})] = e^{-|\widehat{\textbf{p}}|^2 t}\,\widetilde{\phi}(\textbf{p})\,,
\end{align}
but with non-trivial variance in the sense that $\widetilde{\phi}(\textbf{p})$ can be stochastically mapped to other nearby momentum modes.  In this sense, the Carosso scheme provides a realization of~\eqref{E:stochanalog1}.

We emphasize that we can view~\eqref{E:f1eq1} as being a continuous RG map, where the continuum parameter is $t$.  Increasing $t$ means we are further along the RG flow.  A key feature of~\eqref{E:f1eq1} is that the lattice remains the same size for any $t$.  That is, unlike our previous RG schemes which change the lattice size, here we are instead affecting the profiles of the fields that can live on a lattice of fixed size.  In fact, our lattice analyses in the previous subsections can be carried over to continuous setting with appropriate modifications.

There are many different kinds of stochastic RG flows, for instance the Polchinski RG flow~\cite{polchinski1984renormalization}, which we will discuss later.  We note that the Carosso RG flow and its cousins are all lattice discretizations of RG flows for continuous fields on $\mathbb{R}^d$.  These RG flows for continuous fields are part of the formalism of the Exact Renormalization Group (ERG), which we review in Appendix~\ref{App:ERGreview}.  In the ERG context, the stochastic ODE defined in~\eqref{E:Carossointro1} becomes a stochastic PDE.  The reason is that in the ERG context the fields are defined for all spatial points $\textbf{y} \in \mathbb{R}^d$ as opposed to on a finite lattice.

\subsection{The concept of Effective Field Theory}

\subsubsection{General comments}

So far, we have explored various manifestations of RG flow in the lattice setting.  In particular, we have emphasized the role of \textit{RG fixed points}, which can be viewed as endpoints of RG flows that can possess a type of scale-invariance.  But there is another key concept which RG flows enable us to formulate, namely Effective Field Theory.

The basic idea is as follows.  Suppose we have some \textit{microscopic theory}, meaning a probability distribution describing degrees of freedom which are not arbitrarily close together in space.  Our lattice theories for finite $\epsilon$ are an example.  In the parlance of physics, such theories are also referred to as \textit{UV theories}, stemming from the terminology of UV light being short-wavelength (at least relative to visible light).  We often suppose that the UV theories in question possess certain symmetries, e.g.~translation-invariance on the lattice, and have spatially local interactions.  Then when we perform RG flow on the theory, the log-probability (i.e.~the negative of the \textit{action} of the theory) will change and possibly gain new terms; the RG flow has the effect of zooming out and coarsening our description of the probability distribution.  That is, the RG flow only preserves data from the probability distribution that allows us to compute expectation values of functions which are smooth on large distance scales.  What is perhaps a priori surprising is that, in many circumstances, the RG-flowed theory can be well-described by a \textit{finite} number of parameters, which manifest as couplings in the log-probability.  This is to say that we can truncate the log-probability to a finite number of terms.  Interestingly, there are many distinct microscopic theories which flow to log-probabilities with the same finite set of terms (albeit with coefficients which depend on the microscopic theory); this class of truncated log-probabilities defines a \textit{universality class}.  They are akin to the stable distributions in classical probability theory, i.e.~the Gaussian distributions, Cauchy distributions, etc.  We emphasize that in field theory, the effective truncation comes into play at \textit{intermediate distance scales}; this means that we do not have to RG flow our UV theory until it hits a fixed point.  When we have a UV theory that has been RG flowed so that it approximately sits in a universality class, we call the resulting probability distribution an \textit{IR theory}, coming from the terminology of IR light being long-wavelength (relative to visible light).

The universality classes themselves, each specified by a finite number of parameters, are called \textit{Effective Field Theories} or \textit{EFTs}~\cite{weinberg1979phenomenological, georgi1993effective}.  They provide an accurate description of the long-distance behavior of many physically important microscopic theories.  In fact, the Standard Model of particle physics is an example of an EFT, which can be understood through the following perspective.  Suppose we know the underlying symmetries and types of particles observed at colliders.  Then we imagine that the particles we observe are themselves built out of much smaller, more microscopic degrees of freedom which are inaccessible to us.  As such, we stipulate that the particles we do observe should be described by a universality class corresponding to the long-distance behavior of the inaccessible, microscopic degrees of freedom.  Then we can write down the EFT consistent with the data we can observe; the EFT has a finite number of parameters which can be determined by looking at the detailed experimental data.  Remarkably, this procedure yields the Standard Model as we know it.

\subsubsection{A heuristic example}

To gain some intuition for how EFT can work, let us go back to our favorite example of scalar field theory, here in $d$ spatial dimensions.  We will make some heuristic arguments which work equally well in a lattice or continuum formulation; we opt for the continuum formulation since it makes the arguments slightly easier to phrase.  We accordingly consider the negative of the log-probability (i.e.~the action)
\begin{equation}
\label{E:scalarsagain1}
\int_{\mathbb{R}^d} dy \left(\frac{1}{2} \nabla \phi \cdot \nabla \phi + \frac{1}{2}\, m^2 \phi^2 + \frac{\lambda}{4!}\, \phi^4\right)\,.
\end{equation}
This theory possesses symmetry under rotations and translations, as well as under $\phi \to - \phi$.  Let us suppose that our RG flow (approximately) respects these symmetries, and as such only generates terms which are (approximately) consistent with said symmetries.  If we think of $y$ as corresponding to a position coordinate which has units of length, then $dy$ has units of length to the $d$th power.  We denote this fact by $[dy] = [\ell]^d$, in evident notation.  Since the log-probability, e.g.~\eqref{E:scalarsagain1}, is unitless, we can work out the dimensions of $\phi$.  Since $[\nabla] = [\ell]^{-1}$, it follows that $[\phi] = [\ell]^{- \frac{d-2}{2}}$ in order for $\int_{\mathbb{R}^d} dy \,\frac{1}{2} \nabla \phi \cdot \nabla \phi$ to be dimensionless.  Similarly, we determine that $[m] = [\ell]^{-1}$ and $[\lambda] = [\ell]^{d-4}$.  Let us define the dimensionless parameters $\widehat{m} := \ell\, m$ and $\hat{\lambda} := \ell^{4-d} \lambda$, the dimensionless coordinate $\hat{y} = y/\ell$, and the dimensionless field $\hat{\phi} := \ell^{\frac{d-2}{2}}\phi$.  In these definitions, $\ell$ can be regarded as a convenient length scale at which we perform measurements.  Then we can rewrite~\eqref{E:scalarsagain1} as
\begin{equation}
\label{E:scalarsagain2}
\int_{\mathbb{R}^d} d\hat{y} \left(\frac{1}{2} \widehat{\nabla} \hat{\phi} \cdot \widehat{\nabla} \hat{\phi} + \frac{1}{2}\,\ell^2\,\widehat{m}^2 \hat{\phi}^2 + \ell^{4-d}\,\frac{\hat{\lambda}}{4!}\,\hat{\phi}^4\right)\,.
\end{equation}

Before proceeding with our arguments, we note that we have chosen in~\eqref{E:scalarsagain1} and thus in~\eqref{E:scalarsagain2} not to put a (dimensionful) coupling in front of the kinetic term $\nabla \phi \cdot \nabla \phi$.  This is akin to our previous discussion surrounding~\eqref{eq:renormalized-density_2} where we discussed removing factors of $a_{NN'}$ in the kinetic term in the lattice action by rescaling $\phi$.  In short, we omit the coupling in front of the kinetic term in order to canonically normalize $\phi$ so that its fluctuations (which are dictated by the kinetic term) have unit size independent of the length scale at which we measure $\phi$\,; this is tantamount to a definition of the scalar field $\phi$, and in particular calibrates what it means to measure it.

Returning our attention to~\eqref{E:scalarsagain2}, we use the following heuristic.  Note that in the equation we have not accounted for any RG flow which would smooth out to a length scale $\ell$.  However, we can use the $\ell$-scaling of the couplings $\widehat{m}$ and $\hat{\lambda}$ as a heuristic proxy for how important the corresponding terms in the log-density might be to measurements at length scale $\ell$.  This heuristic is not entirely reliable, but it often indicates the right answer.  Note that if $\ell$ is a large length, then the $\frac{1}{2}\,\ell^2\,\widehat{m}^2 \hat{\phi}^2$ term is large.  This indicates that this quadratic term in $\hat{\phi}$ is important at long distances.  On the other hand, the importance of the quartic term $\ell^{4-d} \hat{\lambda} \hat{\phi}^4$ depends on the dimension; for $d < 4$ the term is large for large $\ell$, for $d = 4$ the term does not scale with $\ell$, and for $d > 4$ the term shrinks with increasing $\ell$.

More broadly, suppose we write down the most general version of~\eqref{E:scalarsagain2} that is a sum of polynomials of $\widehat{\nabla}\hat{\phi}$'s and $\hat{\phi}$'s, and which satisfies translational and rotational symmetry as well as a symmetry under $\phi \to - \phi$.  Then we would have
\begin{equation}
\label{E:scalarsagain3}
\int_{\mathbb{R}^d} d \hat{y} \left(\frac{1}{2} \widehat{\nabla} \hat{\phi} \cdot \widehat{\nabla} \hat{\phi} + \frac{1}{2}\,\ell^2\,\widehat{m}^2 \hat{\phi}^2 + \ell^{4-d}\,\frac{\hat{\lambda}}{4!}\,\hat{\phi}^4 + \ell^{6 - 2d}\,\hat{\alpha} \, \phi^6 + \ell^{2-d}\,\hat{\beta} \,\hat{\phi}^2 (\widehat{\nabla} \hat{\phi} \cdot \widehat{\nabla} \hat{\phi}) + \cdots \right)\,.
\end{equation}
For $d > 4$, only the first two terms survive for large $\ell$; all of the other terms go to zero.  As such, one expects that EFT in $d > 4$ dimensions for the scalar with our chosen symmetries is characterized by a single parameter, namely $\widehat{m}$.  For $d = 4$, the only surviving terms for large $\ell$ are the first three; as such one expects that EFT in $d = 4$ is characterized by the two parameters $\widehat{m}$ and $\hat{\lambda}$.  In a similar vein, we expect that EFT in $d = 3$ is characterized by the parameters $\widehat{m}$, $\hat{\lambda}$, and $\hat{\alpha}$.

One can check, however, that for $d = 2$, infinitely many parameters are important when $\ell$ is large; in particular, we can have terms like $\int_{\mathbb{R}^2} dy\,\ell^2\,\hat{\gamma}\,\hat{\phi}^{2n}$ and $\int_{\mathbb{R}^2} dy\, \hat{\delta}\,\hat{\phi}^{2n}(\widehat{\nabla} \hat{\phi} \cdot \widehat{\nabla} \hat{\phi})$.  While there are infinitely many terms, they have a somewhat constrained form.  For $d = 1$ we likewise have infinitely many terms which are important for large $\ell$, but the types of terms are more varied.

We see from the above heuristics that EFT is most constraining in higher dimensions; in $d \geq 3$ there are only finitely many terms which survive for large $\ell$.  The EFT paradigm remains somewhat useful even in $d = 2$ where the terms surviving at large $\ell$ are usefully constrained.

The above analysis can be generalized to incorporate multiple scalar fields or other kinds of fields beyond scalars (e.g.~fermions, vector fields, tensor fields, etc.).  Moreover, the heuristic analysis can be refined to more carefully account for the effects of RG.  One subtlety is that there are circumstances in which RG produces terms in the log-probability which are non-polynomial in the $\widehat{\nabla} \hat{\phi}$'s and $\hat{\phi}$'s; however, in many circumstances such non-polynomial terms can be approximated at long distance scales by polynomial terms.  Even recognizing such subtleties, the above gives a flavor for the universality enjoyed by a wide variety of theories at long distances (i.e., in the IR).

\subsubsection{Concluding remarks}

We emphasize that not all UV theories are RG flowed into universality classes described by a finite number of parameters.  This means that not all theories fit into the EFT paradigm.

Let us remark that while our account of the EFT perspective of the Standard Model is the contemporary viewpoint, the Standard Model was developed in a more circuitous manner.  It was only near the end, or perhaps slightly after, the Standard Model had come into being that its justification in terms of EFT was post facto formulated (see e.g.~\cite{weinberg1979phenomenological}).  This formulation was and is viewed as a grand synthesis.

Indeed, the EFT paradigm is extremely powerful, and has a wide range of applicability including fundamental particle physics (e.g.~\cite{weinberg1979phenomenological, georgi1993effective}), condensed matter systems (e.g.~\cite{altland2010condensed, brauner2022snowmass}), fluids (e.g.~\cite{dubovsky2012effective, crossley2017effective}), and even the collective behavior of certain kinds of biological systems (e.g.~\cite{ramaswamy2010mechanics, marchetti2013hydrodynamics}).  From a practical point of view, EFT can be thought of as an organization principle that is highly useful for modeling in the circumstances in which it is valid.

\section{Renormalizing diffusion models and multiscale modeling}
\label{sec:renormalizing-diffusion-models}

\subsection{Overview}

The renormalization group organizes physical models into a hierarchy of theories associated to physics at different length scales.  It is often the case that a UV theory flows to an IR theory described by EFT, and thus at long distances is approximately parameterized by a finite number of couplings.  As previously discussed, these EFTs can accurately capture coarse-grained statistics; in addition, the simplified equations they provide are often efficient to simulate on a computer. While EFTs can be useful, to capture the full range of relevant physical phenomena one must account for additional terms in the log-probability which are suppressed but non-zero when we probe features of the theory at intermediate-to-large distance scales.  The challenge of appropriately adapting coarse-grained simulations to take into account corrections arising from more fine-grained physics is the purview of \emph{multiscale modeling} \cite{ams-notices-multiscale-modeling}.

The subject of multiscale modeling is rather capacious.  It includes as a special case EFTs, which as previously discussed are essential to our understanding of particle physics, statistical field theories, and condensed matter systems~\cite{Cardy2015-qt}. Similar phenomena arise in fluid dynamics, where e.g.~correction terms to simplified models are needed to understand fusion physics and stellar plasmas. For example, there are important corrections to the ideal MHD equations describing low-frequency, large-scale plasma dynamics, which arise from the Maxwell-Vlasov equations -- a higher-dimensional PDE that is comparatively more difficult to simulate \cite{califano_manfredi_valentini_2016, sulem_passot_2015}.  In another example, the large-scale structure of the cosmic web can be understood through through the Zeldovich approximation or through Lagrangian Perturbation Theory \cite{10.1093/mnras/stx1976, PhysRevD.92.123517}, which in principle is an uncontrolled approximation to cosmological models involving N-body simulations \cite{Vogelsberger2020,10.1093/mnras/stab1855}. It is natural to try to learn better coarse-grained approximations from comparatively expensive fine-grained simulations using machine-learning (ML) techniques (see e.g.~\cite{Liu2022, He2019}).

\begin{figure}[t!]
    \centering
    \includegraphics[scale = .38]{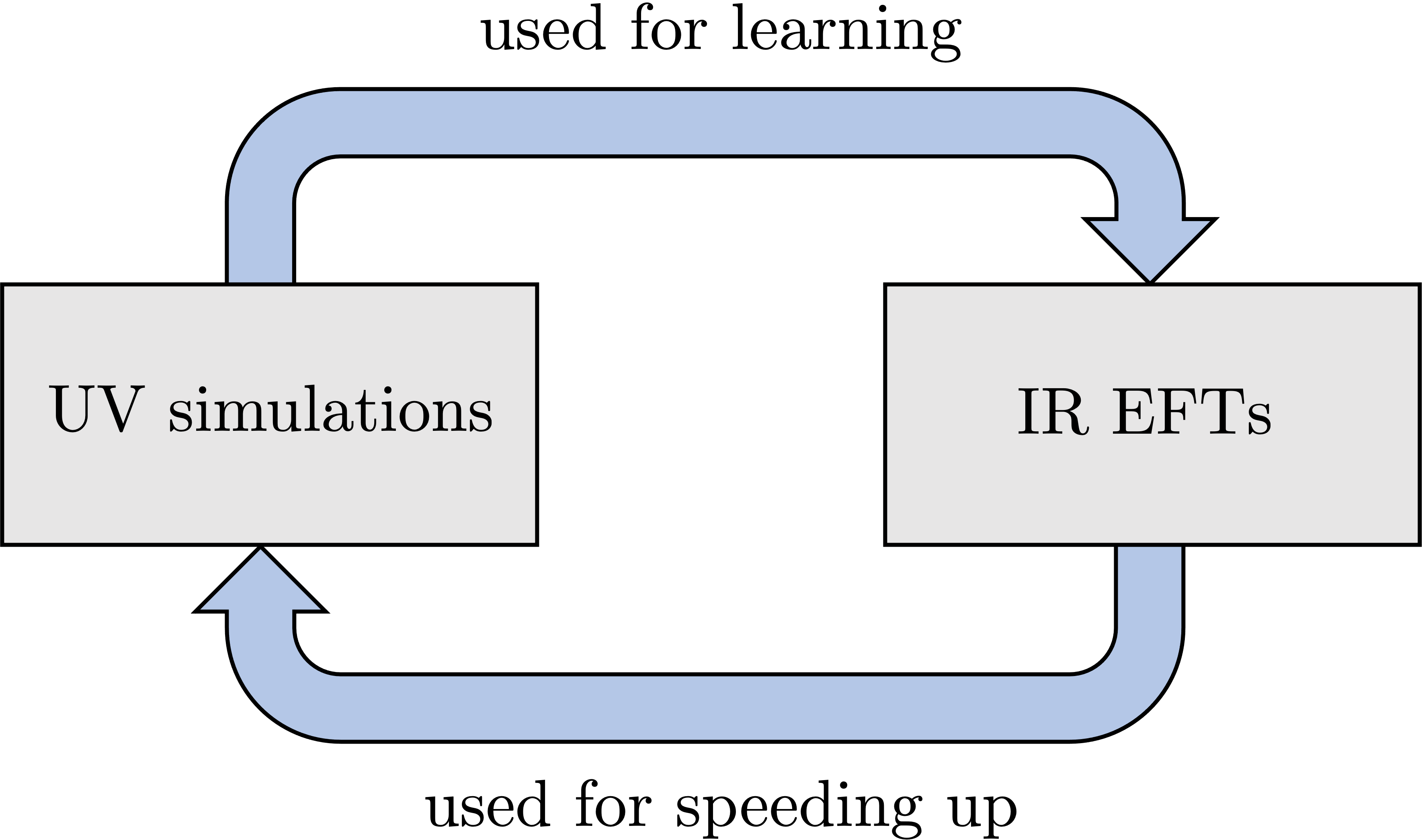}
    \caption{A diagram outlining how our algorithms leverage UV simulations to learn IR EFTs and in turn leverage IR EFTs to speed up UV simulations.}
    \label{fig:UVcycle1}
\end{figure}

While coarse-grained and fine-grained physics are rigorously connected through the renormalization group, approaches to improve coarse-grained models using machine learning can lack interpretability and be difficult to evaluate due to a lack of physical interpretation of the model parameters \cite{grojean2022lessons}. In our view, it is natural to look for principles for designing ML-based multiscale models for which model parameters are interpretable through the unifying formalism of EFT and the renormalization group, by analogy to existing work on physics-constrained ML models \cite{lemos2022rediscovering,Hansen2023 }  which operate on a single scale. In other words, we should be able to rigorously \emph{learn IR EFTs} from UV microscopic simulations, and dually use learned EFTs to \emph{speed up UV simulations}, as schematized in Figure~\ref{fig:UVcycle1}. 

While we view the development of a general rigorous methodology for multiscale ML-physics simulations as an exciting research problem, in this Section we focus on a concrete setting where the renormalization group formalism is fully developed, and has been studied with ML-based techniques: the problem of \emph{sampling from a statistical field theory}. Specifically,
we will turn to the problem of building improved Monte Carlo samplers for densities $\mu_{N,\{\lambda_{i,N}\}}[\phi_N]$ like~\eqref{eq:discretized-path-integral-measure} or~\eqref{E:morecomplicatedmu} (where $\{\lambda_{i,N}\}$ are the parameters in the action) associated to statistical field theories. 
In the setting of statistical field theory the log-density is known, and so it is possible to draw samples using the methods reviewed in Section~\ref{sec:inference-background}.  The basic challenge is to design the transition probabilities $Q(\phi | \phi')$ such that the Markov chain mixes well. This becomes very challenging as one makes the lattice parameter $N = L/\epsilon$ large due to the $N^d$-dependence of the dimension of $\mu_{N,\{\lambda_{i,N}\}}[\phi_N]$.

Various measures of mixing times can have a power-law dependence on $N$, especially when parameters $\{\lambda_{i,N}\}$ are tuned to match an RG fixed point; this dependence goes by the name of \emph{critical slowing down}~\cite{WOLFF199093}. Moreover, in certain kinds of physically relevant models, there are long-distance modes governed by topological quantities like the \emph{topological charge} which contribute in complex ways to physical processes such as instanton condensation \cite{t1976symmetry, callan1976structure}; these topological charges can be essential for understanding and characterizing the critical slowing down.  Another way to characterize mixing times is to compute the \emph{integrated autocorrelation time} of the sampler, which can be estimated via time-averages of rejection rates. Various sophisticated methods, many of them based on RG, have been used by the lattice field theory community to design samplers which improve on the aforementioned measures~\cite{PhysRevD.40.2035, endres2015multiscale, swendsen1987nonuniversal, wolff1989collective, prokof1998exact}. Some of the methods are based around bridge sampling and the renormalization group, wherein one introduces several parallel Markov chains for models sampling from $\mu_{N,\{\lambda_{i,N}\}}[\phi_N]$, such that the bridge used in the parallel tempering algorithm is an approximation to the RG flow of a field theory \cite{PhysRevD.40.2035, endres2015multiscale}.

Alternatively, instead of trying to use physically-inspired ans\"{a}tze for the proposal distribution $Q_\theta(\phi | \phi')$, one can try to \emph{learn} the proposal distribution using a variational family $Q(\phi | \phi')$.  We turn to this perspective below.

\subsection{Variationally optimizing the proposal distribution: normalizing flows}

In the rest of this Section, we will use the notation $\phi$ and $\phi_{N}$ interchangeably when the context is clear.  We will similarly use $d\phi$ and $\mathcal{D} \phi_N$ interchangeably.  Let us fix, for the remainder of this subsection, the dimension of the lattice to be $N_0$.  Suppose we desire to learn the distribution
 \begin{equation}
 \mu_{N_0, \{\lambda_{i,N_0}\}}(\phi) := \frac{1}{Z_{\lambda_{N_0}}}\, e^{-S_{N_0, \{\lambda_{i,N_0}\}}\!(\phi)} \mathcal{D}\phi_{N_0} =: p(\phi)\, d\phi\,.
 \end{equation}
Since the minus log-density $-\log p(\phi)$ is given by the action $S_{N_0, \{\lambda_{i,N_0}\}}$ of the field theory and thus explicitly known up to a normalizing constant $\log Z_{N_0, \{\lambda_{i,N_0}\}}$, one can use Metropolis-Hastings to produce samples from this distribution. Given a $\psi$-independent family of proposal distributions $Q_\theta(\phi | \psi) =  Q_\theta(\phi)$, it is natural to try to minimize one of the KL divergences 
\begin{equation}
\text{KL}(Q_\theta(\phi) | p(\phi)) \quad \text{or} \quad \text{KL}(p(\phi) | Q_\theta(\phi))
\end{equation}
over the parameters $\theta$.  Optimizing either of these quantities has different trade-offs for Monte Carlo sampling, as optimizing the former tends to find distributions which underestimate the support of the target distribution $p(\phi)$ while optimizing the latter tends to find distributions which overestimate the support of the target distribution~\cite{murphy2012machine}. 

Let us focus for the time being on the problem of optimizing $\text{KL}(Q_\theta | p)$. In this case, a straightforward way to define a family of proposal distributions $Q_\theta(\phi)$ is to (i) fix an an easy-to-sample distribution $p_\infty(\phi')\, d\phi'$ 
 such as a Gaussian, (ii) define a family of diffeomorphisms
 $\phi \mapsto \phi' = f_\theta(\phi)$, and (iii) set
\begin{equation}
\label{eq:pushforward-generative-model}
     Q_\theta(\phi) \,d \phi := (f^{-1}_{\theta})_*(p_\infty(\phi')\,d\phi')  
 = p_\infty(f_\theta(\phi)) \left| \det \frac{\partial f_\theta(\phi)}{\partial \phi} \right|\,d\phi.
 \end{equation}
Then so long as the Jacobian factor in the above formula takes a simple analytical form for the parameterization $\theta \mapsto f_\theta$, one can used automatic differentiation and stochastic gradient descent to minimize the loss
\begin{equation}
    \label{eq:normalizing-flow-loss}
    L(\theta) = \text{KL}(Q_\theta | p) - \log Z_{N_0, \{\lambda_{i,N_0}\}} = \int d\phi \,Q_\theta(\phi) \left(\log Q_\theta(\phi) + S_{N_0, \{\lambda_{i,N_0}\}}\!(\phi)\right). 
\end{equation}
From \eqref{eq:kl-gradient-identity-1}, this loss admits a straightforward Monte Carlo gradient estimator, since
 \begin{equation}
     \label{eq:normalizing-flow-gradient-estimator}
     \Grad_\theta L(\theta) = \mathbb{E}_{\phi \sim Q_\theta(\phi)}\!\left[\left(\Grad_\theta \log Q_\theta(\phi)\right)\!\left(\log Q_\theta(\phi) + S_{N_0, \{\lambda_{i,N_0}\}}\!(\phi)+1\right) \right] 
 \end{equation}
where as usual we use the fact that $Z_{N_0,\{\lambda_{i,N_0}\}}$ and $S_{N_0,\{\lambda_{i,N_0}\}}$ are $\theta$-independent. We can readily estimate the gradients arising on the right-hand side by sampling from $Q_\theta = (f^{-1}_{\theta})_*(p_\infty(\phi')\,d\phi')$:
we plug the samples into the analytically computed quantity in the brackets on the right-hand side of~\eqref{eq:normalizing-flow-gradient-estimator}.  The machine learning community has produced a wealth of function approximators, such as the \emph{real NVP} flow \cite{real_nvp}, which have the feature that the Jacobian factor in \eqref{eq:pushforward-generative-model} as well as $f^{-1}_\theta$ are both efficiently computable.  

The approach described above has been explored extensively in a series of papers  \cite{flows_for_field_theory, flows_for_gauge_theory, intro_to_normalizing_flows_for_field_theory}. 
In these papers it is shown that on small lattice sizes ($N < 20$) the aforementioned sampler improves upon HMC with respect to various measures of critical slowing down. 
However, unlike in HMC, there is a pre-training cost to any such flow-based sampler due to the variational inference needed to learn $\theta$, and a systematic quantification of how much pre-training is needed to improve performance is not yet available for reasonable lattice sizes.

Any reasonable parameterization of the family of diffeomorphisms $f_\theta$ is built up by iterative compositions, e.g. 
\begin{equation}
    f_\theta = f^T_{\theta_{T}} \circ f^{T-1}_{\theta_{T-1}} \circ \cdots \circ f^1_{\theta_1}\,.
\end{equation}
Thus, writing $p^\theta_T = p_\infty$, such a model in principle defines a sequence of distributions $p^\theta_{t}$ for $t=T, \ldots, 0$, interpolating between $p_\infty = p_T$ and $p = p_0$, via 
\begin{equation}
    \label{eq:flow-based-model}
    p^\theta_t = (f^{t+1}_{\theta})^{-1}_* p^\theta_{t+1}\,.
\end{equation} 
One can take the continuous-time limit and instead use a continuous family of distributions $p^\theta_t$ as the pushforwards $(f^{t, T}_{\theta})^{-1}_* p_\infty$, where $\phi \mapsto f^{t, T}_\theta(\phi)$ is the flow obtained by solving an ODE 
\begin{equation}
    \label{eq:neural-ode}
    \frac{d\phi_\tau}{d\tau} = b_\theta(\phi_\tau, \tau)
\end{equation}
from $\tau=t$ to $\tau = T$. In such a framework, one can get estimates of $\log p^\theta_t$ and $\Grad_\theta \log p^\theta_t$ via Neural ODE methods \cite{chen2018neural}. This class of models has improved the scaling performance with respect $N$ for sampling in $\phi^4$ theory~\cite{miranda_sampling}. 

\subsection{Learning the renormalization group}

One challenge with the class of models discussed above \cite{flows_for_field_theory, flows_for_gauge_theory, miranda_sampling} is that the flow $p_t^\theta$ learned by the models has no physical meaning, and thus it can be challenging to debug or tune the models using physical intuition about the field theories being studied.  In this subsection, we propose to modify the objective functions such that \emph{the flow learned by the models will be the RG flow of the field theory}. We will describe a more flexible perspective on this class of models in Section~\ref{sec:other-rg-schemes}; for now, we will focus on the use of a single RG scheme, and make the smallest possible modification to the objective~\eqref{eq:normalizing-flow-loss}.

Recall from equations~\eqref{E:Carossointro1} and~\eqref{E:Carossoexpect2} that the Carosso RG scheme is governed by the SDE
\begin{equation}
\label{E:Carossointro1v2}
\partial_t \phi_t(\textbf{n}) = \Delta \phi(\textbf{n}) + \eta_t(\textbf{n})\,,\qquad \phi_0(\textbf{n}) = \phi(\textbf{n})\,,
\end{equation}
where $t$ is a fictitious time parameter associated with scale, $\Delta$ is the discrete Laplacian, and $\eta_t$ is mean-zero Gaussian noise satisfying
\begin{align}
\label{E:Carossoexpect2v2}
\mathbb{E}[\eta_t(\textbf{n})] = 0\,, \qquad \mathbb{E}[\eta_t(\textbf{n}) \eta_s(\textbf{m})] := \left(\frac{N}{L}\right)^{\! d}\delta(t-s)\, \delta_{\textbf{n},\textbf{m}}\,.
\end{align}
The probability flow $p_t$ of $p(\phi) =: p_0(\phi)$ under~\eqref{E:Carossointro1v2} can be computed explicitly, as we now explain.  Recall the formula~\eqref{E:latticeFourier1} for the discrete Fourier transform
\begin{equation}
\label{E:latticeFourier1v2}
\widetilde{\phi}(\textbf{p}) := \frac{1}{N^d} \sum_{\textbf{n} \in \mathbb{Z}_N^d} e^{-i \frac{2\pi}{N}\textbf{p}\cdot \textbf{n}}\,\phi(\textbf{n})
\end{equation}
which can be computed in time $O((N \log N)^d)$.  Then taking the Fourier transform of~\eqref{E:Carossointro1v2} we obtain the equation
\begin{equation}
\label{E:Carossomomentum1}
\partial_t \widetilde{\phi}_t(\textbf{p}) = -|\widehat{\textbf{p}}|^2 \widetilde{\phi}(\textbf{p}) + \widetilde{\eta}_t(\textbf{p})\,,\qquad \widetilde{\phi}_0(\textbf{p}) = \widetilde{\phi}(\textbf{p})\,,
\end{equation}
where we recall the definition~\eqref{E:phat1}
\begin{equation}
\label{E:phat2}
\widehat{p}_i := \frac{2N}{L} \sin\!\left(\frac{2\pi}{N}\,\frac{p_i}{2}\right),\quad i=1,...,d\,,
\end{equation}
so that $|\widehat{\textbf{p}}|^2 = \sum_{i=1}^d \widehat{p}_i^2$, and the Gaussian noise satisfies
\begin{align}
\label{E:Carossoexpect2v3}
\mathbb{E}[\widetilde{\eta}_t(\textbf{p})] = 0\,, \qquad \mathbb{E}[\widetilde{\eta}_t(\textbf{p}) \widetilde{\eta}_s(\textbf{k})] := \frac{1}{L^d}\,\delta(t-s)\, \delta_{\textbf{p},-\textbf{k}}\,.
\end{align}
Notice that~\eqref{E:Carossomomentum1} naturally contains $|\widehat{\textbf{p}}|^2$ instead of $|\textbf{p}|^2$, and so as such our RG scheme most naturally suppresses modes with large $|\widehat{\textbf{p}}|$.  We further observe that~\eqref{E:Carossomomentum1} is diagonal in $\widetilde{\phi}(\textbf{p})$.  It will be convenient to use the notation $\widetilde{\eta}_t(\textbf{p})\,dt := \Omega\,dB_t(\textbf{p})$ where $B_t$ is standard Gaussian noise and $\Omega = \frac{1}{L^d}$.  Recall that the stationary distribution of the $1$-dimensional Ornstein-Uhlenbeck process 
\begin{equation}
    dX_t = -\kappa X_t + \sigma dB_t
\end{equation}
is given by
\begin{equation}
    X_\infty \sim \mathcal{N}(0, \sigma^2/2 \kappa)\,.
\end{equation}
As such, we see that we can sample $\phi(\textbf{n})$ from  $p_\infty = \lim_{t \to \infty} p_t$ by sampling 
\begin{align}
    \text{Re}\Big\{\widetilde{\phi}(\textbf{p})\Big\},\,\text{Im}\Big\{\widetilde{\phi}(\textbf{p})\Big\} &\sim \mathcal{N}\!\left(0, \frac{\Omega}{4 |\widehat{\textbf{p}}|^2}\right)\,,\qquad \textbf{p} \not = \textbf{0}\,\text{ or }\,(N/2, N/2,...,N/2) \\ \nonumber \\
    \widetilde{\phi}(\textbf{0}),\,\widetilde{\phi}((N/2, N/2,...,N/2)) &\sim \mathcal{N}\!\left(0, \frac{\Omega}{2 |\widehat{\textbf{p}}|^2}\right)\,,
\end{align}
as well as using $\phi(-\textbf{p}) = \phi(\textbf{p})^*$, and then further applying the inverse of the transformation~\eqref{E:latticeFourier1v2}. Similarly, after performing a discrete Fourier transform, the transition kernel for \eqref{E:Carossomomentum1} is also explicitly computable:
\begin{equation}
\label{eq:carosso-kernel}
    p_{t, 0}(\widetilde{\phi}_t| \widetilde{\phi}_0) \propto \prod_{\textbf{p} \in \mathbb{Z}_N^d} \,\exp\!\left(- \frac{1}{2}\, \frac{2|\widehat{\textbf{p}}|^2}{\Omega(1 - e^{-2 |\widehat{\textbf{p}}|^2 t})} \left|\widetilde{\phi}_t(\textbf{p}) - \widetilde{\phi}_0(\textbf{p}) \,e^{-|\widehat{\textbf{p}}|^2 t}\right|^2\right)\,.
\end{equation}
In particular, given samples from $p_0$, one can efficiently sample from $p_t$, and one has an efficient analytic formula for $\Grad_{\phi'}p_{t, 0}(\phi'|\phi)$. 

In some instances, it is useful to modify the Carosso flow by deforming with a mass term.  It is natural to let $\Omega = \frac{2}{L^d}$ and send $|\widehat{\textbf{p}}|^2 \to |\widehat{\textbf{p}}|^2 + M^2$ so that the kernel above becomes
\begin{equation}
\label{eq:carosso-kernel-v2}
    p_{t, 0}(\widetilde{\phi}_t| \widetilde{\phi}_0) \propto \prod_{\textbf{p} \in \mathbb{Z}_N^d} \,\exp\!\left(- \frac{L^d}{2}\, \frac{|\widehat{\textbf{p}}|^2 + M^2}{(1 - e^{-2 (|\widehat{\textbf{p}}|^2 + M^2) t})} \left|\widetilde{\phi}_t(\textbf{p}) - \widetilde{\phi}_0(\textbf{p}) \,e^{-(|\widehat{\textbf{p}}|^2 + M^2) t}\right|^2\right)\,.
\end{equation}
Note that as $t \to \infty$, we obtain the probability distribution over $\widetilde{\phi}_t$ for a free scalar field with bare mass $M$.  An important feature of~\eqref{eq:carosso-kernel-v2} is that when $\textbf{p} = \textbf{0}$, the argument of the exponential is non-zero; this is not true for~\eqref{eq:carosso-kernel}.  We find in numerical experiments that~\eqref{eq:carosso-kernel-v2} is more numerically stable than~\eqref{eq:carosso-kernel}, and so we use~\eqref{eq:carosso-kernel-v2} in our numerical experiments in Section~\ref{sec:numerics}.   

We now turn our attention to learning distributions $p_t^\theta$ which approximate $p_t$.  Accordingly, we should minimize a $t$-integral of divergences between $p_t^\theta$ and $p_t$. The functional
\begin{equation}
\int_0^T dt\,\lambda(t) \,\text{KL}(p_t^\theta | p_t)
\end{equation}
does not work well for this task, because the gradient estimator~\eqref{eq:normalizing-flow-gradient-estimator} for the term $\text{KL}(p_t^\theta | p_t)$ would require access to $\log p_t$, while we only have sample access to $p_t$. Instead, we consider the objective 
\begin{equation}
\label{eq:modified-objective}
    \text{KL}(p^\theta_0 | p_0) + \int_0^T dt\,\lambda(t)\,\text{KL}(p_t | p^\theta_t)\,.
\end{equation}
We can then parameterize $p^\theta_t$ for $t=0,...,T$ using a flow-based model~\eqref{eq:flow-based-model} or a Neural ODE~\eqref{eq:neural-ode}.  This gives us access to the log-densities $p^\theta_t$ and their $\theta$-gradients, which is necessary since we need to compute $\Grad_\theta \text{KL}(p_t | p^\theta_t) = \mathbb{E}_{\phi_t \sim p_t}[\Grad_\theta \log p^\theta_t]$. We can then run an algorithm which improves $\theta$ while drawing samples from $p_0$. For example, let us parameterize the flow as in~\eqref{eq:neural-ode}, and let $\theta = (\theta^1, \ldots, \theta^M)$. Then we can run the following loop:
\\ \\ 
\textbf{initialize} $\theta = (\theta^1,...,\theta^M)$, $\phi_0$, $i = 0$ \\
\textbf{if} $i < i_{\max}$ \textbf{do} \\
\indent sample $\phi'_T \leftarrow p^\theta_T = p_\infty$ \\
\indent set $\phi'_0$ by solving \eqref{eq:neural-ode} from $t=T$ to $t=0$ with initial condition $\phi_T'$ \\
\indent set $\theta_0 \leftarrow \epsilon \left\{\Grad_\theta \log Q_\theta(\phi'_0)\left(\log Q_\theta(\phi'_0) - S_{N_0,\{\lambda_{i,N_0}\}}\!(\phi'_0) + 1\right)\right\}$ \\
\indent \textit{optional}: draw $\phi''_0$ from an auxiliary distribution $Q'(\phi'' | \phi')$, e.g.~simulate several steps of \\
\indent \qquad \qquad \quad Hamiltonian Monte Carlo for $p_0$ with initial position $\phi'_0$, and then set $\phi'_0 \leftarrow \phi''_0$ \\
\indent with probability $\alpha = \exp\!\big(\!-S_{N_0,\{\lambda_{i,N_0}\}}\!(\phi_0')+ S_{N_0,\{\lambda_{i,N_0}\}}\!(\phi_0)\big) = p(\phi'_0) / p(\phi_0)$, set $\phi_0 \leftarrow \phi'_0$, replace $i \to i+1$ \\
\indent \textbf{for} $j=1, \ldots, \tau$  \\
\indent \qquad sample $t_j \leftarrow [0,T]$ \\
\indent \qquad sample $\phi_j \leftarrow p_{t_j}$ using \eqref{eq:carosso-kernel} \\
\indent \qquad compute $\Grad_\theta \log p^\theta_{t_j}(\phi_j)$ by calling an ODE solver, namely:\\
\indent \qquad \qquad (i) solve $dx/dt = b_\theta(x, t)$ from $t=t_j$ to $t=T$ with initial condition $x_t = \phi_t$\\
\indent \qquad \qquad (ii) set $x_T \leftarrow x(T)$\\
\indent \qquad solve the following system from $t = T$ to $t = t_j$ for $k = 1,...,M$\,: 
\begin{align*}
\qquad \quad \frac{dx}{dt} &= b_\theta(x,t)\,, \,\,
\frac{d (\delta_{\theta^k} x)}{dt} = \frac{\partial}{\partial \theta^k}\, b_\theta(x,t)\,,\,\, \frac{d (\delta \theta^k)}{dt} = \nabla_x \!\cdot\! \left(\frac{\partial}{\partial \theta^k}\, b_\theta(x,t)\right) + \sum_\ell \sum_m b^\ell(x,t) \,\delta_{\theta^m}x\,,
\end{align*}
\indent \qquad \qquad with initial conditions $\delta_{\theta^k} x(T) = 0,\,\,\delta \theta^k(T) = 0$ for all $k$ where $\delta \theta^k(t) := \frac{\partial}{\partial \theta^k}\log p_t^\theta(x(t))$ \\
\indent set $\theta_j \leftarrow (\delta\theta^1(t_j), \ldots, \delta \theta^M(t_j))$\\
\indent \textbf{end for} \\
\indent set $\theta \leftarrow \theta - \sum_{j=0}^{\tau} \theta_j $ \\
\textbf{end if} \\
\textbf{return} $\theta$
\\ \\
Having completed the above algorithm, we can now sample $\phi$'s from $p_T^\theta$ using our outputted $\theta$.  In the course of running the algorithm, we have produced a number of samples of $\phi$, which we can also leverage if we so desire (with the understanding that convergence of expectation values with respect to $\phi$'s sampled earlier on in the algorithm may have a higher variance).

\subsection{Variational inference for sampling and diffusion models}

\subsubsection{Description of the algorithm}

A major downside of the objective~\eqref{eq:modified-objective} is that it requires the computation of $\Grad_\theta \log p^\theta_t$ for optimization, which can be expensive as it involves repeated solution of a differential equation.  Instead, taking inspiration from Section~\ref{sec:diffusion-models}, we notice that we are trying to learn the inverse process to the diffusion process~\eqref{E:Carossointro1v2} which implements the Carosso RG scheme. Thus, instead of minimizing~\eqref{eq:modified-objective}, we can minimize a weighted sum of score function objectives like~\eqref{eq:continuum-loss}:
\begin{align}
    L(\theta) &= \int_0^T dt \,\lambda(t)\,\mathbb{E}_{\phi_t \sim p_t}\!\left[|\Grad \log p_t(\phi_t) - \Grad \log p^\theta_t(\phi_t)|\right] \\
    &= \int_0^T dt \,\lambda(t)\,\mathbb{E}_{\phi_0 \sim p_0} \mathbb{E}_{\phi_t \sim p_{t, 0}(\phi_t | \phi_0)}\!\left[|\Grad_{\phi_t} \log p_{t, 0}(\phi_t | \phi_0)- \Grad_{\phi_t} \log p^\theta_t(\phi_t)|\right] + C \nonumber
\end{align}
for a constant $C$ \cite{benton2022denoising}.

The rewriting of the objective in the second line is helpful because $\Grad_{\phi_t} \log p_{t, 0}(\phi_t | \phi_0)$ is explicit due to~\eqref{eq:carosso-kernel}, and the optimization of such an objective can be parallelized across different values of $t$. Moreover, with the above objective, one only needs access to the score functions $s^\theta_t(\phi_t) =  \Grad_{\phi_t} \log p^\theta_t(\phi_t)$, which can now be parameterized using a neural net architecture like a U-Net~\cite{ronneberger2015u} that have been developed for highly-successful applications in the generative modeling of images.  Writing~\eqref{E:Carossointro1v2} in the notation
\begin{equation}
\label{E:Carossointro1v3}
d\widetilde{\phi}_t = - \Delta \phi\,dt + \sigma\,dB_t
\end{equation}
where for us $\sigma = 1$, the corresponding time-reversed process is
\begin{equation}
\label{eq:finite-dimensions-inverse-Carosso-sde}
d\widetilde{\phi}_t = \left(- \Delta \phi + \sigma^2 s_t[\phi_t]\right)dt + \sigma\,dB_t
\end{equation}
where $s_t = \Grad_\phi \log p_t$ is the score function. In this setting, the training algorithm is as follows:
\\ \\
\textbf{initialize} $\theta$, $\phi_0$ \\
\textbf{if} $i < i_{\max}$ \textbf{do} \\
\indent sample $\phi_T \leftarrow p^\theta_T = p_\infty$ \\
\indent set $\phi'_0$ by solving~\eqref{eq:finite-dimensions-inverse-Carosso-sde} from $t=T$ to $t=0$ with initial condition $\phi_T$ and substituting $s_t \to s^\theta_t$\\
\indent \textit{optional}: draw $\phi''_0$ from an auxiliary distribution $Q'(\phi'' | \phi')$, e.g.~simulate several steps of \\
\indent \qquad \qquad \,\,\, Hamiltonian Monte Carlo for $p_0$ with initial position $\phi'_0$, and then set $\phi'_0 \leftarrow \phi''_0$ \\
\indent with probability $\alpha = \exp\!\big(\!-\! S_{N_0,\{\lambda_{i,N_0}\}}\!(\phi'_0) + S_{N_0,\{\lambda_{i,N_0}\}}\!(\phi_0)\big) = p(\phi'_0)/p(\phi_0)$ set $\phi_0 \leftarrow \phi'_0$, replace $i \to i+1$ \\
\indent \textbf{for} $j=1, \ldots, \tau$\\
\indent \qquad sample $t_j \leftarrow [0,T]$ \\
\indent \qquad sample $\phi_j$ from $p_{t_j}$ using \eqref{eq:carosso-kernel} \\
\indent \qquad set $\theta_j \leftarrow \lambda(t_j)\,(\Grad_{\phi'} p_{t_j}(\phi'|\phi_0) - \Grad_{\phi'} s^\theta_{t_j}(\phi', t_j))\big|_{\phi'= \phi_j}$\\
\indent \textbf{end for} \\
\indent set $\theta \leftarrow \sum_{j=1}^\tau \theta_j$ \\
\textbf{end if} \\
\textbf{return} $\theta$
\\ \\
Below we will discuss some aspects of the above method in further detail.

\subsubsection{Comments on the method}

Since we have provided a formal connection between a specific diffusion model and a particular RG scheme (i.e.~the Carosso scheme), our method produces a physically interpretable class of ML architectures for sampling from field theories. We elaborate more on variations of this method in Section~\ref{sec:other-rg-schemes}, including generalizations to RG schemes other than Carosso's.  Before pursuing such generalizations, there are several conceptual points about our method that are already evident.

While we have initially focused on the Carosso scheme due to the simplicity of its SDE formulation, it has the seemingly unusual property that the induced RG flow for long times maps all probability functionals to a fixed functional with a \emph{fixed} cutoff at the lattice scale. This may seem in tension with the usual discussion about (Wilsonian-type) RG schemes wherein IR fixed points of certain field theories do not have to be free (i.e.~Gaussian). The way to resolve this apparent tension is to note that in the Carosso scheme, the convergence rate of the Fourier transform of an arbitrary initial distribution to the final distribution $p_\infty$ is \emph{frequency-dependent}, and thus by performing an appropriate \emph{field renormalization} one can extract a modified flow which does not have to converge to a trivial (Gaussian) fixed point but instead can converge to nontrivial RG fixed points.  We further discuss field renormalization and how to incorporate it into our modeling framework in Section~\ref{sec:other-rg-schemes}.

\subsubsection{Towards score-based EFT}

In our method, we have identified the score function $s^\theta_t$ with the negative $\phi$-gradient of the effective action.  Here $t$ parameterizes the cutoff scale $\Lambda_t$. In physics applications, much is known about the most important terms contributing to the effective action by using the methodologies of EFT, and so this information can be incorporated explicitly into the structure of the neural networks parameterizing $s^\theta_t$.  Note that the reverse SDE~\eqref{eq:finite-dimensions-inverse-Carosso-sde} is a non-parameteric version of the inverse of the renormalization group equation and involves $s^\theta_t$.  It may be convenient to parameterize the score function by e.g.
\begin{equation}
\label{eq:eft-parametrization}
    s^\theta_t(\phi) = \left(\sum_{i=1}^r f_i(\theta^i,t)\,g_i(\phi)\right) + \tilde{s}^{\tilde{\theta}}_t(\phi)  
\end{equation}
where the right-hand side can be understood as follows:
\begin{itemize}
    \item $\theta = (\theta_1, \ldots \theta_r, \tilde{\theta})$ are the parameters of the neural network.
    \item The terms $g_i(\phi)$  are the $\phi$ gradients of the most significant terms in the effective action as dictated by EFT. It is natural to take, for example,  $g_1(\phi) = \Delta \phi$, which is proportional to the $\phi$ gradient of kinetic terms in the effective lattice action. Here $\Delta$ is the discrete Laplacian \eqref{eq:discrete-laplacian}. It is natural to further take $g_2(\phi) =  \phi$, which is the $\phi$ gradient of the mass term in the effective action; and similarly $g_3(\phi) = \phi^3$ which is proportional to the $\phi$ gradient of the $\phi^4$ term in the effective action.  We emphasize that $\phi$ should be viewed as a $N^d$-dimensional vector, and similarly $g_1(\phi)$, $g_2(\phi)$, and $g_3(\phi)$ define $N^d$-dimensional vectors.
    \item The terms $f_i(\theta^i, t)$ are scalar-valued neural networks. Upon training, these terms will \emph{learn the coefficients} of the terms $g_i$ in the effective action. As such, the estimated functions $f_i$ are physically meaningful, and give qualitatively novel estimators of fundamental physical quantities, e.g.~the bare couplings at scale $\Lambda$ and the wave function renormalization.
    \item The term $\tilde{s}^{\tilde{\theta}}_t(\phi)$ is a general neural network  which captures the remaining terms needed to describe the (gradient of the) effective action. To decouple the training of  $\tilde{s}^{\tilde{\theta}}_t(\phi)$ from the training of the $f_i(\theta^i, t)$'s, it is natural to require that for all $\theta$ and $t$, the vector field $\tilde{s}^{\tilde{\theta}}_t(\phi)$ is orthogonal to the vector fields $g_i(\phi)$, $i=1, \ldots, r$, a condition that is easy to implement when designing the neural network $\tilde{s}^{\tilde{\theta}}_t$ by adding a fixed linear projection as the last layer. 
\end{itemize}  

Renormalizing diffusion models can thus be designed to naturally \emph{estimate the flows of parameters of the effective action.} This provides both a new class of estimators for these important physical quantities, as well as a series of heuristics for further designing the architectures and optimization processes for the neural networks $f_i$ and $\tilde{s}^{\tilde{\theta}}$, since much is known about the coefficients of the effective action as well as the behavior of the error term under RG flow of many field theories.

Because the (neural network) quantities $f_i$ have a physical interpretation, plots of their values during training may be used as diagnostics for the training, to see e.g.~if the training is being slowed down by an RG critical point. Moreover, since phase transitions and the onset of spontaneous symmetry breaking are meant to be captured by the behavior of critical points of the effective action, there are natural methods which use the estimated quantities $f_i$ to detect phase transitions in lattice field theories, which can otherwise be a difficult problem requiring searches for appropriate order statistics. Finally, our method may be combined with transfer learning methods~\cite{miranda_sampling}, which have been found to speed up training, to learn a single score function or effective action estimator for field theories with a whole range of UV parameters at once. This opens up the exciting possibility of learning a single representation of the entire RG flow diagram of a field theory, or even to automatically search for phase transitions in field theories in a physically justified way using ML techniques.  

We will elaborate on our above score-based schemes for EFT and provide numerical examples in~\cite{score-based}.

\subsection{More history}

We now place the discussion of the previous Subsection into a broader context.  The first observation to be made is that the noising process~\eqref{E:Carossointro1v3}  as well as the inverse process~\eqref{eq:finite-dimensions-inverse-Carosso-sde} already appear in the machine learning literature for image generative modeling under the name of \emph{Blurring Diffusion Models} \cite{blurring_diffusion_models}. More generally, these models fits into a large class of models \cite{subspace_diffusion, mallat_diffusion, cascased_diffusion, iterative_refinrement, spectral_diffusion_processes}, which we call \emph{multiscale diffusion models}, which all share the feature that they generate images in a multi-step process, where earlier steps generate a \emph{coarse approximation to the image} and then later steps \emph{refine the coarse approximation} to produce the final, high-fidelity image. The first motivation for considering such models for image generation is that images naturally have features across a hierarchy of scales (in fact, the power frequency spectrum of natural image distributions tends to follow power laws \cite{ruderman1997origins}), and thus it seems natural to allocate separate model variables for feature generation at different scales. The second motivation is \emph{computational}: it has been found that repeatedly `upsampling' the image is simply computationally more efficient than trying to generate a high-resolution image from scratch. Leveraging the multiscale structure of image models has been found to consistently improve model performance \cite{subspace_diffusion, iterative_refinrement}. 

In fact, the original paper on diffusion modeling \cite{sohl-dickstein_deep_2015} is loosely motivated by ideas about the renormalization group. While there has been much work aimed at connecting specific neural network architectures to special cases of RG transformations \cite{mehta2014exact}, as well as works that draw heuristic, non-physically-grounded  analogies \cite{dlrg1, dlrg3, dlrg4, dlrg5, dlrg6, dlrg7}, the introduction of multiscale diffusion models was done as a practical matter, without connection to physics. In our method detailed in the previous Subsection, we showed how to rigorously interpret an existing diffusion model architecture (when applied to a problem in lattice field theory) as a pre-existing RG scheme, namely the Carosso scheme. In Section~\ref{sec:other-rg-schemes} below, we expand on this connection, indicating how one can \emph{design} diffusion models to implement different classes of RG schemes, and explaining how to compare the results from different multiscale diffusion models when applied to field-theoretic problems by identifying the appropriate rescaling transformations.

The Carosso scheme, which has been our focus so far, is connected to an interesting series of works in the lattice field theory community. 
Specifically, Carosso \cite{carosso_thesis, Carosso2020} introduces his RG scheme as a formal renormalization group counterpart of a widely-used method dubbed ``gradient flow'', which descends\footnote{The earlier-cited work on normalizing flows for lattice field theory \cite{flows_for_field_theory, flows_for_gauge_theory, miranda_sampling} is also loosely motivated by the idea of learning L\"uscher's trivializing flows; explicit statements to this end can be found in \cite{learning_trivializing_flows}.} from L\"uscher's pioneering \cite{luscher_trivializing_maps}. This latter paper pointed out that when trying to sample from lattice $SU(N)$ gauge theory, which is a distribution $\mu$ over a compact manifold, there is an implicit characterization of a flow $f_t$ called the \emph{Wilson Flow} such that $(f_t)_*\mu$ limits to a uniform (Haar measure) distribution. In fact, this flow ends up being a lattice analog of the gradient flow of the Yang-Mills functional \cite{atiyah_bott}. As such, it is a PDE for the fields with a gradient flow interpretation which has as its highest order term the Laplacian operator acting on the fields. Evidently the ``gradient flow'' smooths out the fields and gives rise to an operation akin to the renormalization group; this has led to a significant amount of numerical work in lattice field theory \cite{luscher2010properties}, including a widely used method for scale setting \cite{borsanyi2012high}. Carosso \cite{Carosso2020} introduces his scheme as a stochastic analog of the gradient flow, such that certain long-range correlators for the scheme can be computed directly without the stochastic component of the renormalization scheme. In Section~\ref{subsubsec:WMlattice}, we show that the Carosso scheme~\cite{Carosso2020} can be written in the framework of the Wegner-Morris equation~\cite{wegner1974some, Morris:1999px, Latorre:2000qc, Morris:2005tv}, and thus put on a common footing with more familiar schemes such as Polchinski renormalization~\cite{polchinski1984renormalization}.

The methods described in the present paper unify the ideas behind L\"uscher's gradient flow, normalizing flows for field theories, and variational characterizations of the renormalization group \cite{cotler2022renormalization} into a framework for building physically-interpretable ML models for field-theoretic applications.  Our work also has synergies with the framework of ``Bayesian Renormalization'' developed in~\cite{berman2023bayesian}, and may also have applications in the related mathematical subject of \textit{stochastic localization} which is reviewed in e.g.~\cite{el2022information, bauerschmidt2023stochastic}.

\subsection{Other RG schemes}
\label{sec:other-rg-schemes}

\subsubsection{Wegner-Morris RG flow on the lattice}
\label{subsubsec:WMlattice}

So far we have focused on Carosso's RG scheme (see e.g.~\eqref{E:Carossointro1v2} and~\eqref{E:Carossoexpect2v2}) for its simplicitly.  However, there are a wide variety of other RG schemes available which have different features and tradeoffs.

First we must comment on what qualifies as an RG scheme.  Since RG as a conceptual framework is rather capacious, there are few general rules for which RG schemes are allowed versus disallowed; the praxis is to be inclusive of any scheme which produces results which can be justified on the grounds of quantitative reasoning and conceptual appeals to universality.  However, there do exist desirable properties for RG flows to posses, which can be achieved by special subclasses of RG schemes.  Let us give an example of a desirable property following Wilson~\cite{Wilson:1973jj} and Polchinski~\cite{polchinski1984renormalization}.

Suppose we have a distribution $p_0(\phi)$ on which we would like to enact an RG flow.  On the lattice, the natural short-distance cutoff scale is $\epsilon = L/N$; then we say that the initial probability density $p_0(\phi)$ captures correlations at distances larger than $L/N$.  Equivalently, we can say that $p_0(\phi)$ captures correlations at momenta smaller than $\sim 1/\epsilon = N/L$.  For convenience, let us take $\Lambda_0 = 2\sqrt{d}\,N/L$.  Now let $f(t)$ be a strictly monotonically decreasing function of $t$ such that $f(0) = 1$.  We desire an RG flow such that $p_t(\phi)$ captures all correlations at momenta smaller than $\Lambda_0\, f(t)$.  In other words, as we flow $p_t(\phi)$ for increasing $t$, it continues to capture correlations at momentum scales smaller than $\Lambda_0\,f(t)$, but correlations at larger momenta may not be preserved.  This is the sense in which the RG flow preserves only correlations at progressively smaller momentum scales, corresponding to progressively larger distance scales.

More precisely, we want our RG flowed distribution $p_t(\phi)$ to satisfy
\begin{equation}
\label{E:RGproperty1}
\mathbb{E}_{\phi \sim p_t(\phi)}[\widetilde{\phi}(\textbf{p}_1) \widetilde{\phi}(\textbf{p}_2) \cdots \widetilde{\phi}(\textbf{p}_r)] \approx \mathbb{E}_{\phi \sim p_0(\phi)}[\widetilde{\phi}(\textbf{p}_1) \widetilde{\phi}(\textbf{p}_2) \cdots \widetilde{\phi}(\textbf{p}_r)]\,, \quad \text{for all }\,\,\,|\widehat{\textbf{p}}_i| \leq \Lambda_0 f(t)
\end{equation}
for some strictly monotonically decreasing $f(t)$ with $f(0) = 1$.  Above, the number $r$ of $\widetilde{\phi}$'s in the correlator is arbitrary.  The property~\eqref{E:RGproperty1} means that $p_t(\phi)$ has the same long-distance correlators (dictated by small momentum) as $p_0(\phi)$.  As $t$ grows, the momentum scale below which correlators of $p_0$ and $p_t$ agree becomes smaller and smaller; equivalently, the distance scale above which $p_0$ and $p_t$ agree becomes larger and larger.  We would also like for $p_t(\phi)$ to induce a flow on samples $\widetilde{\phi}(\textbf{p})$ which mixes modes in momentum space, as discussed in detail in Section~\ref{subsec:continuoustime}.
There is a very nice class of RG schemes which satisfy the property~\eqref{E:RGproperty1}.  One of the more famous example is Polchinski's scheme~\cite{polchinski1984renormalization}, which can be viewed as a special case of the Wegner-Morris flow equation~\cite{wegner1974some, Morris:1999px, Latorre:2000qc, Morris:2005tv} (which is discussed in more detail in Appendix~\ref{App:ERGreview}).  Indeed, the Carosso scheme~\cite{Carosso2020} can also be viewed as a special case of the Wegner-Morris flow equation.

Let us briefly explain the lattice version of the Wegner-Morris flow equation here.  We will specialize to a particular class of Wegner-Morris flows reviewed in e.g.~\cite{rosten2012fundamentals, cotler2022renormalization}.  To define the flow, we require two basic objects, which are prescribed functions of $t$:
\begin{enumerate}
    \item \textit{A cutoff function} $\widetilde{B}_t(|\widehat{\textbf{p}}|)$.  There is a non-increasing function $g(t)$ for $t \geq 0$ with $g(0) = 1$ such that $B_t(|\widehat{\textbf{p}}|)$ goes rapidly to zero for $|\widehat{\textbf{p}}| \geq g(t)$.  We further require that $\widetilde{B}_t(|\widehat{\textbf{p}}|)$ is $O(1)$ for $|\widehat{\textbf{p}}| \leq \Lambda_0 g(t)$.  Note that this $g(t)$ is distinct from the $f(t)$ discussed above; in fact we will later see that $f(t) \leq g(t)$. 
    \item \textit{A seed probability density $q_t(\phi)$}.  This probability density has the property that 
    \begin{equation}
    \mathbb{E}_{\phi \sim q_t(\phi)}[\widetilde{\phi}(\textbf{p}_1) \, \widetilde{\phi}(\textbf{p}_2) \cdots \widetilde{\phi}(\textbf{p}_r)] \approx 0 \quad \text{for any}\,\,\,|\widehat{\textbf{p}}_i| \geq \Lambda_0 g(t)\,,
    \end{equation}
    where $g(t)$ is the same function which controls the cutoff function.  This means that correlation functions with momenta $|\widehat{\textbf{p}}|$ greater than $\Lambda_0 g(t)$ are suppressed.  By convention we write the seed probability density as $q_t(\phi) = \frac{1}{Z_q}\,e^{- 2\widehat{S}_t(\phi)}$, where $\widehat{S}_t(\phi)$ is called the \textit{seed action}.
\end{enumerate}
With the above ingredients, the lattice Wegner-Morris flow equation is
\begin{equation}
\label{E:WegnerMorrisLattice1}
\boxed{\partial_t p_t(\phi) = \frac{1}{2} \sum_{\textbf{n},\textbf{m} \in \mathbb{Z}_N^d} B_t(|\textbf{n}-\textbf{m}|_{\mathbb{Z}_N^d}) \left(\frac{\partial^2 p_t(\phi)}{\partial \phi(\textbf{n})\, \partial \phi(\textbf{m})} + 2 \,\frac{\partial}{\partial \phi(\textbf{n})}\!\left(\frac{\partial \widehat{S}_t(\phi)}{\partial \phi(\textbf{m})}\,p_t(\phi)\right)\right)}
\end{equation}
where we have used
\begin{equation}
B_t(|\textbf{n}-\textbf{m}|_{\mathbb{Z}_N^d}) := \sum_{\textbf{p} \in \mathbb{Z}_N^d} e^{i \frac{2\pi}{N}\,\textbf{p}\cdot(\textbf{n}- \textbf{m})} \widetilde{B}_t(|\textbf{p}|)\,.
\end{equation}
We notice that~\eqref{E:WegnerMorrisLattice1} is a lattice version of a \textit{convection-diffusion equation}, where $\frac{\partial^2 p_t(\phi)}{\partial \phi(\textbf{n})\, \partial \phi(\textbf{m})}$ is the diffusive term and $2 \,\frac{\partial}{\partial \phi(\textbf{n})}\!\left(\frac{\partial \widehat{S}_t(\phi)}{\partial \phi(\textbf{m})}\,p_t(\phi)\right)$ is the convective term.  We can rewrite~\eqref{E:WegnerMorrisLattice1} in momentum space as
\begin{equation}
\label{E:WegnerMorrisLattice2}
\boxed{\partial_t p_t(\widetilde{\phi}) = \frac{1}{2} \sum_{\textbf{p} \in \mathbb{Z}_N^d} \widetilde{B}_t(|\textbf{p}|) \left(\frac{\partial^2 p_t(\widetilde{\phi})}{\partial 
\widetilde{\phi}(\textbf{p})\, \partial \widetilde{\phi}(-\textbf{p})} + 2 \,\frac{\partial}{\partial \widetilde{\phi}(\textbf{p})}\!\left(\frac{\partial \widehat{S}_t(\widetilde{\phi})}{\partial \widetilde{\phi}(-\textbf{p})}\,p_t(\widetilde{\phi})\right)\right)}
\end{equation}

It is pleasing that the Wegner-Morris flow is an RG flow scheme taking the form of a convection-diffusion equation for $p_t(\phi)$.  This makes intuitive sense: a convection-diffusion equation smooths out a probability density.  Interestingly, our requirements on the cutoff function and seed probability density in fact give the convection-diffusion equation special structure, in particular so that~\eqref{E:RGproperty1} is satisfied.  The connection between RG flows and convection-diffusion equations, as well as their direct interplay with the theory of optimal transport, is analyzed in detail in~\cite{cotler2022renormalization}.  A recent study~\cite{berman2023bayesian} further connects these kinds of convection-diffusion equations to the subject of Bayesian inference.

It may appear that~\eqref{E:WegnerMorrisLattice1} is rather different than e.g.~the Carosso scheme, which we formulated using an SDE for $\phi_t$.  However, there is in fact an SDE formulation of~\eqref{E:WegnerMorrisLattice1}, which as a special case reproduces the Carosso scheme.  In particular, consider the SDE
\begin{equation}
\label{E:generalSDE1}
\partial_t \phi_t(\textbf{n}) = - \sum_{\textbf{m} \in \mathbb{Z}_N^d} B_t(|\textbf{n}-\textbf{m}|_{\mathbb{Z}_N^d}) \, \frac{\partial \widehat{S}_t(\phi)}{\partial \phi(\textbf{m})} + \eta_t(\textbf{n})\,, \qquad \phi_0(\textbf{n}) = \phi(\textbf{n})\,,
\end{equation}
with the noise being Gaussian and satisfying
\begin{equation}
\label{E:generalSDE2}
\mathbb{E}[\eta_t(\textbf{n})] = 0\,, \qquad \mathbb{E}[\eta_t(\textbf{n})\eta_s(\textbf{m})] = \delta(t-s)\,B_t(|\textbf{n}-\textbf{m}|_{\mathbb{Z}_N^d})\,.
\end{equation}
Letting $\phi_t[\psi, \eta_t]$ be a solution~\eqref{E:generalSDE1} with initial condition $\psi$ and for a fixed sample of the noise $\eta_t$, the Wegner-Morris flow equation~\eqref{E:WegnerMorrisLattice1} is equivalent to
\begin{equation}
p_t(\phi) = \int \!\prod_{\textbf{n} \in \mathbb{Z}_N^d} \! d\psi(\textbf{n})\,\,\mathbb{E}_{\eta_t}\!\left[\delta(\phi - \phi_t[\psi, \eta_t])\right]\,p_0(\psi)\,.
\end{equation}
In other words, to sample from a $p_t(\phi)$ at time $t$ which satisfies the Wegner-Morris equation~\eqref{E:generalSDE1}, we simply have to sample $\phi \leftarrow p_0(\phi)$ and flow $\phi$ to time $t$ according to the SDE given by~\eqref{E:generalSDE1},~\eqref{E:generalSDE2}.  This idea has been emphasized recently in the literature by Carosso~\cite{Carosso2020} in the context of his scheme.

For completeness, we include the momentum space version of~\eqref{E:generalSDE1} and~\eqref{E:generalSDE2}, namely
\begin{equation}
\label{E:generalSDE3}
\partial_t \widetilde{\phi}_t(\textbf{p}) = - \widetilde{B}_t(|\widehat{\textbf{p}}|) \, \frac{\partial \widehat{S}_t(\widetilde{\phi})}{\partial \widetilde{\phi}(-\textbf{p})} + \widetilde{\eta}_t(\textbf{p})\,, \qquad \widetilde{\phi}_0(\textbf{p}) = \widetilde{\phi}(\textbf{p})\,,
\end{equation}
where the noise is Gaussian and satisfies
\begin{equation}
\label{E:generalSDE4}
\mathbb{E}[\widetilde{\eta}_t(\textbf{p})] = 0\,, \qquad \mathbb{E}[\widetilde{\eta}_t(\textbf{p})\widetilde{\eta}_s(\textbf{k})] = \delta(t-s)\,\widetilde{B}_t(|\widehat{\textbf{p}}|)\,\delta_{\textbf{p},-\textbf{k}}\,.
\end{equation}

Below we explain how the Wegner-Morris equation~\eqref{E:generalSDE1}, or equivalently its SDE formulation in~\eqref{E:generalSDE1},~\eqref{E:generalSDE2}, specializes to the Carosso scheme and the Polchinski scheme.  Then we make more comments about Wegner-Morris flows more generally, including how to compare between different RG schemes.

\subsubsection{Recovering the Carosso scheme}

By comparing~\eqref{E:generalSDE1},~\eqref{E:generalSDE2} with~\eqref{E:Carossointro1v2},~\eqref{E:Carossoexpect2v2}, we see that the Carosso scheme corresponds to
\begin{align}
B_t(|\textbf{n}-\textbf{m}|_{\mathbb{Z}_N^d}) = \left(\frac{N}{L}\right)^{\! d} \delta_{\textbf{n},\textbf{m}}\,, \qquad \widehat{S}_t(\phi) = - \frac{(L/N)^d}{2} \sum_{\textbf{n} \in \mathbb{Z}_N^d} \phi(\textbf{n}) \Delta \phi(\textbf{n})\,. 
\end{align}
In particular, both the cutoff function $B_t$ and the seed action $\widehat{S}_t$ (coming from the seed probability density $q_t(\phi)$) are $t$-independent in this scheme.  This corresponds to the special case of $g(t) = 1$.

In momentum space, the Carosso scheme suppresses the mean value of the $\widetilde{\phi}(\textbf{p})$'s when $|\widehat{\textbf{p}}|$ is sufficiently large; in particular from~\eqref{eq:carosso-kernel} we saw that $\mathbb{E}_{\eta_t}[\widetilde{\phi}_t[\widetilde{\phi}_0, \eta_t](\textbf{p})] = e^{- |\widehat{\textbf{p}}|^2 t} \,\widetilde{\phi}_0(\textbf{p})$.  As such, examining~\eqref{eq:carosso-kernel}, we approximately have $f(t) \sim \frac{L}{2 N \sqrt{d}}\,\frac{1}{\sqrt{t}}$ for large $t$.

\subsubsection{Recovering the Polchinski scheme}

In~\cite{polchinski1984renormalization}, Polchinski designed a nice RG scheme for a scalar $\phi^4$ theory in the continuum. Here we give a latticized version of Polchinski's flow.  Polchinski begins by writing out the scalar $\phi^4$ theory with a cutoff in momentum space, which we recapitulate on the lattice.  Defining $K_t(|\widehat{\textbf{p}}|) := e^{- |\widehat{\textbf{p}}|^2/(e^{-t} \Lambda_0)^2}$, we have
\begin{align}
\label{E:Polchinskiinitial1}
p_0(\phi) &= \frac{1}{Z}\,\exp\!\Bigg(- L^d \sum_{\textbf{p} \in \mathbb{Z}_N^d} \frac{1}{K_0(|\widehat{\textbf{p}}|)}\,\frac{1}{2}\,\widetilde{\phi}(\textbf{p}) (|\widehat{\textbf{p}}|^2 + m^2)  \widetilde{\phi}(-\textbf{p}) \nonumber \\
& \qquad \qquad \qquad - L^d \sum_{\textbf{p}_1,\textbf{p}_2,\textbf{p}_3,\textbf{p}_4 \in \mathbb{Z}_N^d} \! \frac{\lambda}{4!}\,\widetilde{\phi}(\textbf{p}_1) \widetilde{\phi}(\textbf{p}_2) \widetilde{\phi}(\textbf{p}_3) \widetilde{\phi}(\textbf{p}_4)\, \delta_{\textbf{p}_1 + \textbf{p}_2 + \textbf{p}_3 + \textbf{p}_4, 0} \bigg),
\end{align}
where $\Lambda_0 = 2 \sqrt{d} N/L$ as usual.  Notice the $\frac{1}{K_0(|\widehat{\textbf{p}}|)} = e^{|\widehat{\textbf{p}}|^2/\Lambda_0^2}$ term appearing in front of the $\frac{1}{2}\,\widetilde{\phi}(\textbf{p}) (|\widehat{\textbf{p}}|^2 + m^2)  \widetilde{\phi}(-\textbf{p})$ part of the action.  At the moment it may appear superfluous; it forces modes $\widetilde{\phi}(\textbf{p})$ with $|\widehat{\textbf{p}}| \gtrsim \Lambda_0$ to be zero with small variance, but this is already accomplished by the lattice discretization since momenta with $|\widehat{\textbf{p}}| > \Lambda_0$ simply do not exist.  This being said, the role of the $\frac{1}{K_0(|\widehat{\textbf{p}}|)} = e^{|\widehat{\textbf{p}}|^2/\Lambda_0^2}$ term will become clear shortly.

Considering the Polchinski scheme, in momentum space we have
\begin{equation}
\label{E:Polchinskidefs1}
\widetilde{B}_t(|\widehat{\textbf{p}}|) = - \frac{1}{L^d}\frac{1}{|\widehat{\textbf{p}}|^2 + m^2}\, \partial_t K_t(|\widehat{\textbf{p}}|)\,, \qquad \widehat{S}_t(\phi) = \frac{L^d}{2} \sum_{\textbf{p} \in \mathbb{Z}_N^d} \frac{1}{K_t(|\widehat{\textbf{p}}|)}\,\widetilde{\phi}(\textbf{p}) (|\widehat{\textbf{p}}|^2 + m^2) \widetilde{\phi}(-\textbf{p})\,.
\end{equation}
This scheme is designed so that the RG flow of $p_0(\phi)$ for $\lambda = 0$ is simply
\begin{align}
\label{E:simpleflow1}
p_t(\phi) &= \frac{1}{Z_t}\,\exp\!\Bigg(- \frac{L^d}{2} \sum_{\textbf{p} \in \mathbb{Z}_N^d} \frac{1}{K_t(|\widehat{\textbf{p}}|)}\,\widetilde{\phi}(\textbf{p}) (|\widehat{\textbf{p}}|^2 + m^2)  \widetilde{\phi}(-\textbf{p})\Bigg).
\end{align}
That is, the \textit{free theory} (i.e.~having a Gaussian initial probability density $p_0(\phi)$ since $\lambda = 0$) has a simple flow, in which modes with $|\widehat{\textbf{p}}| \gtrsim e^{-t}\Lambda_0$ are exponentially suppressed.  In a sense, this nice feature of the Polchinski flow is the reason for the $\frac{1}{K_0(|\widehat{\textbf{p}}|)}$ term in~\eqref{E:Polchinskiinitial1}; this term conspires with~\eqref{E:Polchinskidefs1} to produce the simple flow in~\eqref{E:simpleflow1} of the free theory.

Even for general $\lambda$, the Polchinski flow (with the particular choice of $K_t(|\widehat{\textbf{p}}|)$ given here), the Polchinski scheme suppresses the mean value of $\widetilde{\phi}(\textbf{p})$'s as $\mathbb{E}_{\eta_t}[\widetilde{\phi}_t[\widetilde{\phi}_0, \eta_t](\textbf{p})] = \exp\!\left(- (e^{2t}-1)\frac{|\widehat{\textbf{p}}|^2}{\Lambda_0^2}\right) 
 \widetilde{\phi}_0(\textbf{p})$.  Accordingly, we have that $f(t) = g(t) = e^{-t}$.  Thus the notion of time $t$ in the Polchinski scheme is exponentially different than in the Carosso scheme: the Polchinski scheme suppresses high-momentum modes exponentially faster.

 The transition kernel $\mathbb{E}_{\eta_t}[\delta(\phi_t - \phi_t[\phi_0, \eta_t])]$ in the Polchinski setting is given by
\begin{align}
\label{E:polchinski_kernel}
p_{t,0}^{\text{Polchinski}}(\widetilde{\phi}_t | \widetilde{\phi}_0) \propto \prod_{\textbf{p} \in \mathbb{Z}_N^d} \exp\!\left(- \frac{L^d}{2} \frac{|\widehat{\textbf{p}}|^2 + m^2}{K_t(|\widehat{\textbf{p}}|^2) - \frac{K_t(|\widehat{\textbf{p}}|^2)^2}{K_0(|\widehat{\textbf{p}}|^2)}}\,\left|\widetilde{\phi}_t(\textbf{p}) - \frac{K_t(|\widehat{\textbf{p}}|^2)}{K_0(|\widehat{\textbf{p}}|^2)} \,\widetilde{\phi}_0(\textbf{p})\right|^2 \right)
\end{align}
This expression holds for any $K_t(|\widehat{\textbf{p}}|^2)$, i.e.~not merely $K_t(|\widehat{\textbf{p}}|^2) = e^{-|\widehat{\textbf{p}}|^2/(e^{-t}\Lambda_0)^2}$.  For purposes of slowing down the flow, we will later use the kernel
\begin{align}
\label{E:modifiedK1}
K_t(|\widehat{\textbf{p}}|^2) = e^{- (|\widehat{\textbf{p}}|^2 + M^2)\,t}
\end{align}
where $M$ is a mass parameter satisfying $M/\Lambda_0 \ll 1$.  The main advantage of~\eqref{E:modifiedK1} is that the exponential of~\eqref{E:polchinski_kernel} does not blow up for $\textbf{p} = \textbf{0}$.  For our previously-defined kernel this blow-up does occur, which means that the momentum zero mode is preserved by the flow.  Using~\eqref{E:modifiedK1} is more numerically stable, and so we will opt to use it in Section~\ref{sec:numerics}.

\subsubsection{More general features}

Any Wegner-Morris type schemes, which satisfy~\eqref{E:RGproperty1}, can be readily compared at low-momentum.  In particular, suppose we have two such schemes with different $f(t)$'s, which we will call $f_1(t)$ and $f_2(t)$.  Given an initial probability distribution $p_0(\phi)$, suppose its flow under the first scheme is denoted by $p_{t}^{(1)}(\phi)$, and its flow under the second scheme is denoted by $p_{t}^{(2)}(\phi)$.  Then we have
\begin{align}
\mathbb{E}_{\phi \sim p_{t}^{(1)}(\phi)}[\widetilde{\phi}(\textbf{p}_1) \widetilde{\phi}(\textbf{p}_2) \cdots \widetilde{\phi}(\textbf{p}_r)] \approx \mathbb{E}_{\phi \sim p_{t}^{(2)}(\phi)}[\widetilde{\phi}(\textbf{p}_1) \widetilde{\phi}(\textbf{p}_2) \cdots \widetilde{\phi}(\textbf{p}_r)]\,, \,\,\, \text{for all }\,\,\,|\widehat{\textbf{p}}_i| \leq \Lambda_0\min\{f_1(t), f_2(t)\}\,.
\end{align}
The minimum on the right-hand side accounts for the fact that the two schemes perform RG at different rates as a function of $t$.  This is all fine, but the more interesting setting for comparison is when both flows reach the same RG fixed point and we compare their properties at \textit{all} momentum scales.  To understand this, we need to first address how in the Wegner-Morris flow schemes we can find an RG fixed point in the first place.

Let us begin by discussing the Carosso and Polchinski schemes in particular.  In both the Carosso and Polchinski schemes, as $t$ increases, the flow of $p_t(\phi)$ erases more and more of the high-frequency information about the initial distribution specified by $p_0(\phi)$.  This erasure might na\"{i}vely seem problematic since these flows appear to preclude the possibility of finding a non-trivial RG fixed point, importantly even in more general settings beyond $\phi^4$ theory.  However, the key is that to access a fixed point, we need to judiciously rescale the fields $\phi$ and couplings in a $t$-dependent manner, as discussed in the setting of a discrete-time flow in~\eqref{E:criticalcondition0},~\eqref{E:criticalcondition1}.  In other words, we need to appropriately `zoom in' on the interesting fluctuations in our distribution, or else they will be lost to us.   

More explicitly, suppose we have a distribution $p_{t,\{\lambda_i(t)\}}(\phi)$ where $\{\lambda_i(t)\}$ are the couplings in the action after an amount of RG flow $t$.  To be maximally explicit, let us reprise our notation from Section~\ref{subsec:phasetransitions} with some slight modifications.  Suppose $p_{t,\{\lambda_i(t)\}}(\phi)$ is well approximated by
\begin{align}
\label{E:morecomplicatedpt}
p_{t, \{\lambda_i(t)\}}(\phi) \approx \frac{1}{Z_{t}} \, \exp\!\left(- \left(\frac{L}{N}\right)^{\! d}\sum_i\,\lambda_{i}(t)\,M_{i,N}[\phi]\right)
\end{align}
where as before the $M_{i,N}[\phi]$'s given by
\begin{align}
\label{E:Mdef2}
M_{i,N}[\phi] = \sum_{\textbf{n} \in \mathbb{Z}_N^d}\prod_{k = 0}^{k_{\max, i}} (\Delta^k \phi(\textbf{n}))^{q_{i,k}}
\end{align}
such that each $q_{i,k} \in \mathbb{Z}_{\geq 0}$.  It is prudent for us to define
\begin{align}
\delta_i := 2 \sum_{k = 0}^{k_{\max,i}} k\, q_{i,k}
\end{align}
which counts the powers of $\epsilon = L/N$ appearing in $M_{i,N}[\phi]$ due to the Laplacians.  Moreover, suppose that $t$ corresponds to an effective cutoff scale $\Lambda_t$\,, which depends on the RG scheme employed.  Then at some time $t_*$, the couplings $\{\lambda_i(t_*)\}$ define an (approximate) RG fixed point if for $t' \geq 0$ there exists a monotonic function $b_{t'}$ with $b_0 = 1$ such that
\begin{equation}
p_{t_* + t',\,\{\lambda_i(t_* +t')\}}(\phi)\, \prod_{\textbf{n} \in \mathbb{Z}_N^d} d\phi(\textbf{n}) \approx p_{t_* ,\,\{(\frac{\Lambda_{t_*+t'}}{\Lambda_{t_*}})^{d + \delta_i}\lambda_i(t_*)\}}\!(b_{t'}\phi) \, \prod_{\textbf{n} \in \mathbb{Z}_N^d} b_{t'}\,d\phi(\textbf{n})\,.
\end{equation}
An equivalent formulation is that there is a scaling function $c_t$ so that
\begin{equation}
p_{t,\,\{(\frac{\Lambda_t}{\Lambda_0})^{d + \delta_i}\lambda_{i}(t)\}}(c_t \phi)\,\prod_{\textbf{n} \in \mathbb{Z}_N^d} c_{t}\,d\phi(\textbf{n}) \,\,\longrightarrow\,\, p^{\text{fixed pt}}_{\{\hat{\lambda}_i\}}(\phi)\,\prod_{\textbf{n} \in \mathbb{Z}_N^d} d\phi(\textbf{n})
\end{equation}
as $t$ becomes large (but not so large that $\Lambda_t/\Lambda_0 \approx 0$), where $\{\hat{\lambda}_i\}$ is a fixed set of couplings.

We can easily accommodate for such a $c_t$ in the flow equations by modifying the seed action $\widehat{S}_t(\widetilde{\phi})$ as 
\begin{equation}
\widehat{S}_t^{\text{new}}(\widetilde{\phi}) := \widehat{S}_t(\widetilde{\phi}) - \partial_t \log(c_t) \sum_{\textbf{p} \in \mathbb{Z}_N^d} \frac{1}{\widetilde{B}_t(|\textbf{p}|)}\,\widetilde{\phi}(\textbf{p})\,\widetilde{\phi}(-\textbf{p})\,. 
\end{equation}
For instance, the SDE formulation of the Carosso scheme becomes
\begin{equation}
\label{E:Carossobshift}
\partial_t \phi_t(\textbf{n}) = \left(\Delta + \partial_t \log(c_t)\right) \phi(\textbf{n}) + \eta_t(\textbf{n})\,,\qquad \phi_0(\textbf{n}) = \phi(\textbf{n})\,,
\end{equation}
where the distribution of $\eta_t$ is unchanged.  To additionally implement the $\frac{\Lambda_t}{\Lambda_0}$ rescalings of the couplings $\lambda_i(t)$ (which is equivalent to a rescaling of the position or momenta), we can augment~\eqref{E:Carossobshift} by (see e.g.~\cite{rosten2012fundamentals})
\begin{align}
\label{E:Carossobshift2}
\partial_t \phi_t(\textbf{n}) = \left(\Delta + \partial_t \log(c_t)\right) \phi(\textbf{n}) - \partial_t \log(\Lambda_t)\,\textbf{n}\cdot \textbf{D}\phi(\textbf{n}) + \eta_t(\textbf{n})\,,\qquad \phi_0(\textbf{n}) = \phi(\textbf{n})\,,
\end{align}
where
\begin{align}
\textbf{D}\phi(\textbf{n}) := \left(\frac{1}{\left(\frac{L}{N}\right)}(\phi(\textbf{n} + \textbf{e}_1) - \phi(\textbf{n})),\,\frac{1}{\left(\frac{L}{N}\right)}(\phi(\textbf{n} + \textbf{e}_2) - \phi(\textbf{n})),...\,,\frac{1}{\left(\frac{L}{N}\right)}(\phi(\textbf{n} + \textbf{e}_d) - \phi(\textbf{n}))\right)\,,
\end{align}
and again the distribution of $\eta_t$ is unchanged.

We emphasize that the $b_t$ or $c_t$ required to access a fixed point is scheme-dependent; they will be different for e.g.~the Carosso scheme versus the Polchinski scheme.  The simplest way to see this is that the Carosso scheme and Polchinski scheme can suppress high-momentum modes at different rates.

A key problem is that, unless we have detailed analytic control over some combination of our RG scheme and our RG fixed point of interest, it is difficult to write down a suitable $b_t$ or $c_t$ a priori.  This being said, if we suspect that a $p_{t,\,\{\lambda_i(t)\}}(\phi)$ is nearby a fixed point, we can use heuristics to search for an appropriate scaling function $b_t$\,, which if found would corroborate the presence of an RG fixed point.  For the some, assume (somewhat unrealistically) that we have access to the explicit form of $p_{t,\{\lambda_i(t)\}}$.  Then a straightforward heuristic is to fix a small time $\delta t' > 0$ and find a constant $b$ such that
\begin{equation}
\label{E:twopointheuristic1}
\mathbb{E}_{\phi \sim p_{t_* + \delta t',\{\lambda_i(t_* +\delta t')\}}(\phi)}[\phi(\textbf{n}) \phi(\textbf{m})] \approx \frac{1}{b^2}\,\mathbb{E}_{\phi \sim p_{t,\{(\frac{\Lambda_{t_* + \delta t'}}{\Lambda_{t_*}})^{d + \delta_i}\lambda_i(t)\}}\!(\phi)}[\phi(\textbf{n}) \phi(\textbf{m})]\,,
\end{equation}
where both sides are functions of $|\textbf{n}-\textbf{m}|_{\mathbb{Z}_N^d}$ due to translation-invariance.  If such a $b$ exists, then it is reasonable to guess that $b = b_{\delta t'}$\,.  One can also look at higher-point analogs of~\eqref{E:twopointheuristic1}.

In more realistic settings, we do not have direct access to the explicit form of $p_{t,\{\lambda_i(t)\}}$.  Then it is standard to rely on \textit{order parameters}, which are certain combinations of correlation functions of $p_{t,\{\lambda_i(t)\}}(\phi)$ (with no rescalings of the couplings $\{\lambda_i(t)\}$ or field $\phi$) which diagnose the presence of a critical point.  One drawback is that order parameters must be tailored to the critical point of interest; a priori, if we do not know anything about the critical point that an RG flow may land on, then it is unclear how to find an order parameter that detects it.  For a review of order parameters in standard statistical field theories, see~\cite{kardar2007statistical}.

Now suppose we have two different Wegner-Morris RG flow schemes for the same initial $p_0(\phi)$; we denote their respective flows by $p_{t,\,\{\hat{\lambda}_i^{(1)}\}}^{(1)}(\phi)$ and $p_{t,\,\{\hat{\lambda}_i^{(2)}\}}^{(2)}(\phi)$, where $\{\hat{\lambda}_i^{(1)}\}$ denotes the fixed-point couplings for the first flow, and $\{\hat{\lambda}_i^{(2)}\}$ denotes the fixed-point couplings for the second flow.  We do not in general expect $\{\hat{\lambda}_i^{(1)}\} \approx \{\hat{\lambda}_i^{(2)}\}$, nor do we expect $p_{t,\,\{\hat{\lambda}_i^{(1)}\}}^{(1)}(\phi) \approx p_{t,\,\{\hat{\lambda}_i^{(2)}\}}^{(2)}(\phi)$.  However, universality of RG fixed point suggests that the two fixed points reached by two distinct RG flow schemes on the same initial $p_0(\phi)$ are related by a rescaling, namely that there is a constant $c$ such that
\begin{equation}
p_{t,\,\{\hat{\lambda}_i^{(1)}\}}^{(1)}(\phi) \prod_{\textbf{n} \in \mathbb{Z}_N^d} d\phi(\textbf{n}) \approx p_{t,\,\{\hat{\lambda}_i^{(2)}\}}^{(2)}(c\,\phi)\,\prod_{\textbf{n} \in \mathbb{Z}_N^d} c\,d\phi(\textbf{n})\,.
\end{equation}
The constant $c$ can be ascertained by comparing the second moments of each distribution akin to our heuristic algorithm in~\eqref{E:twopointheuristic1}; in particular:
\begin{equation}
\label{E:twopointheuristic2}
\mathbb{E}_{\phi \sim p_{t,\,\{\hat{\lambda}_i^{(1)}\}}^{(1)}(\phi)}[\phi(\textbf{n}) \phi(\textbf{m})] \approx \frac{1}{c^2}\,\mathbb{E}_{\phi \sim p_{t,\,\{\hat{\lambda}_i^{(2)}\}}^{(2)}(\phi)}[\phi(\textbf{n}) \phi(\textbf{m})]\,.
\end{equation}
Having discovered this $c$, which amounts to a choice of normalization of the $\phi$ field, we can now explore the same aspects of the RG fixed point with each scheme in a manner such that the results will (approximately) agree.

In other applications of RG on the lattice, one considers theories which are well-approximated by a finite number of terms in the action at long distances (i.e.~a \textit{renormalizable} EFT).  Let the couplings associated to those terms be called $\{\lambda_i(t)\}$.  One desires to pick initial values of those couplings $\{\lambda_i(0)\}$ for the theory at short distances (in the UV) so that they flow to a desired set of couplings $\{\lambda_i(T)\}$ at the long distance scale corresponding to $T$ (in the IR); all other possible couplings not in the set can be initially set to zero since they will be suppressed at long distances anyway.  More specifically, suppose that for a particular renormalization grouop scheme, all momentum modes with $|\widehat{\textbf{p}}_i| \geq \Lambda_0 f(T)$ are suppressed, and all momentum modes with $|\widehat{\textbf{p}}_i| \leq \Lambda_0 f(T)$ are unsuppressed.  Let us say that we have a physical system for which we have measured (some of) the effective couplings at the momentum scale corresponding to $\Lambda_0 f(T)$, for a fixed $T$.  Then, to match this to lattice data, we would like to tune the $\{\lambda_i(0)\}$ so that they flow to the right couplings at $t = T$.  Having identified the appropriate $\{\lambda_i(0)\}$, our lattice model can now make predictions about the values of the couplings at $t \not = T$, for instance correlation functions at momentum scales other than $\Lambda_0 f(T)$.

Some of the ideas and techniques for RG in this section also apply to models outside the purview of Effective Field Theory, such as image models.  Perhaps a useful insight in this more general context is as follows.  We know that RG flows are sensitive to recaling $\phi$ along the flow, especially if want to access interesting correlations.  As such, it could be prudent to consider the pushforward of $p_t(\phi)$ under rescalings $\phi \to b\,\phi$ by a judicious constant $b$ which may allow one to `zoom in' on the interesting correlations.  Such a $b$ could be identified in the following way.  Let $F_b(\phi) = b \, \phi$, and define $\overline{\phi}(\textbf{n}) := \mathbb{E}_{\phi \sim F_{b\,*} p_t(\phi)}$.  Then we may desire to pick a $b$ such that e.g.
\begin{equation}
\Phi(\textbf{n}):=\sum_{\textbf{m}} \mathbb{E}_{\phi \sim F_{b\,*}  p_t(\phi)}\!\left[(\phi(\textbf{n}) - \overline{\phi}(\textbf{n}))(\phi(\textbf{m})-\overline{\phi}(\textbf{m}))\right]
\end{equation}
is $\approx 1$ for $\textbf{n}$ near the center of the lattice (if it does not have periodic boundary conditions, as in the case of an image).  In other words, we let $b$ set the scale of fluctuations in the model.

Another general feature which may be interesting in general models is to choose a $\widehat{S}_t(\phi)$ which contains terms non-quadratic in $\phi$, for instance quartic.  This was discussed in~\cite{Fujikawa:2016qis} (see also comments in~\cite{carosso_thesis}), and may ameliorate the ambiguity of the overall scale of $\phi$ near fixed points; that is, the nonlinearity of the RG flow in some instances may pick out a particular scaling.  At the present time, this avenue of non-quadratic $\widehat{S}_t(\phi)$'s has been less explored.

In sum, we emphasize that the Wegner-Morris-type RG schemes offer broad flexibility for RG flows with desirable properties in momentum space.  Their usage, in physical applications or more generally, needs to be augmented by novel physical intuitions coming from RG theory in order to robustly access desired correlations in the ensuing $p_t(\phi)$'s.

\section{Finding ground states of quantum field theories}
\label{sec:groundstates}

\subsection{Review of difficulties with variational methods in quantum field theory}

A fundamental problem in the study of quantum mechanics is finding the ground state of a system.  At a formal level, quantum systems are in part described by a Hermitian operator $H$ called the \textit{Hamiltonian}, which has bounded spectrum from below.  Suppose, for simplicity, that there is a unique ground state of $H$, i.e.~the eigenvector $|\Psi_{\text{gs}}\rangle$ of $H$ with the smallest eigenvalue $E_{\text{gs}}$.  The eigenvalue $E_{\text{gs}}$ is called the \textit{ground state energy}.  As such, there is a variational principle that recovers the ground state, namely
\begin{equation}
\label{E:RR1}
|\Psi_{\text{gs}}\rangle = \argmin_{|\Psi\rangle} \frac{\langle \Psi | H |\Psi\rangle}{\langle \Psi | \Psi\rangle}\,.
\end{equation}
This is called the \textit{Rayleigh-Ritz} variational principle.  (Above, $\langle \Psi|$ is the Hermitian conjugate of the vector $|\Psi\rangle$.)

The Rayleigh-Ritz variational principle is often used in the following manner.  Suppose we parameterize some submanifold of the space of $|\Psi\rangle$'s with some parameters $\theta$.  The $|\Psi\rangle$ corresponding to a fixed $\theta$ is denoted by $|\Psi^\theta\rangle$.  Then a proxy for the minimization in~\eqref{E:RR1} is
\begin{equation}
\label{E:RR2}
\widetilde{\theta} := \argmin_{\theta} \frac{\langle \Psi^\theta | H |\Psi^\theta\rangle}{\langle \Psi^\theta | \Psi^\theta\rangle}\,,
\end{equation}
where $|\Psi^{\widetilde{\theta}}\rangle$ is an approximation to the true ground state $|\Psi_{\text{gs}}\rangle$, and $\frac{\langle \Psi^{\widetilde{\theta}} |H|\Psi^{\widetilde{\theta}}\rangle}{\langle \Psi^{\widetilde{\theta}}|\Psi^{\widetilde{\theta}}\rangle}$ is an approximation to (and in fact an upper bound for) the true ground state energy $E_{\text{gs}}$.

In practical applications, one attempts to formulate a judicious parameterization $|\Psi^\theta\rangle$ so that, for a given class of Hamiltonians $H$, the optimal $|\Psi^{\widetilde{\theta}}\rangle$ should be close to the true ground state $|\Psi_{\text{gs}}\rangle$.  However, it is most often intractable to prove that a particular parameterization of convenience can be optimized so that the resulting state is in close proximity to a desired ground state.  As such, variational methods are often used as heuristics, and compared with other methods and forms of data including experiments coming from natural systems.  A highly-successful specialization of these methodologies is Density Functional Theory (DFT), which is a workhorse of calculations in quantum chemistry~\cite{hohenberg1964inhomogeneous, kohn1965self, kohn1996density}.

Here we will focus our attention on the setting of quantum field theory, where the application of variational methods is both highly desirable and difficult.  To understand the source of the essential difficulty, we recall some basic facts about quantum field theory via a standard example.  Earlier in this paper, we considered scalar $\phi^4$ theory as a \textit{Euclidean field theory}, e.g.~as a \textit{statistical field theory}.  Now we consider its quantum mechanical counterpart, and focus on the lattice setting in particular.

As usual, we consider a $d$-dimensional lattice in the hypercube with side length $L$, and lattice sites $\frac{L}{N}\,\textbf{n}$ for $\textbf{n} \in \{0,1,...,N\}^d$. We periodically identify opposing sides of the hypercube so that it becomes a torus, and thus $\textbf{n} \in \mathbb{Z}_N^d$.  To each lattice site we associate the Hilbert space $L^2(\mathbb{R})$, so that the total Hilbert space is $\mathcal{H} \simeq \bigotimes_{\textbf{n} \in \mathbb{Z}_N^d} \mathcal{H}_{\textbf{n}} \simeq (L^2(\mathbb{R}))^{\otimes \left(\frac{L}{N}\right)^d}$.  Here each $\mathcal{H}_{\textbf{n}} \simeq L^2(\mathbb{R})$ is a space of $L^2$ functions, which we associate with the variable $x_{\textbf{n}}$.  So, for instance, a function on $\mathcal{H}_{\textbf{n}}$ will be denoted by $f(x_{\textbf{n}})$, a derivative of such a function is denoted with a $\frac{\partial}{\partial x_{\textbf{n}}}$, and so on.  Now the Hamiltonian $H$ operator acts on $\mathcal{H}$, where the Hamiltonian is given by
\begin{align}
\label{E:Heq1}
H := \left(\frac{L}{N}\right)^{\! d}  \sum_{\textbf{n} \in \mathbb{Z}_N^d} \left(- \frac{1}{2} \frac{\partial^2}{\partial x_{\textbf{n}}^2} + \frac{1}{2\!\left(\frac{L}{N}\right)^{\! 2}}\sum_{i=1}^d(x_{\textbf{n} + \textbf{e}_i} - x_{\textbf{n}})^2 + \frac{1}{2}\,m^2\,x_{\textbf{n}}^2 + \lambda\, x_{\textbf{n}}^4\right)\,.
\end{align}
We can view a state $\Psi(\{x_{\textbf{n}}\})$ in the Hilbert space $\mathcal{H}$ as living on a $d$-dimensional spatial lattice, and the Schr\"{o}dinger equation tells us that it evolves in time $t$ as $e^{- i H t} \Psi(\{x_{\textbf{n}}\})$.  For physically-relevant systems, we essentially always have $d=1,2,3$.

With our Hilbert space and Hamiltionian at hand, it is useful to understand the dimensionality of said Hilbert space.  Each $L^2(\mathbb{R})$ tensor factor is infinite-dimensional, but in practice we can truncate each into a $k$-dimensional vector space for some suitable $k$.  With this truncation, the total Hilbert space dimension is $k^{\left(\frac{L}{N}\right)^{\,d}}$, which is enormous.  This has grave implications for applying~\eqref{E:RR1} to our Hamiltonian $H$: if we parameterize an arbitrary state in the Hilbert space, it requires $k^{\left(\frac{L}{N}\right)^{\,d}}$ numbers to specify, and so the minimization would be over $k^{\left(\frac{L}{N}\right)^{\,d}}$ parameters.  This is clearly intractable even for modest system sizes.  As such, to apply the Rayleigh-Ritz variational principle to the setting of quantum field theory, it is imperative to find an efficient parameterization $\Psi^\theta(\{x_{\textbf{n}}\})$ involving a \textit{sub-exponential} number of parameters $\theta$, such that a minimization like~\eqref{E:RR2} over $\theta$ will lead to a good approximation for the ground state of the quantum field theory.  However, there are interesting technical obstructions to finding such an efficient parameterization.

As pointed out by Feynman in~\cite{feynman1987proceedings}, a central difficulty in the setting of quantum field theory is that the minimization in the Rayleigh-Ritz variational principle is overly sensitive to changes in the wavefunction $\Psi^\theta(\{x_{\textbf{n}}\})$ at the lattice scale.  For example, this sensitivity can be gleaned by examination of the $\left(\frac{L}{N}\right)^{\! d}  \frac{1}{2\left(\frac{L}{N}\right)^{\! 2}} \sum_{\textbf{n} \in \mathbb{Z}_N^d} \sum_{i=1}^d(x_{\textbf{n} + \textbf{e}_i} - x_{\textbf{n}})^2$ part of the Hamiltonian~\eqref{E:Heq1}.  In physical terms, changes of the parameters $\theta$ that affect $\Psi^\theta(\{x_{\textbf{n}}\})$ at the lattice scale have an outsized energetic cost, whereas changes of $\theta$ that affect longer-range correlations have comparatively smaller energetic cost.  This presents a problem because, in practice, the medium-range and longer-range correlations are essential for comparing to physical experiments, whereas the short-range correlations tied to the lattice scale can be viewed as non-universal artifacts.  For instance, if our lattice theory is viewed as an approximation to a continuum theory at long distances, then we desire that the lattice nature of our theory (often implemented for numerical convenience) does not hold hostage the accuracy of long-range correlations when a variational optimization is performed.

One strategy to ameliorate Feynman's roadblock is to choose hierarchical ans\"{a}tze for the $\theta$-dependence of $\Psi^\theta(\{x_{\textbf{n}}\})$ so that $\theta$ more equitably controls correlations in the wavefunction across all distance scales.  At a heuristic level, we can imagine performing RG flow in reverse: we start with a fiducial state which parameterizes long-distance correlations and we allot some part of the $\theta$-parameters to describe this state; then we perform some kind of reverse-RG flow to build up shorter-distance correlations on top of the longer-distances ones, wherein the new shorter-distance correlations are described by another part of the $\theta$ parameters.  This is then repeated over multiple rounds until we have a state $\Psi^\theta(\{x_{\textbf{n}}\})$ for which the parameters $\theta$ equitably parameterize correlations across all scales.  This strategy has been concretely implemented in the context of \textit{tensor networks}~\cite{orus2014practical}, which have been remarkably successful for providing tractable and viable ans\"{a}tze for the Rayeleigh-Ritz variational principle in the context of field theories in one spatial dimension~\cite{white1992density, perez2006matrix, vidal2008class} (i.e.~$d = 1$).  However, the setting of two and three spatial dimensions (i.e.~$d = 2,3$) remains mostly out of reach with current tensor network methods (see e.g.~\cite{schuch2007computational} for a discussion of difficulties in $d = 2$), and so new ans\"{a}tze are required.  In particular, existing tensor network ans\"{a}tze in higher spatial dimensions are either not computationally practical, or if they are then they are not capacious enough to well-approximate the true desired ground state wavefunction.  In the Subsections below we describe a class of potentially useful ans\"{a}tze inspired by the reverse-RG logic explained here, and anticipate that this class may have great utility in describing ground states of field theories for all of $d = 1,2,3$.

\subsection{Real-valuedness of the ground state wavefunction}
\label{Subsec:realvalued}

Here we explain a useful fact about the ground state wavefunction $\Psi_{\text{gs}}(\{x_{\textbf{n}}\})$ of a Hamiltonian $H$ with a unique ground state.  The fact will be useful in our variational algorithm described in the next Subsection.

Suppose that $\Psi_{\text{gs}}(\{x_{\textbf{n}}\})$ is the ground state wavefuction of $H$, and that it is complex-valued.  If the energy of the ground state is $E_{\text{gs}}$, then we have
\begin{equation}
(H - E_{\text{gs}}) \, \text{Re}\{\Psi_{\text{gs}}(\{x_{\textbf{n}}\})\} + i \, (H - E_{\text{gs}}) \, \text{Im}\{\Psi_{\text{gs}}(\{x_{\textbf{n}}\})\} = 0\,,
\end{equation}
and due to the Hermiticity of $H$ we find that $(H - E_{\text{gs}}) \, \text{Re}\{\Psi_{\text{gs}}(\{x_{\textbf{n}}\})\}  = 0$ and $(H - E_{\text{gs}}) \, \text{Im}\{\Psi_{\text{gs}}(\{x_{\textbf{n}}\})\}  = 0$ individually.  But if $H$ has a unique ground state, then $\text{Re}\{\Psi_{\text{gs}}(\{x_{\textbf{n}}\})\}$ must be proportional to $\text{Im}\{\Psi_{\text{gs}}(\{x_{\textbf{n}}\})\}$.  But this means that, if $\Psi_{\text{gs}}(\{x_{\textbf{n}}\})$ is $L^2$-normalized, we must have
\begin{equation}
\Psi_{\text{gs}}(\{x_{\textbf{n}}\}) = e^{i \varphi} \,\text{Re}\{\Psi_{\text{gs}}(\{x_{\textbf{n}}\})\}
\end{equation}
for some phase $\varphi$.  But since wavefunctions are only defined up to a global phase (since a global phase $e^{i \varphi}$ does not affect any measurable quantity in a quantum-mechanical theory), it follows that we can take $\varphi = 0$ so that $\Psi_{\text{gs}}(\{x_{\textbf{n}}\})$ is purely real-valued.

In summary, if we have a quantum field theory with a unique ground state wavefunction, then it can be taken to be real.  This holds for the example of~\eqref{E:Heq1} above, and for many other examples.

\subsection{Learning ground states of QFTs with diffusion models}

In this Subsection we develop a variational algorithm for learning ground states of quantum lattice systems such as~\eqref{E:Heq1}.  The basic idea is to leverage the stochastic formalism of the Exact Renormalization Group from Section~\ref{sec:renormalizing-diffusion-models}, and in particular to variationally perform ERG in reverse on a fiducial state to build up correlations at progressively smaller distance scales in such a way that we ultimately arrive at a good approximation to our desired ground state wavefunction.  Before describing our algorithm, let us develop some further notation.

Let us reprise the Hamiltonian in~\eqref{E:Heq1}, writing it in a slightly more compact form as
\begin{equation}
\label{E:Heq2}
H = \left(\frac{L}{N}\right)^{\! d}  \left(- \frac{1}{2} \sum_{\textbf{n} \in \mathbb{Z}_N^d} \frac{\partial^2}{\partial x_{\textbf{n}}^2} + \mathcal{F}(\{x_{\textbf{n}}\})\right)\,.
\end{equation}
We previously denoted the ground state wavefunction by the notation $\Psi_{\text{gs}}(\{x_{\textbf{n}}\})$, which is a function of $N^d$ variables.  But now let us define a function
\begin{equation}
\phi : \mathbb{Z}_N^d \longrightarrow \mathbb{R}\,,\qquad \phi(\textbf{n}) = x_{\textbf{n}}\,,
\end{equation}
so that we have
\begin{equation}
\frac{\partial}{\partial \phi(\textbf{n})} \, \longleftrightarrow \, \frac{\partial}{\partial x_{\textbf{n}}}\,, \qquad d\phi \, \longleftrightarrow\, \prod_{\textbf{n} \in \mathbb{Z}_N^d} dx_{\textbf{n}}\,.
\end{equation}
With these scalar field notations, we can rewrite the Hamiltonian~\eqref{E:Heq2} as
\begin{equation}
\label{E:Heq3}
H = \left(\frac{L}{N}\right)^{\! d}  \left(- \frac{1}{2} \,\Delta_\phi + \mathcal{F}(\phi)\right)\,,
\end{equation}
where $\Delta_\phi := \nabla_\phi \cdot \nabla_\phi = \sum_{\textbf{n} \in \mathbb{Z}_N^d} \frac{\partial}{\partial \phi(\textbf{n})}$, and we write the ground state wavefunction as $\Psi_{\text{gs}}(\phi)$.  We will similarly write our variational wavefunction as $\Psi^\theta(\phi)$.

For our purposes, let us write the variational wavefunction as
\begin{equation}
\label{E:Psit1}
\Psi_t^{\theta}(\phi) = \frac{1}{\sqrt{Z_t^\theta}}\,e^{- S_t^\theta(\phi)/2}
\end{equation}
where $\Psi_{t = 0}^\theta(\phi)$ will ultimately correspond to our best variational approximation to the ground state of $H$.  Above, $S_t^\theta(\phi)$ is not an action as in the statistical field theory context, but is simply a convenient family of functions.  Note, however, that $S_t^\theta(\phi)$ can be chosen to be real on account of our discussion in Section~\ref{Subsec:realvalued}.  The factor $Z_t^\theta$ is a normalizing constant that we will not need explicitly, but such that the wavefunction is normalized as
\begin{equation}
\int d\phi \, |\Psi_t^{\theta}(\phi)|^2 = \frac{1}{Z_t^{\theta}} \int d\phi \, e^{-S_t^\theta(\phi)} = 1\,.
\end{equation}
It will often be convenient to write
\begin{equation}
p_t^\theta(\phi) := |\Psi_t^{\theta}(\phi)|^2\,.
\end{equation}
The energy of the state $\Psi_t^{\theta}(\phi)$ can be written as 
\begin{equation}
\label{E:energyPsi1}
    \frac{1}{Z_t^\theta}\int d\phi\,e^{-S_t^\theta/2} H e^{-S_t^\theta/2} = \mathbb{E}_{\phi \sim p_t^\theta(\phi)}\!\left[-\frac{1}{2}\,\Delta_\phi  S_t^\theta + \frac{1}{4}\,(\Grad_\phi S_t^\theta)^2 +\mathcal{F}(\phi) \right].
\end{equation}
One can compute the $\theta$-gradient of this function using the formula $\Grad_\theta \,e^{-S_t^\theta/2} = e^{-S_t^\theta/2} (-\Grad_\theta S_t^\theta/2)$. Unfortunately, this means that we need estimates for $S_t^\theta$ to compute the gradients of the function; fortunately, such estimates are provided by using normalizing flows and Neural ODE methods. We present one possible approach below. Recall that the probability flow ODE associated to \eqref{eq:finite-dimensions-inverse-Carosso-sde} is 
\begin{equation}
\label{eq:reverse-carosso-ode}
\frac{d\phi}{dt} = \Delta \phi - \frac{1}{2} \sigma^2 s^\theta_t(\phi) =: b_\theta(\phi,t)\,,
\end{equation}
with $\epsilon = L/N$, and $s_t^\theta := \nabla_\phi \log p_t^\theta$.  Supposing that $p_\infty(\phi)$ is some distribution of our choice which is easy to sample from, we have the following variational algorithm for determining $\Psi_t^\theta(\phi)$: 
\\ \\
\textbf{initialize} $\theta = (\theta^1,...,\theta^M)$, $i = 0$ \\
\textbf{if} $i < i_{\max}$ \textbf{do} \\
\indent sample $\phi'_T \leftarrow p^\theta_T = p_\infty$ \\
\indent compute $\phi'_0$ as well as  $\delta\theta(0)$ via an ODE solver, namely, by solving \eqref{eq:reverse-carosso-ode} as well as the system \\
\indent \indent \indent from $t=T$ to $t=0$ with initial condition $x = \phi_T'$, $\delta_{\theta^k} x(T) = 0,\,\,\delta \theta^k(T) = 0$ for all $k$: \\
\begin{align*}
\qquad \quad \frac{dx}{dt} &= b_\theta(x,t)\,, \,\,
\frac{d (\delta_{\theta^k} x)}{dt} = \frac{\partial}{\partial \theta^k}\, b_\theta(x,t)\,,\,\, \frac{d (\delta \theta^k)}{dt} = \nabla_x \!\cdot\! \left(\frac{\partial}{\partial \theta^k}\, b_\theta(x,t)\right) + \sum_\ell \sum_m b^\ell(x,t) \,\delta_{\theta^m}x\,,
\end{align*}
\indent set $\bar{\theta}_0 \leftarrow (\delta\theta^1(0), \ldots, \delta \theta^M(0))$\\
\indent set
    $\theta_0 \leftarrow \bar{\theta}_0 \left(-\Grad_\phi \cdot s_0^\theta(\phi)/2 + |s_0^\theta(\phi)|^2 + \mathcal{F}(\phi)\right)|_{\phi = \phi'_0} +  \left(-\Grad_\theta \Grad_\phi \cdot s_0^\theta(\phi) - \Grad_\theta |s_0^\theta(\phi)|^2\right)|_{\phi = \phi'_0}$ \\
\indent \textbf{for} $j=1, \ldots, \tau$  \\
\indent \qquad sample $t_j \sim [0,T]$ \\
\indent \qquad sample $\phi_j \leftarrow p_{t_j}$ using \eqref{eq:carosso-kernel} \\
\indent \qquad compute $\Grad_\theta \log p^\theta_{t_j}(\phi_j)$ by calling an ODE solver, namely:\\
\indent \qquad \qquad (i) solve $dx/dt = b_\theta(x, t)$ from $t=t_j$ to $t=T$ with initial condition $x_t = \phi_t$\\
\indent \qquad \qquad (ii) set $x_T \leftarrow x(T)$\\
\indent \qquad solve the following system from $t = T$ to $t = t_j$ for $k = 1,...,M$\,: 
\begin{align*}
\qquad \quad \frac{dx}{dt} &= b_\theta(x,t)\,, \,\,
\frac{d (\delta_{\theta^k} x)}{dt} = \frac{\partial}{\partial \theta^k}\, b_\theta(x,t)\,,\,\, \frac{d (\delta \theta^k)}{dt} = \nabla_x \!\cdot\! \left(\frac{\partial}{\partial \theta^k}\, b_\theta(x,t)\right) + \sum_\ell \sum_m b^\ell(x,t) \,\delta_{\theta^m}x\,,
\end{align*}
\indent \qquad \qquad with initial conditions $\delta_{\theta^k} x(T) = 0,\,\,\delta \theta^k(T) = 0$ for all $k$ where $\delta \theta^k(t) := \frac{\partial}{\partial \theta^k}\log p_t^\theta(x(t))$ \\
\indent \indent set $\theta_j \leftarrow (\delta\theta^1(t_j), \ldots, \delta \theta^M(t_j))$\\
\indent \textbf{end for} \\
\indent set $\theta \leftarrow \theta - \sum_{j=0}^{\tau} \theta_j $ \\
\textbf{end if} \\
\textbf{return} $\theta$
\\ \\
Note that while the algorithm provides us with $\Psi_{t = 0}^\theta(\phi)$ as an approximation to the ground state $\Psi_{\text{gs}}(\phi)$, we also end up with $\Psi_{t}^\theta(\phi)$ for various values of $t$, which can be thought of as (an approximation to) the ground state RG-flowed by different amounts.

\section{Numerically learning RG flows}
\label{sec:numerics}

\subsection{Overview}

We tested the methods of Section \ref{sec:renormalizing-diffusion-models} on the simple examples of 2D lattice scalar field theories. Specifically, we tested that the normalizing-flow based method based on optimizing~\eqref{eq:modified-objective} can learn the RG flow of such theories under the Carosso scheme~\eqref{E:Carossointro1v2}. We parameterized the normalizing flow $b^\theta_t$ of~\eqref{eq:neural-ode} using the neural network architecture proposed by~\cite{miranda_sampling}. We used a batch size of $64$, namely we computed gradient updates for $64$ independent samples treated as in the algorithm described below~\eqref{eq:modified-objective}, and fed the averaged gradients into the Adam optimizer~\cite{kingma2014adam}. For all experiments, we used hyperparameters $b_1 = 0.8$, $b_2=0.9$ for the optimizer provided by the Optax package~\cite{deepmind2020jax}, and used an exponentially decaying learning rate with initial value $0.005$ followed by  $8000$ transition steps and decay rate $0.1$.

We show that this method learns the flows of basic physical quantities like the renormalized mass. We present our lattice field theory conventions, as well as the estimators we use for the relevant physical quantities, in Section~\ref{sec:lattice-field-theory-numerics-conventions}. We plot the flows of these quantities using the Carosso and Polchinski SDEs in Section~\ref{sec:polchinski-vs-carosso-numerics} to get a sense of the kinds of differences that are possible when using these two radically different schemes. 

We found that it is difficult to learn the RG flows of field theories when the RG flow is defined using the Polchinski scheme. This is likely because the flow of the Polchinski scheme rapidly compresses the support of the distribution $p_t$ to a very low-dimensional manifold, and the inverse flow has trouble spreading out this low-dimensional prior to the UV distribution over lattice fields, which is supported everywhere.  These issues with the Polchinski scheme might be ameliorated by directly incorporating field rescalings and spatial/momentum rescalings into the flow equations, as discussed around~\eqref{E:Carossobshift2}.

With the Carosso scheme, we found that learning the RG flow was not possible unless a small ``mass'' term was added to modify the RG SDE to 
\begin{equation}
\label{E:CMmodified1}
\partial_t \phi_t(\textbf{n}) = (\Delta - M^2)\phi(\textbf{n}) +  \eta_t(\textbf{n})\,,\qquad \phi_0(\textbf{n}) = \phi(\textbf{n})\,,
\end{equation}
which the noise distribution unchanged.  We previously discussed this modification around~\eqref{eq:carosso-kernel-v2}.  Note that the unmodified Carosso scheme does not change the \emph{mean value of the field}, i.e.~the distribution over paths in field-space $\phi_t$ induced by~\eqref{E:Carossointro1v2} is invariant under $\phi_t \mapsto \phi_t + C$ for any constant $C$. As such, there is a $1$-dimensional line of limiting distributions $p_\infty$; in contrast, with the above modification in~\eqref{E:CMmodified1}, there is only a single limiting distribution. We use $M = 1$ for the experiments below; we find that the qualitative features of all plots are not sensitive to the value of $M$ so long as it is sufficiently small.  We also make a similar $M$-modification to the Polchinski scheme.  The corresponding transition kernels of the $M$-modified Carosso and Polchinski schemes are~\eqref{eq:carosso-kernel-v2} and~\eqref{E:polchinski_kernel} with the $K_t(|\widehat{\textbf{p}}|^2)$ in~\eqref{E:modifiedK1}.

\subsection{Conventions and estimators}
\label{sec:lattice-field-theory-numerics-conventions}

Let us recapitulate our conventions for scalar $\phi^4$ theory in two dimensions.  We have the probability distribution
\begin{align}
\label{E:phi4_v1}
p_0(\phi) &= \frac{1}{Z}\,\exp\!\Bigg(- L^2 \sum_{\textbf{p} \in \mathbb{Z}_N^2} \frac{1}{2}\,\widetilde{\phi}(\textbf{p}) (|\widehat{\textbf{p}}|^2 + m^2)  \widetilde{\phi}(-\textbf{p}) \nonumber \\
& \qquad \qquad \qquad - L^2 \sum_{\textbf{p}_1,\textbf{p}_2,\textbf{p}_3,\textbf{p}_4 \in \mathbb{Z}_N^2} \! \frac{\lambda}{4!}\,\widetilde{\phi}(\textbf{p}_1) \widetilde{\phi}(\textbf{p}_2) \widetilde{\phi}(\textbf{p}_3) \widetilde{\phi}(\textbf{p}_4)\, \delta_{\textbf{p}_1 + \textbf{p}_2 + \textbf{p}_3 + \textbf{p}_4, \textbf{0}} \bigg),
\end{align}
where $m$ and $\lambda$ are the initial bare mass and bare coupling, respectively.  In our numerics below, we present statistical estimates for four quantities along the RG flow of a scalar $\phi^4$ theory: (i) the renormalized mass $m_R$, (ii) the wave function renormalization $Z$, (iii) the renormalized coupling $\lambda_R$, and (iv) a diagnostic of field excursions $\max_{\textbf{n}} |\phi(\textbf{n})|$.

We will define the renormalized couplings $m_R$ and $\lambda_R$ in the standard way, which we review here.  In the $M$-modified Carosso scheme (and the $M$-modified Polchinski scheme using the $K_t(|\widehat{\textbf{p}}|^2)$ in~\eqref{E:modifiedK1}), an effective cutoff scale corresponding to $t$ is
\begin{align}
\Lambda_t = \max\left\{\min\left\{\sqrt{\frac{1}{t}-M^2}, \,\Lambda_0\right\},0\right\}\,.
\end{align}
Defining $r_t := \frac{\Lambda_t}{\Lambda_0}$, we have
\begin{align}
r_t &= \frac{\Lambda_t}{\Lambda_0} := \max\left\{\min\left\{\sqrt{\frac{1}{\Lambda_0^2 t}-\frac{M^2}{\Lambda_0^2}}, \,1\right\},0\right\}\,.
\end{align}
The Carosso scheme does not explicitly account for rescaling momenta by $\textbf{p} \to r_t\, \textbf{p}$ and fields by $\widetilde{\phi} \to \widetilde{\phi}/r_t^{2}$ (i.e.~accounting for non-anomalous dimensions of $\widetilde{\phi}$ in $d=2$).  However, these rescalings are necessary to provide the usual definitions of the renormalized couplings, so we need to include these rescalings by hand.  We incorporate these rescalings below.

To estimate our renormalized couplings, we need to define a few correlation functions.  The first is the momentum space 2-point function, namely
\begin{align}
\widetilde{G}_{2}^{\text{Carosso}}(\textbf{p}_1, \textbf{p}_2) := \langle \widetilde{\phi}(\textbf{p}_1) \widetilde{\phi}(\textbf{p}_2)\rangle\,.
\end{align}
The next one is the momentum space connected 4-point function, given by
\begin{align}
\widetilde{G}_{4,\,\text{conn}}^{\text{Carosso}}(\textbf{p}_1, \textbf{p}_2, \textbf{p}_3, \textbf{p}_4) &:= \langle \widetilde{\phi}(\textbf{p}_1) \widetilde{\phi}(\textbf{p}_2) \widetilde{\phi}(\textbf{p}_3) \widetilde{\phi}(\textbf{p}_4)\rangle - \langle \widetilde{\phi}(\textbf{p}_1) \widetilde{\phi}(\textbf{p}_2) \rangle \langle \widetilde{\phi}(\textbf{p}_3) \widetilde{\phi}(\textbf{p}_4)\rangle \nonumber \\
& \quad - \langle \widetilde{\phi}(\textbf{p}_1) \widetilde{\phi}(\textbf{p}_3) \rangle \langle \widetilde{\phi}(\textbf{p}_2) \widetilde{\phi}(\textbf{p}_4)\rangle  - \langle \widetilde{\phi}(\textbf{p}_1) \widetilde{\phi}(\textbf{p}_4) \rangle \langle \widetilde{\phi}(\textbf{p}_2) \widetilde{\phi}(\textbf{p}_3)\rangle\,.
\end{align}
For small $|\widehat{\textbf{p}}|$, the 2-point function goes as
\begin{align}
\widetilde{G}_2^{\text{Carosso}}(\textbf{p},-\textbf{p}) \approx \frac{Z}{|\widehat{\textbf{p}}|^2 + r_t^2 \, m_R^2}\,.
\end{align}
Then we can formulate estimators for $m_R$ and $Z$ by
\begin{align}
\boxed{\frac{1}{m_R^2} = \xi^2 \approx \frac{r_t^2}{4 L^2} \sum_{\textbf{p} \in \{(1,0), (0,1), (-1,0), (0,-1)\}} \frac{1}{\frac{4 N^2}{L^2} \sin^2(\pi/N)} \!\left(\frac{\widetilde{G}_2^{\text{Carosso}}(\textbf{0},\textbf{0})}{\widetilde{G}_2^{\text{Carosso}}(\textbf{p},-\textbf{p})} - 1\right)}
\end{align}
and
\begin{align}
\boxed{Z \approx r_t^2\, m_R^2\,\widetilde{G}_2^{\text{Carosso}}(\textbf{0},\textbf{0})}
\end{align}
where $\xi = 1/m_R$ is known as the correlation length.  For the renormalized quartic coupling we use the standard estimator (see e.g.~\cite{wolff2014triviality} or~\cite{vierhaus2010simulation} and references therein)
\begin{align}
\boxed{\lambda_R \approx - r_t^2\,(r_t m_R L)^2 \,\frac{\widetilde{G}_{4,\,\text{conn}}^{\text{Carosso}}(\textbf{0},\textbf{0},\textbf{0},\textbf{0})}{\widetilde{G}_2^{\text{Carosso}}(\textbf{0},\textbf{0})}}
\end{align}
Finally, to estimate extreme field excursions, we will also estimate
\begin{align}
\boxed{\max_{\textbf{n} \in \mathbb{Z}_N^2} |\phi(\textbf{n})|}
\end{align}

\subsection{The Carosso and Polchinski flows}
\label{sec:polchinski-vs-carosso-numerics}

\begin{figure}[t!]
    \centering
\includegraphics[width = \textwidth]{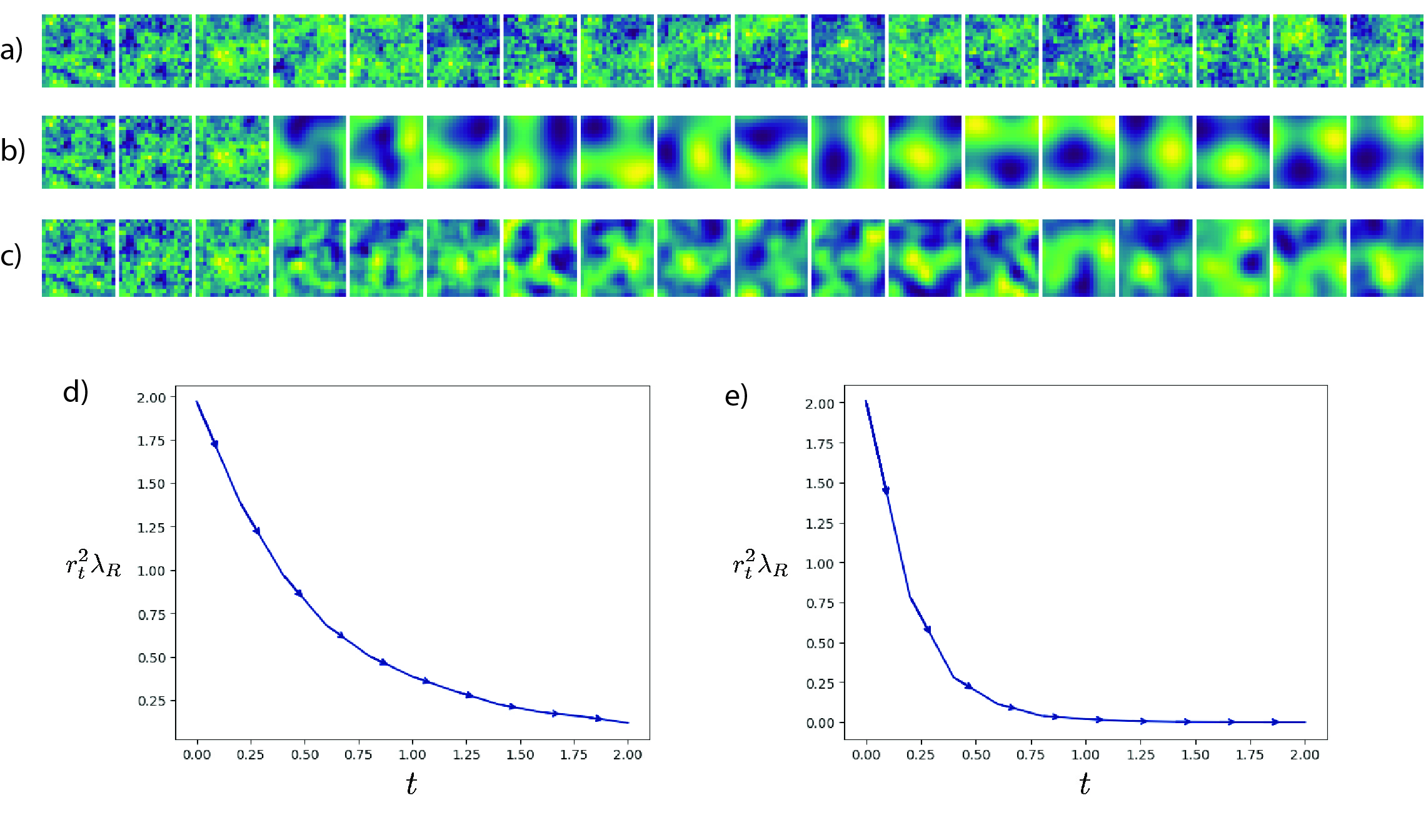}
    \caption{\emph{ERG schemes can be qualitatively and numerically different but physically equivalent.}  \textbf{Top:} Flows of samples from lattice $\phi^4$ theory in two dimensions (with $L=1$, $N = 20$, $m^2 = -2$, $\lambda = 0.0017$) via (a) the Carosso scheme and (b) the Polchinski scheme.  (c) The third row slows down the Polchinski flow, corresponding to the first few frames of (b). Note how the Polchinski flow cuts of high frequencies extremely rapidly, making it difficult to train models that learn the Polchinski flow. \textbf{Bottom:} Plot of renormalized quartic coupling (rescaled by $r_t^2$ for clarity) under the Carosso flow (d) and the Polchinski flow (e), over the entire scale range of flows from the top of Figure in (a) and (b). Flows of couplings are at similar rates to within a small factor, despite the underlying samples being very different in nature, as expected from the ERG formalism.  Note that $r_t^2 \lambda_R$ goes to zero for large $t$ in each scheme, although in the Carosso scheme $t = 2$ is not long enough to see $r_t^2 \lambda_R$ reach near zero.}
    \label{fig:polchinski-and-carosso-flow}
\end{figure}

In Figure~\ref{fig:polchinski-and-carosso-flow}, we show plots of the flows of some of the above estimators using the Carosso and the Polchinski schemes (with the $M$-modifications discussed previously). We see that while UV samples look completely different when flowed using the Carosso scheme or the Polchinski scheme, the resulting flows of estimators of physical quantities are qualitatively similar. In particular, both the mass and the quartic couplings flow at similar rates; the quartic coupling (multiplied by $r_t^2$, for ease of visualization) goes to zero slightly faster in the Polchinski scheme than in the Carosso scheme, but only by a factor of two or so.  In the Carosso scheme, $r_t^2 \lambda_R$ ultimately goes to zero for larger values of $t$ than we show in the plots.

Thus, by picking a convenient exact renormalization group scheme, we are still able to get physically meaningful results while dramatically improving the numerical stability of the training process for the sampler. This highlights the importance of tuning the RG scheme, just as the corresponding tuning of the noising process is necessary to ensure good performance of generative models of images.

$$$$

\newpage
\subsection{Numerics}
\label{sec:learning-carosso-numerics}

\begin{figure}[h!]
    \centering
\includegraphics[width = .9\textwidth]{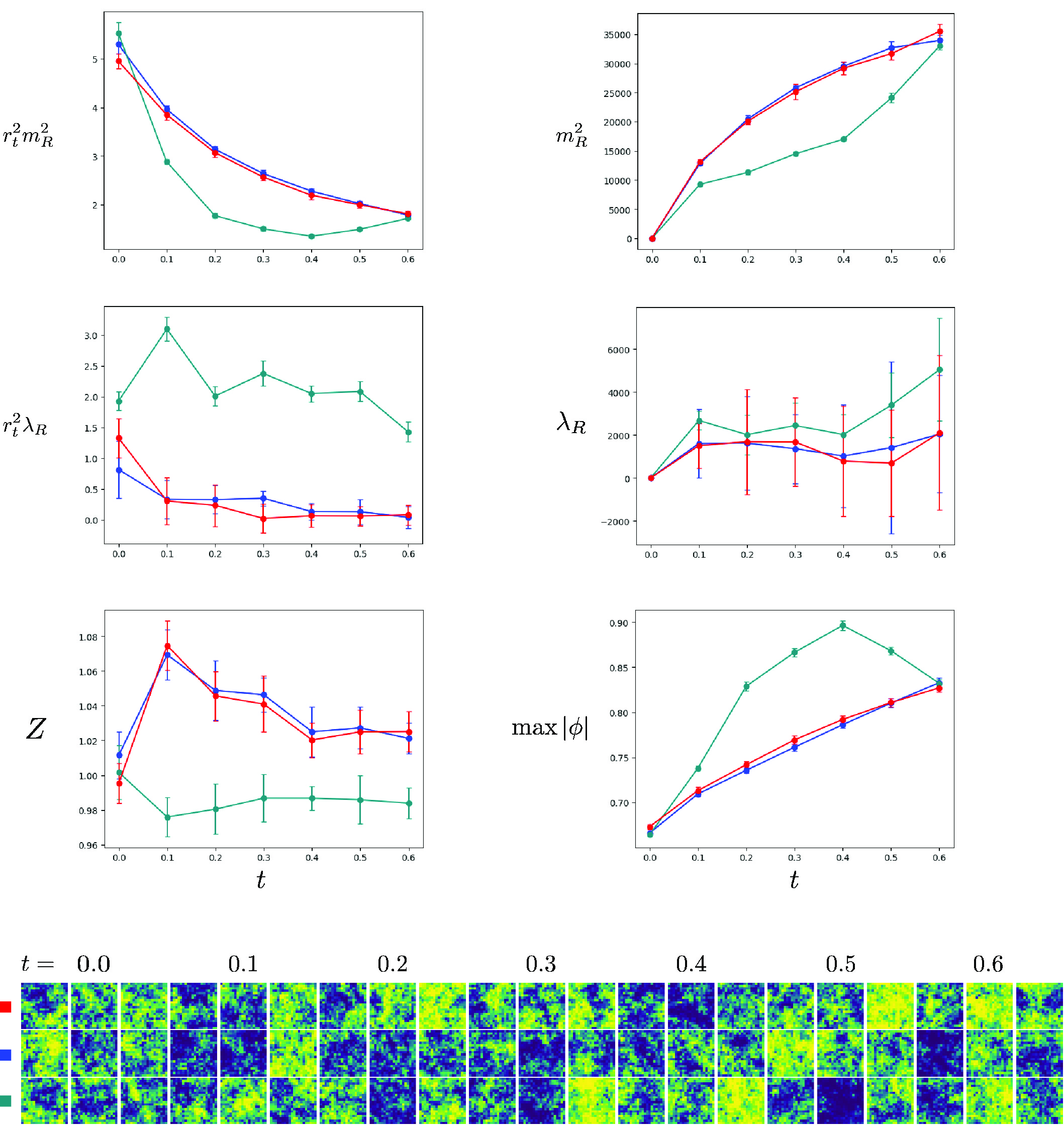}
\vspace{18pt}
    \caption{\emph{Learning RG Flow of $\phi^4$ theory with bare parameters $L=1, N=20, m^2=0.01, \lambda = 0.0017$.} \textbf{Top:} We show the true Carosso RG flow of the estimated parameters of samples from $\phi^4$ theory (\textcolor{red}{red}), as well as the same parameters computed from samples taken  directly from the flow learned using our objective~\eqref{eq:modified-objective} (\textcolor{blue}{blue}) and using the reverse KL objective~\eqref{eq:normalizing-flow-gradient-estimator} (\textcolor{teal}{teal}). Our learned flow is consistently closer to the true flow than the flow learned by the reverse KL objective. The top left and center left denote the physical quantities with $r_t^2$ rescaling removed. Note that renormalized couplings (top right, middle right) grow with RG time, since the couplings are relevant in two dimensions. \textbf{Bottom:} We show samples along the true flow (\textcolor{red}{red}) and the two learned flows (\textcolor{blue}{blue}, \textcolor{teal}{teal}).}
    \label{fig:numerics_alternate_params}
\end{figure}

\begin{figure}[h!]
    \centering
\includegraphics[width = \textwidth]{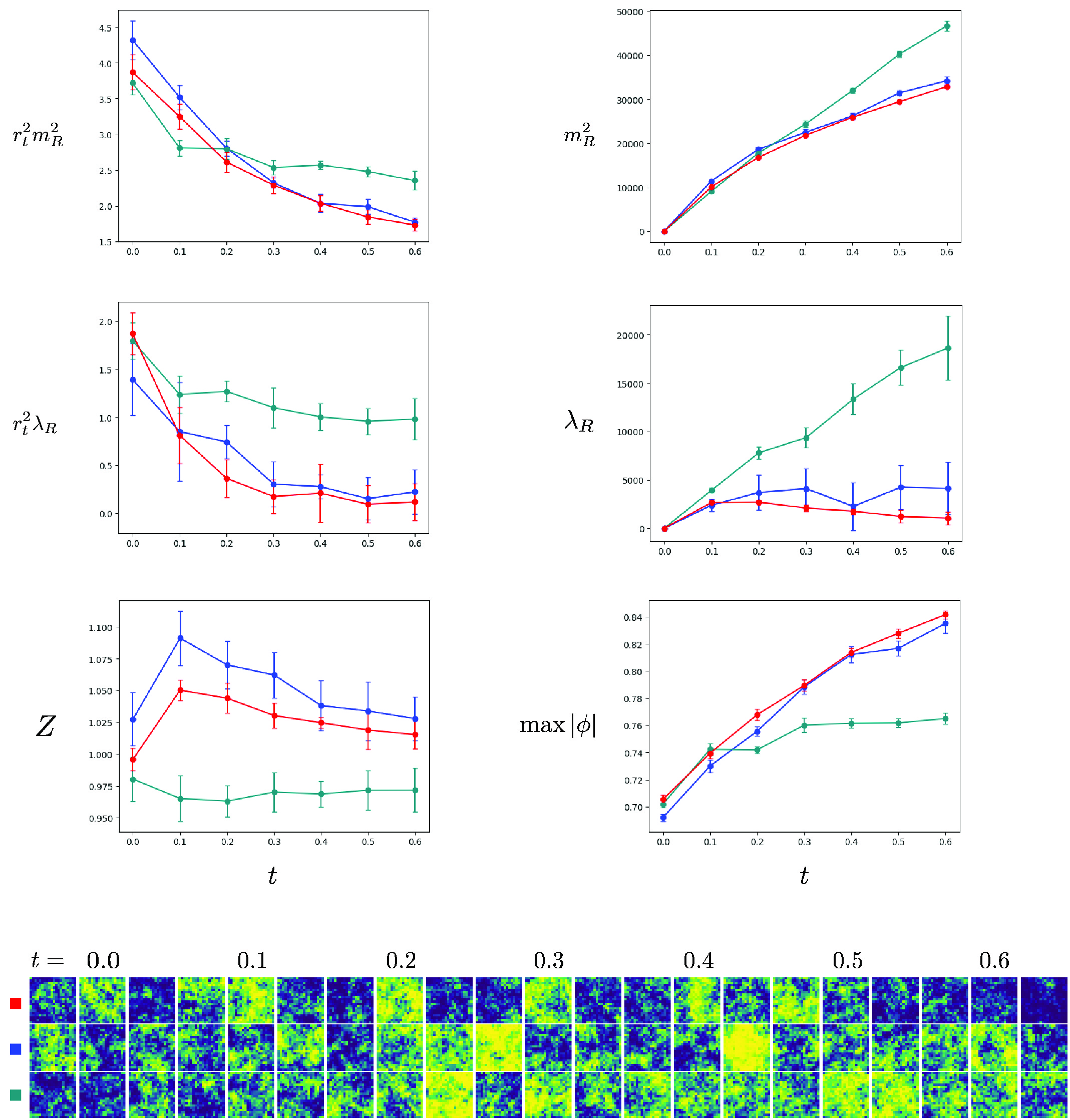}
    \caption{\emph{Learning RG Flow of $\phi^4$ theory with bare parameters $L=1, N=20, m^2=-2, \lambda = 0.0017$. } This Figure is laid out similarly to Figure~\ref{fig:numerics_alternate_params}. Note that the renormalized quartic coupling is a difficult parameter to estimate for all models, as is well-known in lattice field theory.}
    \label{fig:numerics_jordan_params}
\end{figure}

We also compared the normalizing flows trained using the reverse KL objective, as in~\eqref{eq:normalizing-flow-gradient-estimator} (following~\cite{flows_for_field_theory}), with the flows trained using our objective \eqref{eq:modified-objective}. In Figures~\ref{fig:numerics_alternate_params} and~\ref{fig:numerics_jordan_params}, we plot the estimated flows of the physical quantities described above, as well as the flows of the same quantities estimated from samples drawn directly from the learned normalizing flows.  In all cases, we trained the reverse KL model and the model based on ~\eqref{eq:modified-objective} for the same number of gradient steps (1000). To compute estimates of the flows of the physical quantities, we performed sampling from the UV distribution using the NUTS sampler~\cite{hoffman2014no}, as implemented in the Blackjax package~\cite{blackjax2020github}, with an initial warm-up and tuning segment of 4000 time-steps using the ``window adaptation'' method (the standard adaptation method used in Stan~\cite{stan}) implemented in~\cite{blackjax2020github}. We then evolved forwards in $t$ with the Carosso scheme using the kernels derived in this paper. 

We found that according to a variety of estimates of the flows of physically-important quantities, our objective caused the learned normalizing flow to more accurately reflect the physical behavior of the RG scheme, while the method based on optimizing the reverse KL divergence did not reliably lead to any agreement with the RG flow. However, the reverse KL divergence was often able to fit the relevant estimators for the UV distribution, i.e.~at RG time $t = 0$. Moreover, we found that this difference in learned flows, which is evident from the flows of physical quantities, cannot by detected by observing sampled field configurations by eye. This may come as no surprise to those who study lattice field theory, but in the context of machine learning, visual comparisons often play a significant role in identifying a successful model. We found that estimators of physical quantities were very sensitive to the distribution we tried to learn, and that looking at the behavior of these estimators under RG flow allowed us to quickly identify bugs in our training code, since we knew what qualitative behavior to expect from the physics of the $\phi^4$ model. 

We also note that in all of the experiments shown, we used the limiting distribution for the Carosso scheme as the prior distribution for the sampler, even on the samplers trained using the reverse-KL objective. While this is necessary for the consistency of the forwards-KL samplers, it is not required for the reverse-KL samplers. In fact, it is more common~\cite{flows_for_field_theory, miranda_sampling} to use a white noise distribution as a prior for a reverse-KL trained sampler.  If we compared our flows (which have a Carosso-type prior) to flows generated by reverse-KL-trained samplers with a white noise prior, we would find that the reverse-KL-trained samplers would be significantly worse at learning the RG flow of the true field theory than the reverse-KL-trained flows illustrated in Figures~\ref{fig:numerics_alternate_params} and \ref{fig:numerics_jordan_params}; thus our method would appear to perform significantly better.  Instead, we opted to make a more challenging comparison (with respect to the problem of learning the RG flow of the field theory) by comparing our method with reverse-KL samplers trained with a Carosso-type prior.  The Carosso-type prior encourages the reverse-KL-trained normalizing flow to match the `true' RG flows because we are effectively forcing the flows induced by the samplers to have the correct initial and final points; since the learned flows are continuous, they will at least qualitatively look like the `true' RG flows.  Nevertheless, even with our more careful comparison, we still show a significant improvement of our method over previous ones in the context of matching to RG flows.

\newpage
\newpage

\section{Discussion}
\label{sec:discussion}

In this paper, we have described a methodology for designing machine-learning-based methods which \emph{learn the RG flows of field theories}, in order to compute quantities of interest in physics. The basis of the methodology is a precise connection between the mathematics of diffusion models and exact renormalization group equations. As we have demonstrated, one can leverage this connection to design neural networks which correspond to \emph{particular} RG schemes, have \emph{physically-interpretable} model parameters, and yet take the form of \emph{conventional} machine learning models. This class of models offers a powerful platform for combining physically grounded modeling intuitions with the rich numerical methods developed by the machine-learning community, and validates the hypothesis of \cite{cotler2022renormalization} that a variational formulation of the RG flow can be used to build effective and interpretable ML models for field theory. 

It is a fundamental challenge in the scientific use of ML models that ML techniques tend to lack physical interpretability, and thus are difficult to reason about and debug. This is made more challenging by a bewildering collection of numerical techniques which are comparatively difficult to evaluate systematically, and can exhibit a variety of performance characteristics which are hard to extrapolate. For example, although the performance of existing ML-based lattice samplers is promising, they indeed exhibit complex behavior \cite{2211.07541}. Pre-existing lattice methods like the multigrid \cite{PhysRevD.40.2035, endres2015multiscale}, cluster \cite{swendsen1987nonuniversal, wolff1989collective}, and worm \cite{prokof1998exact} algorithms arise from physically-motivated heuristics. We hope that further research on equally physically-grounded ML-based sampling algorithms will lead to improved performance characteristics.

Physically-meaningful machine learning models allow for novel use cases. For example, any of the renormalizing diffusion models described in this paper generates samples from RG flowed field theory configurations during training; as such, one can use estimates of physical observables computed from batches of such configurations produced during training to diagnose how the model is learning, since the RG flows of certain physical quantities are often known from other simulations or from analytical methods. Moreover, one can conceivably modify the parameterization of the score function in a way that is interpretable in Effective Field Theory \eqref{eq:eft-parametrization},  and the coefficients of low-order terms in the Effective Field Theory thus represented by the model may also be used for diagnostics. Beyond this, the fact that the model parameters are physically meaningful should allow for hyperparameter transfer; one can think of e.g.~scale setting in conventional lattice field theory as a physically-motivated method for transferring parameters between disparate models. Moreover, one should be able to use physical insight to engineer RG schemes that are particularly efficient for certain problems, including RG schemes given by nonlinear SDEs \cite{carosso_thesis}. Finally, since phase transitions are \emph{defined} by their behavior under the renormalization group, one can imagine methods which rigorously use an ML model to search for RG fixed points and RG trajectories between fixed points, thus automatically mapping out the phase diagram of a system; this would be in sharp contrast with existing ML methods for finding phase transitions, which use patterns in sampled configurations to train classifiers which can be heuristically thought of as learning order parameters \cite{carrasquilla2017machine}, rather than using the definition of a phase transition from RG. 

More broadly, we see the design of machine learning methods for the physical sciences \emph{built around the ideas of EFT} as an exciting research program. Multiscale modeling is pervasive in domains like cosmology \cite{Vogelsberger2020} and plasma physics \cite{10.1088/978-0-7503-3559-1}, in which effective field theories connect a series of different physical models describing a system at different scales. While our paper focuses on describing and validating the basic properties of a class of ML models which can be applied to statistical field theories, we believe that the mathematical ideas of this work should be adapted to other domains, such as hydrodynamics and molecular simulation. Indeed, the problem of rigorously learning correction terms to coarse-grained models (e.g.~Maxwell-Vlasov corrections to MHD) is central to such simulations. We did not tackle those domains in this work because the formalism of the renormalization group is less fully developed in those domains, but we see this as an exciting next step, which would require developing the mathematical and computational formalism of the renormalization group beyond its most established setting. Such methods would be of considerable practical importance. An initial direction to develop such methods would be to take `limits' or `quantizations' of the methods described in this paper.  This an exciting \emph{theoretical} problem, not just a computational one. We hope that the union of Effective Field Theory and numerical techniques coming from machine learning can help provide a solution for the problem of unifying the traditional equation-based  modeling techniques from theoretical physics and the powerful non-parametric methods of modern machine learning.

\subsection*{Acknowledgements}

We would like to thank Daniel Ranard for valuable discussions, and Thomas Spencer for his interest in this work.  JC is supported by a Junior Fellowship from the Harvard Society of Fellows, as well as in part by the Department of Energy under grant DE-SC0007870.  SR is supported by NSF Mathematical Sciences Postdoctoral Fellowship, DMS-2202959. 

\appendix 

\section{A brief review of literature on diffusion models}
\label{App:diffusion_lit_review}

In this Appendix, we give a few brief pointers to the machine-learning literature on diffusion models for physicists interested in the subject. As mentioned in Section~\ref{sec:intro}, the papers \cite{sohl-dickstein_deep_2015, song2019generative} independently introduced the basic ideas regarding diffusion models. A practical unified perspective is in \cite{ddpg}, which forms the basis of most popular implementations of the methodology, and a more theoretical viewpoint is discussed in \cite{song2020score, song2021maximum}. A compendium of helpful formulae useful for studying diffusion models can be found in \cite{2208.11970}. 

From a practical perspective, the design of the neural network parameterizing the score function is very important, and should not be underestimated, as it encodes an implicit prior; many designs start with the U-Net \cite{ronneberger2015u}. Practictioners usually use comparatively simple SDEs (and it is found that the stochastic flow tends to be more efficient in practice than the deterministic flow \cite{ddpg}), and care is taken with the choice of SDE solver \cite{karras2022elucidating}. Moreover, while structured multiscale methods \cite{blurring_diffusion_models, mallat_diffusion, cascased_diffusion, iterative_refinrement, spectral_diffusion_processes} exist, many of the popular models such as Stable Diffusion \cite{stablediffusion} use methods which diffuse in the ``space of weights of a trained convolutional neural network'' \cite{rombach2022high}, which allows for efficient dimensionality reduction in a way adapted to the distribution of natural images (it is easier to model a distribution over a lower-dimensional space). The literature has evolved rapidly, and commercial implementations use many carefully-tuned optimizations in order to optimize for disparate objectives like inference time and `image quality', which are distinct from an accurate representation of the log-density. It is impossible to review the many applications of diffusion models to disparate domains like image, text, and audio synthesis, to scientific problems like super-resolution and denoising, and to the emergence of multimodal models; for a basic comprehensive review, we point the reader to \cite{2209.00796}. 

We also caution the reader that in fact many possible paths from a background distribution to the unknown distribution can be used to design generative modeling schemes, some with improved numerical properties \cite{liu2022flow}; as such, it is not clear what aspects of the diffusive framework are essential to the effectiveness of these models. A theoretical perspective and a review of alternative `stochastic interpolants' can be found in \cite{2303.08797}. While these alternative methods offer intriguing connections to optimal transport, they have not yet been shown to be superior than simple diffusion models in applications.

\section{Lattice discretization of functional derivatives}
\label{App:functionalderivs}

Here we collect some useful conventions and notations for functional derivatives in continuum field theory, and their analogs in lattice field theory.  Suppose we have a function $\phi : \mathbb{R}^d \to \mathbb{R}$, and a functional $\mathcal{F}[\phi]$.  Then the functional derivative of $\mathcal{F}$ is defined by
\begin{equation}
\frac{\delta}{\delta \phi(x)}\, \mathcal{F}[\phi(y)] := \lim_{\epsilon \to 0} \frac{\mathcal{F}[\phi(x) + \epsilon \,\delta^d(x-y)] - \mathcal{F}[\phi(x)]}{\epsilon}\,.
\end{equation}
For instance, if $g$ is a differentiable function on $\mathbb{R}$, then if $\mathcal{G}[\phi] = \int d x\, g(\phi(x))$ we have $\frac{\delta \mathcal{G}[\phi]}{\delta \phi(y)} = g'(\phi(y))$.

Now suppose we consider a lattice version of $\mathcal{G}[\phi] = \int d x\, g(\phi(x))$, namely
\begin{equation}
\label{E:Glattice1}
\mathcal{G}_{\text{lattice}}[\phi(\textbf{n})] = \left(\frac{L}{N}\right)^{\!d} \sum_{\textbf{n} \in \mathbb{Z}_N^d} g(\phi(\textbf{n}))\,,
\end{equation}
where we are considering a $d$-dimensional latticization of the hypercube with side length $L$, such that the lattice sites are $\frac{L}{N}\,\textbf{n}$ for $\textbf{n} \in \mathbb{Z}_N^d$.  We would like for the lattice analog of the functional derivative, when acting on $\mathcal{G}_{\text{lattice}}[\phi(\textbf{n})]$, to output $g'(\phi(\textbf{n}))$.  In other words, we would like to obtain the integrand of the sum~\eqref{E:Glattice1} (without the $\left(\frac{L}{N}\right)^{\!d}$ factor which is an analog of the integration measure $d x$) with an ordinary derivative acting on the $g$.  Then in our present notation (which we have used throughout the body of the manuscript), the continuum functional derivative must become
\begin{equation}
\label{E:latticefunctionalderiv1}
\frac{\delta}{\delta \phi(x)}\,\,\longrightarrow\,\,\left(\frac{N}{L}\right)^{\!d} \frac{\partial}{\partial \phi(\textbf{n})}\,.
\end{equation}
Then indeed, $\left(\frac{N}{L}\right)^{\!d} \frac{\partial}{\partial \phi(\textbf{n})}\,\mathcal{G}_{\text{lattice}}[\phi(\textbf{n})] = g'(\phi(\textbf{n}))$.  We have implicitly used the correspondence~\eqref{E:latticefunctionalderiv1} in the manuscript to translate between equations in the continuum and equations on the lattice.

We have to be careful when comparing $\frac{\partial}{\partial \phi(\textbf{n})}$ derivatives to their momentum space counterparts $\frac{\partial}{\partial \widetilde{\phi}(\textbf{p})}$.  For instance, consider the lattice functional
\begin{equation}
\mathcal{F}_{\text{lattice}}[\phi] := \sum_{\textbf{n}_1,...,\textbf{n}_k \in \mathbb{Z}_N^d} \phi(\textbf{n}_1) \cdots \phi(\textbf{n}_k)\,f(\textbf{n}_1,...,\textbf{n}_k)
\end{equation}
which can also be expressed in momentum space variables as
\begin{equation}
\mathcal{F}_{\text{lattice}}[\widetilde{\phi}] := \sum_{\textbf{p}_1,...,\textbf{p}_k \in \mathbb{Z}_N^d} \widetilde{\phi}(\textbf{p}_1) \cdots \widetilde{\phi}(\textbf{p}_k)\,\widetilde{f}(\textbf{p}_1,...,\textbf{p}_k)
\end{equation}
where $\widetilde{f}$ is the Fourier transform of $f$.  Then with the Fourier transform conventions in this paper, we have
\begin{equation}
\reallywidetilde{\left[\frac{\partial \mathcal{F}_{\text{lattice}}[\phi]}{\partial \phi}\right]}\!(p)= \frac{1}{N^d}\,\frac{\partial \mathcal{F}_{\text{lattice}}[\widetilde{\phi}]}{\partial \widetilde{\phi}(p)}\,,
\end{equation}
where the big tilde on the left denotes that we are taking the Fourier transform of the entire left-hand side.  We have used the above identity in various parts of the paper.

\section{Review of the Exact Renormalization Group}
\label{App:ERGreview}

In this Appendix we provide a brief review of the Exact Renormalization Group (ERG) in the continuum, with an emphasis on the Wegner-Morris equation.  Our expositions will follow along the lines of~\cite{cotler2022renormalization}.  Consider a probability functional $P_\Lambda[\phi]$ for a scalar field theory, which captures an effective description of the system given that we can only probe momentum scales below $\Lambda$.  An ERG flow addresses precisely how our effective description of the system changes as we change $\Lambda$.  In particular, ERG equations take the form
\begin{equation}
\label{E:appERG1}
- \Lambda \frac{d}{d\Lambda} \, P_\Lambda[\phi] = \mathcal{F}\!\left[P_\Lambda[\phi]\,,\,\frac{\delta P_\Lambda[\phi]}{\delta \phi}\,,\,\frac{\delta^2 P_\Lambda[\phi]}{\delta \phi \, \delta \phi},...\right]\,.
\end{equation}
Often it is convenient to instead work with the variable $t = - \log(\Lambda)$, in which case $-\Lambda\frac{d}{d\Lambda} \to \frac{d}{dt}$.  Moreover, decreasing $\Lambda$ corresponds to increasing $t$.  An equation of the form~\eqref{E:appERG1} with an initial condition $P_{\Lambda_0}[\phi]$ prescribes how the probability functional is RG flowed for all smaller values of $\Lambda$, namely the flow determines $P_\Lambda[\phi]$ for all $0 \leq \Lambda \leq \Lambda_0$.

Consider the moment-generating function (i.e.~the \textit{partition function}) for a Euclidean scalar field theory in $d$ spatial dimensions, namely
\begin{equation}
\label{E:partitionfunction1}
Z_\Lambda[J] := \int \mathcal{D}\phi\,e^{- \frac{1}{2} \int_{\mathbb{R}^d} \frac{d p}{(2\pi)^d}\left(\frac{1}{K_\Lambda(p^2)}\,\phi(p) \phi(-p) (p^2 + m^2) + J(p) \phi(-p)\right) - S_{\text{int},\Lambda}[\phi]}\,,
\end{equation}
where $K_\Lambda(p^2)$ is a soft cutoff function (i.e.~it equals $1$ for $p^2 \lesssim \Lambda^2$ and is approximately $0$ for $p^2 \gtrsim \Lambda^2$), and $S_{\text{int},\Lambda}[\phi]$ contains interaction terms.  We will toggle between the position-space representation of the scalar field (denoted by $\phi(y)$) and its Fourier transform comprising the momentum space representation (denoted by $\phi(p)$) throughout this discussion. 
 Note that the soft cutoff function ensures that correlation functions (at least perturbatively if the interaction terms are small) are regulated at high momentum.  Further observe that by taking the functional derivative of $\log(Z_\Lambda[J])$ with respect to $J$ we can compute moments of the probability distribution
\begin{equation}
\label{E:appP1}
P_\Lambda[\phi] \propto e^{- \frac{1}{2} \int_{\mathbb{R}^d} \frac{d p}{(2\pi)^d}\left(\frac{1}{K_\Lambda(p^2)}\,\phi(p) \phi(-p) (p^2 + m^2)\right) - S_{\text{int},\Lambda}[\phi]}\,.
\end{equation}
It will be convenient for us to define the \textit{action} of the above probability functional as
\begin{align}
\label{E:appS1}
S_\Lambda[\phi] := \frac{1}{2} \int_{\mathbb{R}^d} \frac{d p}{(2\pi)^d}\left(\frac{1}{K_\Lambda(p^2)}\,\phi(p) \phi(-p) (p^2 + m^2)\right) + S_{\text{int},\Lambda}[\phi]\,.
\end{align}

Now suppose we are interested in flowing $P_\Lambda[\phi]$ to some smaller scale $\Lambda_R < \Lambda$, so that we only are interested in correlation functions with momentum scale $p^2 \leq \Lambda_R^2$.  Further suppose that the source $J(p)$ in~\eqref{E:partitionfunction1} satisfies $J(p) = 0$ for $p^2 > \Lambda_R^2 - \varepsilon$ for a small $\varepsilon > 0$.  Then if we do not want the low-momentum correlation functions of $P_\Lambda[\phi]$ to change as we perform RG flow on the probability density, then we require
\begin{equation}
\label{E:applogderiv1}
-\Lambda \frac{d}{d\Lambda} \log(Z_\Lambda[J]) = 0\,.
\end{equation}
A class of ERG flows which achieve this condition is given by the \textit{Wegner-Morris} equation, namely
\begin{equation}
\label{E:appWM1}
- \Lambda \frac{d}{d\Lambda}\,P_\Lambda[\phi] = \int_{\mathbb{R}^d} d x \, \frac{\delta}{\delta \phi(x)}\left(\Psi_\Lambda[\phi,x]\,P_\Lambda[\phi]\right)\,,
\end{equation}
where $\Psi_\Lambda[\phi,x]$ is called the \textit{reparameterization kernel}, which is required to have certain properties which we discuss below.  We first note that~\eqref{E:appWM1} has a simple interpretation: as we change $\Lambda$ infinitesimally by $\delta \Lambda$, the flow induces a field reparameterization as
\begin{equation}
\label{E:WMreparam1}
    \phi'(x) = \phi(x) + \frac{\delta \Lambda}{\Lambda}\,\Psi_\Lambda[\phi,x]\,.
\end{equation}
This is why the reparameterization kernel $\Psi_\Lambda[\phi,x]$ has its name.  We see from~\eqref{E:WMreparam1} that in order for~\eqref{E:applogderiv1} to hold, we must have $\Psi_\Lambda[\phi,x]$ be a functional of $\phi$ that is supported at scales $p^2 \geq \Lambda^2$; that is, we are reparameterizing $\phi$ for momentum scales $p^2 \geq \Lambda^2$.  A standard form of the reparameterization kernel is (see e.g.~\cite{rosten2012fundamentals})
\begin{align}
\label{E:appPsikernel1}
\Psi_\Lambda[\phi,x] = - \int_{\mathbb{R}^d} d y\,\frac{1}{2}\,\dot{C}_\Lambda(x-y)\, \frac{\delta \Sigma_\Lambda[\phi]}{\delta \phi(y)}\,,
\end{align}
where here $\dot{C}_\Lambda(x-y)$ is a positive-definite kernel called the \textit{ERG kernel} which is localized around $p^2 = \Lambda^2$ in momentum space, and
\begin{equation}
\Sigma_\Lambda[\phi] := S_\Lambda[\phi] - 2 \widehat{S}_\Lambda[\phi]\,.
\end{equation}
Here $\hat{S}_\Lambda[\phi]$ is a local action called the \textit{seed action}, which is cut off for $p^2 \geq \Lambda^2$.

With the particular form of $\Psi_\Lambda[\phi,x]$ given~\eqref{E:appPsikernel1}, we can rewrite the Wegner-Morris equation~\eqref{E:appWM1} as a type of convection-diffusion equation, namely:
\begin{align}
- \Lambda \frac{\partial P_\Lambda[\phi]}{\partial \Lambda} = \frac{1}{2}\int_{\mathbb{R}^d} d x\, d y\,\dot{C}_\Lambda(x-y)\left(\frac{\delta^2 P_\Lambda[\phi]}{\delta \phi(x)\,\delta \phi(y)} + 2\,\frac{\delta}{\delta \phi(x)}\left(\frac{\delta \widehat{S}_\Lambda[\phi]}{\delta \phi(y)}\,P_\Lambda[\phi]\right)\right)\,.
\end{align}
A special case of the above gives the Polchinski flow~\cite{polchinski1984renormalization}.  In particular, if $\dot{C}_\Lambda(p^2)$ is the Fourier transform of the ERG kernel $\dot{C}_\Lambda(x-y)$, then the Polchinski flow is given by the specialization
\begin{align}
\dot{C}_\Lambda(p^2) &= (2\pi)^d (p^2 + m^2)^{-1} \Lambda \frac{\partial K_\Lambda(p^2)}{\partial \Lambda} \\
\widehat{S}_\Lambda[\phi] &= \frac{1}{2}\int_{\mathbb{R}^d}\frac{d p}{(2\pi)^d}\,\frac{1}{K_\Lambda(p^2)}\,\phi(p)\phi(-p)\,(p^2 + m^2)\,.
\end{align}
Notice that here, $\dot{C}_\Lambda(p^2)$ and $\widehat{S}_\Lambda[\phi]$ are both built out of the cutoff function $K_\Lambda(p^2)$ appearing in e.g.~\eqref{E:partitionfunction1},~\eqref{E:appP1},~\eqref{E:appS1}.  The Polchinski flow has the feature that the free (i.e.~Gaussian) theory
\begin{equation}
P_\Lambda^{\text{free}}[\phi] \propto e^{- \frac{1}{2} \int_{\mathbb{R}^d} \frac{d p}{(2\pi)^d}\left(\frac{1}{K_\Lambda(p^2)}\,\phi(p) \phi(-p) (p^2 + m^2)\right)}\,.
\end{equation}
satisfies the flow equation.

If we relax the assumptions on the form of the reparameterization kernel~\eqref{E:appPsikernel1}, then we can also write the Carosso scheme~\cite{Carosso2020} in the form of the Wegner-Morris equation~\eqref{E:appWM1}.  However, the modifications to the form of the reparameterization kernel will make it so that~\eqref{E:applogderiv1} only approximately holds, but this is ok.  Suppose that $\Lambda_0$ is the initial cutoff scale at which we begin the RG flow.  The relaxed assumptions on $\Psi_\Lambda[\phi,x]$ amount to relaxed assumptions on $\dot{C}_\Lambda(x-y)$ and $\widehat{S}_\Lambda[\phi]$, namely:
\begin{enumerate}
    \item \textit{The cutoff function} $\dot{C}_\Lambda(p^2)$. There is a non-decreasing function $g(\Lambda)$ for $\Lambda \geq 0$ with $g(\Lambda_0) = \Lambda_0$ such that $\dot{C}_\Lambda(p^2)$ goes rapidly to zero for $|p| \geq \Lambda_0 g(\Lambda)$.  We further require that $\dot{C}_\Lambda(p^2)$ is $O(1)$ for $|p| \leq \Lambda_0 g(\Lambda)$.
    \item \textit{The seed action} $\widehat{S}_\Lambda[\phi]$.  Construct a normalized probability density $Q_\Lambda[\phi] \propto e^{-2 \widehat{S}_\Lambda[\phi]}$.  This probability density has the property that 
    \begin{equation}
    \mathbb{E}_{\phi \sim Q_\Lambda[\phi]}[\phi(p_1) \phi(p_2) \cdots \phi(p_r)] \approx 0 \quad \text{for any}\,\,\,|p_i| \geq \Lambda_0 g(\Lambda)\,,
    \end{equation}
    where $g(\Lambda)$ is the same function which controls the cutoff function.  This means that correlation functions with momenta greater than $\Lambda_0 g(\Lambda)$ are suppressed.
\end{enumerate}
Our discussion here precisely mirrors the one in Section~\ref{subsubsec:WMlattice}, and in fact we have only lightly changed the language.

With these relaxed assumptions, the Carosso scheme is given by
\begin{align}
\dot{C}_\Lambda(p^2) = e^{-p^2/\Lambda_0^2}\,,\quad \widehat{S}_\Lambda[\phi] = \frac{1}{2}\int_{\mathbb{R}^d}\frac{d p}{(2\pi)^d}\,e^{p^2/\Lambda_0^2}\,\phi(p)\phi(-p)\,p^2\,,
\end{align}
which has $g(\Lambda) = \Lambda_0$.  As explained in the body of the text, and in Carosso's paper~\cite{Carosso2020}, these choices allow for a simple SDE formulation of the RG flow (although again at the cost of~\eqref{E:applogderiv1} being only approximately satisfied).

\bibliography{refs}
\bibliographystyle{JHEP}

\end{document}